%% file: draft.tex
\documentclass[aps,prxquantum,reprint,superscriptaddress,floatfix]{revtex4-2}

\usepackage[english]{babel}
\usepackage{amsmath,amssymb,bm,dsfont,graphicx}
\usepackage{booktabs}
\usepackage[table]{xcolor}
\usepackage[colorlinks=true,citecolor=blue,urlcolor=blue,linkcolor=blue]{hyperref}
\usepackage{tikz}
\usetikzlibrary{arrows.meta, shapes.geometric, calc}

\newcommand{\Tr}{\operatorname{Tr}}
\newcommand{\supp}{\operatorname{supp}}
\newcommand{\avg}[1]{\langle #1\rangle}
\newcommand{\tcell}[2]{\parbox[t]{#1}{\raggedright #2\strut}}
\newcounter{thmc}
\newenvironment{thmbox}[1]{\par\refstepcounter{thmc}%
  \begin{quote}\noindent\textbf{#1~\thethmc.}\ \itshape}%
  {\end{quote}\par}
\newcounter{cnjc}
  {\end{quote}\par}

\begin{document}

\title{Frustration-free control for quantum-state preparation:
an exact first-passage theory}

\author{T. D\"orstel}
\affiliation{Universit\"at Innsbruck, Institut f\"ur Theoretische Physik, Technikerstra{\ss}e 21a, 6020 Innsbruck, Austria}
\author{T. C. Lang}
\affiliation{Universit\"at Innsbruck, Institut f\"ur Theoretische Physik, Technikerstra{\ss}e 21a, 6020 Innsbruck, Austria}
\author{T. Iadecola}
\affiliation{Department of Physics, The Pennsylvania State University, University Park, Pennsylvania 16802, USA}
\affiliation{Institute for Computational and Data Sciences, The Pennsylvania State University, University Park, Pennsylvania 16802, USA}
\affiliation{Materials Research Institute, The Pennsylvania State University, University Park, Pennsylvania 16802, USA}
\author{J. H. Wilson}
\affiliation{Department of Physics and Astronomy, Louisiana State University, Baton Rouge, Louisiana 70803, USA}
\affiliation{Center for Computation and Technology, Louisiana State University, Baton Rouge, Louisiana 70803, USA}
\author{M. Buchhold}
\thanks{Contact author: michael.buchhold@uibk.ac.at}
\affiliation{Universit\"at Innsbruck, Institut f\"ur Theoretische Physik, Technikerstra{\ss}e 21a, 6020 Innsbruck, Austria}

\date{\today}

\begin{abstract}
Quantum circuits based on local measurements and feedback can autonomously prepare and stabilize entangled many-body states. In frustration-free control, the target manifold is specified by a set of local constraints, whose violations are detected by measurement and removed by feedback. We develop an exact theory of this mechanism for the preparation of Dicke states of $N_q$ qubits. Dicke states span the fully permutation-symmetric subspace, with a unique Dicke state in each fixed-magnetization sector. Local singlet projectors define the constraints and detect departures from permutation symmetry. The resulting dynamics closes exactly on the operator space spanned by the $N_q(N_q-1)/2$ singlet-pair projectors and is equivalent to a classical first-passage process: measurements move the pair endpoints, while feedback absorbs pairs on measured bonds. The closure reduces the exponentially large many-body Liouvillian to a generator of polynomial size and thereby provides a complete quantification of the preparation process at polynomial complexity. In one dimension we solve this process in closed form. Its slowest rate, $\Lambda_{\rm pair}=2[1-\cos(\pi/N_q)]$ in units of
the measurement rate, is the full Liouvillian gap. For any scheme of this class, the relaxation time is bounded below by the inverse fraction of pair configurations on which feedback acts; local feedback reaches $O(N_q)$ of $O(N_q^2)$ configurations and therefore requires a time at least linear in $N_q$. A second exact closure determines the final state reached from any initial condition. Extensions to modified transport, restricted feedback, and physical errors yield an exact transport-to-absorber crossover, geometric preparation laws, and a bound on the stationary leakage. An exact second-moment closure determines the ensemble fluctuations: the measurement noise is conserving, while feedback generates the shot noise of removal and correlations between pairs. Preparation, acceleration, robustness, and fluctuations are thus governed by one process, the transport and absorption of local constraint violations, which the closures make exactly solvable.
\end{abstract}

\maketitle

\begin{figure}[t]
 \centering
 \input{figures/overview}
\caption{\textbf{Frustration-free control as a first-passage problem.}
(a) A constraint violation is a pair $\hat P_{nm}$ with endpoints on sites
$n$ and $m$. A measurement on a bond moves the endpoint on that bond to its
other site, at half the measurement rate; feedback removes a pair whose two
endpoints form a measured bond, with strength $\lambda$. Transport and
removal together form the closed generator
$\mathsf K=\mathsf M+\mathsf A$, Eq.~\eqref{eq:Kdef}.
(b) The same process viewed in the configuration space of pairs:
measurements generate a random walk (blue), and the pairs whose endpoints
form a feedback bond are the absorber $\mathcal F$ (red). The slowest decay
rate is bounded by the fraction of the $N_p$ configurations that are
absorbing, $\Lambda_0\le\lambda|\mathcal F|/N_p$
(Proposition~\ref{prop:absorber}).
(c) Three ways of changing the process: faster endpoint motion by random
partial-SWAP events (Sec.~\ref{sec:scrambling}), a different absorber by
restricting the feedback geometry (Sec.~\ref{sec:localized}) or by
extending the measurement range, which changes motion and absorber together
(Sec.~\ref{sec:longrange}), and physical errors, which add a source of pairs
and an additional decay (Sec.~\ref{sec:imperfections}).}
 \label{fig:overview}
\end{figure}
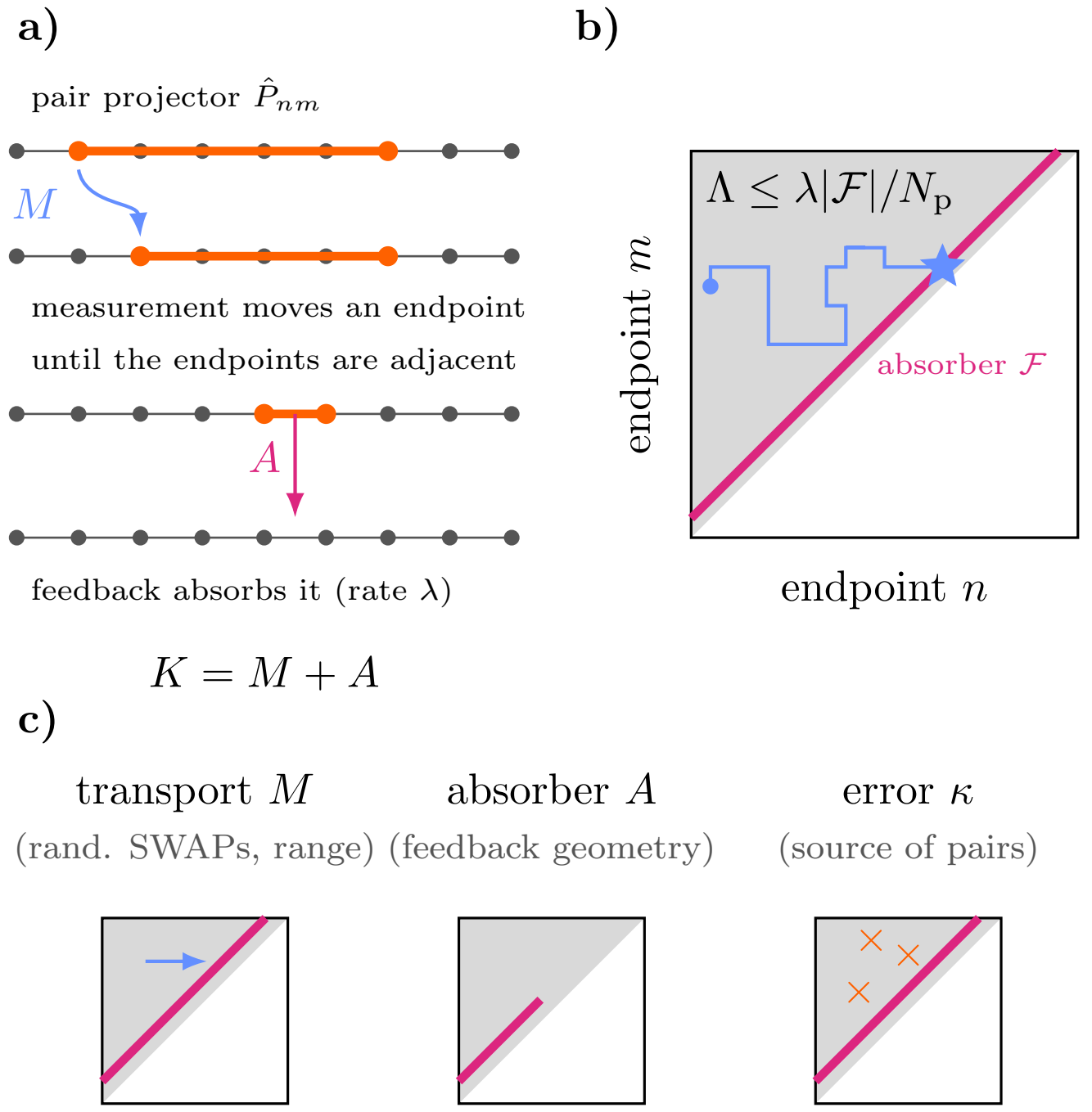

\section{Introduction}
\label{sec:intro}

Adaptive quantum circuits based on local measurements and feedback provide a route to the autonomous preparation and stabilization of entangled many-body states. By making the target state an attractor of the dynamics, the same protocol can prepare the target from arbitrary initial states within an invariant sector and restore it after perturbations~\cite{Kraus2008,Diehl2008,Verstraete2009,Johnson2017,Piroli2021,Harrington2022}. The necessary capabilities include mid-circuit measurements, real-time feedforward, and repeated coupling to local ancillas, all of which are now available in trapped-ion, neutral-atom, and superconducting platforms \cite{Barreiro2011,Lin2013,RyanAnderson2021,Bluvstein2024,Lee2026,Andersen2019}. This raises two questions: how do locality and the structure of measurement and feedback determine the speed and accuracy of preparation, and how do the time and resource requirements compare with those of unitary schemes?

Frustration-free control (FFC) implements such attracting dynamics through local measurements and feedback \cite{Doerstel2026,Hauser2024}. The target manifold is defined as the common kernel of local projectors $\{\hat P_\alpha\}$, and a state with $\langle\hat P_\alpha\rangle\neq0$ violates the local constraint $\alpha$. Measuring these projectors detects such violations, and conditional feedback removes them. FFC is constructed so that the prescribed target manifold is absorbing and is reached with probability one. The same projectors define a frustration-free parent Hamiltonian $\hat H=\sum_\alpha\hat P_\alpha$~\cite{Ticozzi2012}, which fixes the target manifold but not the dynamics by which it is reached. The preparation time depends on how measurement moves constraint violations, where feedback can remove them, and which quantities are conserved. Hence, it can be changed by the measurement or feedback geometry at fixed parent Hamiltonian. 

We apply this framework to the preparation of Dicke states of $N_q$ qubits. These states span the fully permutation-symmetric subspace, with a unique Dicke state in each fixed magnetization sector, and are central resources for quantum metrology and multipartite entanglement \cite{Dicke1954,Pezze2018,Hume2009}. Since the singlet is the only antisymmetric state of two qubits, permutation symmetry can be imposed locally through the singlet projectors $\hat P_{\ell m}=(\mathds{1}-\hat S_{\ell m})/2$, where $\hat S_{\ell m}$ exchanges qubits $\ell$ and $m$. On any connected measurement graph, the common kernel of the measured projectors is the fully symmetric subspace. A measurement on a bond moves one endpoint of a pair projector along that bond, and feedback removes a pair whose two endpoints form a measured bond. Dicke-state preparation is therefore governed by the transport and absorption of constraint violations, Fig.~\ref{fig:overview}. This protocol was introduced in Ref.~\cite{Doerstel2026}, which showed that the dynamics closes exactly on the operator space spanned by the pair projectors and derived the diffusive relaxation time $\tau\sim N_q^2$ of the nearest-neighbor chain together with its extension to long-range measurements; a complementary scaling picture for frustration-free targets was subsequently obtained from quasiparticle cooling \cite{Feldmeier2026}.

Here we further develop the operator closure picture of FFC: the central finding is that the approach to the target manifold, the final state, and the ensemble fluctuations can all be completely characterized through operator spaces of polynomial size that close under the dynamics. Closure here means that measurement and feedback map each operator space into itself. These closures are operator identities, so the resulting equations hold for arbitrary states. Three such structures determine the preparation process: (i) The $N_p=N_q(N_q-1)/2$ pair projectors $\{\hat P_{\ell m}\}$ close linearly, and their expectation values obey a classical first-passage process that governs the approach to the Dicke manifold. (ii) The kernel of the adjoint Lindbladian is spanned by $(N_q+1)^2$ collective-spin polynomials, whose conserved expectation values fix the unique stationary state reached from each initial condition, including coherences between magnetization sectors. (iii) We further establish the closure on the operator space containing products of pair projectors, which determines the ensemble covariance and the noise generated by measurement and feedback. The three spaces have dimensions $O(N_q^2)$, $O(N_q^2)$, and $O(N_q^4)$, respectively, replacing the exponentially large many-body problem with linear equations of polynomial size, and define the fluctuating absorbing hydrodynamics of constraint violations.

This fully characterizes the preparation dynamics in terms of two paradigmatic regimes of absorbing hydrodynamics: a transport-limited and an absorber-limited regime. In the transport-limited regime, measurements or additional unitary operations must move the pair endpoints until the pair reaches a bond on which feedback acts. In the absorber-limited regime, this motion is fast and removal by feedback sets the rate. We prove that the fraction of pair configurations accessible to feedback bounds the relaxation rate, regardless of how rapidly measurements or additional unitary operations move the endpoints. Local feedback acts on $O(N_q)$ of $O(N_q^2)$ pair configurations, so the relaxation time is at least linear in $N_q$. The bound is saturated in the absorber-limited regime. Accelerating pair motion drives a crossover from the transport-limited to the absorber-limited regime, at which the preparation time reaches the bound set by the feedback geometry; this crossover marks the optimal use of transport-generating resources, e.g., additional unitaries.

We compute exact solutions that connect these relaxation rates to preparation times through fidelity bounds. Extensions to single-spin errors determine the stationary leakage and recovery rate, which quantifies the accuracy of autonomous stabilization. We compare the resulting time and resource requirements with unitary and adaptive preparation schemes in Sec.~\ref{sec:broader}. Section~\ref{sec:summary} and Table~\ref{tab:roadmap} summarize the main results.

\begin{table*}[t]
\caption{\textbf{Overview of principal results.}}
\label{tab:roadmap}
\begingroup
% column widths: adjust here (sum should stay below about 0.96\textwidth)
\newcommand{\wRes}{0.52\textwidth}   % Result
\newcommand{\wEq}{0.19\textwidth}    % Equation or theorem
\newcommand{\wSec}{0.1\textwidth}   % Section
\newcommand{\wSt}{0.14\textwidth}    % Status
\newcommand{\cres}[1]{\tcell{\wRes}{#1}}
\newcommand{\ceq}[1]{\tcell{\wEq}{#1}}
\newcommand{\csec}[1]{\tcell{\wSec}{#1}}
\newcommand{\cst}[1]{\tcell{\wSt}{#1}}
\setlength{\tabcolsep}{4pt}
\renewcommand{\arraystretch}{1.08}
\begin{ruledtabular}
\small
\begin{tabular}{@{}llll@{}}
\cres{\textbf{Result}} & \ceq{\textbf{Equation or theorem}} & \csec{\textbf{Section}} & \cst{\textbf{Status}}\\
\hline
\cres{-- Pair closure on any connected measurement graph; mapping to an absorbing random walk}
& \ceq{Eqs.~\eqref{eq:exactpair}, \eqref{eq:Kdef}} & \csec{Sec.~\ref{sec:endpoint}} & \cst{exact solution}\\[0.3em]
\cres{-- Two-sided infidelity bound from the witness $\hat W$; state-independent preparation time}
& \ceq{Eqs.~\eqref{eq:fidelitybound}, \eqref{eq:Teps}} & \csec{Sec.~\ref{sec:certificate}} & \cst{exact bounds}\\[0.3em]
\cres{-- Liouvillian gap, $\Delta_{\rm Liou}=\Lambda_0$}
& \ceq{Eq.~\eqref{eq:gapidentity}} & \csec{Sec.~\ref{sec:liouvilliangap}} & \cst{exact identity}\\[0.3em]
\cres{-- Pair spectrum of the periodic chain, $\Lambda_{\rm pair}=2[1-\cos(\pi/N_q)]$}
& \ceq{Eq.~\eqref{eq:closedgap}} & \csec{Sec.~\ref{sec:spectrum}} & \cst{exact solution}\\[0.3em]
\cres{-- Preparation laws for bulk, boundary, and finite-region feedback in $d$ dimensions, transverse dimension $n=2d-d_{\rm fb}$}
& \ceq{Eqs.~\eqref{eq:codim}, \eqref{eq:capacity}; Table~\ref{tab:geometry}} & \csec{Sec.~\ref{sec:localized}} & \cst{asymptotic laws}\\[0.3em]
\cres{-- Transport-to-absorber crossover under random partial-SWAP transport, $z_{\rm prep}=\max\{2-\alpha,1\}$; coherent exchange leaves the one-singlet rates unchanged}
& \ceq{Eqs.~\eqref{eq:Dtotal}, \eqref{eq:zalpha}} & \csec{Secs.~\ref{sec:scrambling}, \ref{sec:coherent}} & \cst{exact spectrum; asymptotic exponent}\\[0.3em]
\cres{-- All-to-all monitoring, $\Lambda^{\rm all}_{\rm pair}=2/(N_q-1)$}
& \ceq{Eq.~\eqref{eq:alltoallgap}} & \csec{Sec.~\ref{sec:resourcecomparison}} & \cst{exact solution}\\[0.3em]
\cres{-- Speed limit: the relaxation rate cannot exceed the fraction of configurations on which feedback acts}
& \ceq{Prop.~\ref{prop:absorber}, Eq.~\eqref{eq:uniformabsorberbound}} & \csec{Sec.~\ref{sec:absorberbound}} & \cst{exact bound}\\[0.3em]
\cres{-- Unique stationary states selected through conserved collective-spin operators}
& \ceq{Prop.~\ref{prop:conservedalgebra}, Eq.~\eqref{eq:stationaryform}} & \csec{Sec.~\ref{sec:stationaryensemble}} & \cst{exact solution}\\[0.3em]
\cres{-- Stabilization against isotropic single-qubit errors: stationary profile, recovery rate, leakage $\epsilon^*_{\rm cert}\sim\kappa N_q^{z_{\rm prep}+1}$}
& \ceq{Eqs.~\eqref{eq:coshprofile}, \eqref{eq:recoveryrate}, \eqref{eq:fidelityz}} & \csec{Sec.~\ref{sec:imperfections}} & \cst{exact solution; asymptotic law}\\[0.3em]
\cres{-- Ensemble covariance: second-moment closure, conserving measurement noise, shot noise of removal}
& \ceq{Thm.~\ref{thm:closure2}, Eqs.~\eqref{eq:coveq}, \eqref{eq:noiseconservation}, \eqref{eq:reactionnoise}} & \csec{Sec.~\ref{sec:fluctuations}} & \cst{exact solution}
\end{tabular}
\end{ruledtabular}
\endgroup
\end{table*}

\section{Summary of main results}
\label{sec:summary}
Below we summarize the main results, which follow from the three exact operator closures and together characterize Dicke-state preparation through FFC. Throughout this summary, feedback is maximal, converting every detected singlet into a triplet, and rates are measured in units of the measurement rate per bond; the general case with feedback strength $\lambda$ is treated in the text. We determine how quickly and accurately the target manifold is reached, which stationary state is selected, and how physical errors and ensemble fluctuations affect preparation. To understand what limits the rate and how it can be improved, we compare measurement and feedback geometries in different spatial dimensions, calculating the time required for pairs to reach feedback bonds and be removed. In one spatial dimension, we obtain an exact closed-form solution for a periodic chain with nearest-neighbor bond measurements. For other measurement geometries and higher dimensions, we construct the exact generators and derive asymptotic scaling laws.

\subsubsection*{Pair closure and preparation bounds}
On any connected measurement graph of $N_q$ qubits, the dynamics of the pair projectors $\hat P_{nm}$ closes exactly, Eq.~\eqref{eq:exactpair}: a measurement on a bond moves one endpoint of a pair along that bond, and feedback removes a pair whose endpoints form a measured bond. The exact evolution of the $N_p=\tfrac12 N_q(N_q-1)$ pair projectors is equivalent to a classical random walk with absorption, governed by a symmetric positive-definite matrix $\mathsf K$, Eq.~\eqref{eq:Kdef}; see Fig.~\ref{fig:overview}.
The smallest eigenvalue $\Lambda_0>0$ of $\mathsf K$ is the slowest pair decay rate. The witness $\hat W=\sum_{n<m}\hat P_{nm}$, the total pair operator, vanishes on the Dicke manifold. Its expectation value is at least $N_q/2$ for every state orthogonal to that manifold and never exceeds $N_q(N_q+2)/8$. Consequently, $W(t)=\langle\hat W\rangle_t$ gives exact upper and lower bounds on the infidelity $\epsilon(t)$, the weight outside the target manifold:
\begin{equation}
\frac{8W(t)}{N_q(N_q+2)}
\leq\epsilon(t)\leq
\frac{2W(t)}{N_q}.
\tag{\ref{eq:fidelitybound}}
\end{equation}
Let $T_\epsilon$ denote the earliest time beyond which the infidelity stays below $\epsilon$. The closed pair dynamics and the upper infidelity bound give the state-independent preparation bound
\begin{equation}
T_\epsilon
\leq\frac{1}{\Lambda_0}
\log\left[\frac{\sqrt{(N_q-1)(N_q+2)}}{2\epsilon}\right].
\tag{\ref{eq:Teps}}
\end{equation}
The slowest pair decay rate also gives the Liouvillian gap, the slowest
nonzero decay rate of the full many-body dynamics~\cite{Zhou2026private}:
\begin{equation}
 \Delta_{\rm Liou}=\Lambda_0.
 \tag{\ref{eq:gapidentity}}
\end{equation}

\subsubsection*{Preparation as a first-passage problem: Configuration space and absorber}
The random walk does not take place on the lattice but in the configuration space of pairs. A configuration is an unordered pair of sites, the two endpoints of a pair projector $\hat P_{nm}$, and there are $N_p=N_q(N_q-1)/2$ of them. On a lattice of linear size $L$ in $d$ dimensions this space is $2d$-dimensional, $N_p\sim L^{2d}$, because both endpoints move. Feedback acts only on pairs whose two sites form a measured bond on which a correction is applied. We call these pairs absorbing and their set the absorber, $\mathcal F$. With local feedback on nearest-neighbor bonds, $|\mathcal F|=dN_q$, so feedback acts on $O(N_q)$ of the $O(N_q^2)$ configurations, and a pair has to be transported to the absorber before it can be removed. This mismatch between the size of the configuration space and the size of the absorber is the origin of the preparation time. Monitoring more pairs enlarges the absorber but does not raise this limit: when the chain is compared with all-to-all monitoring at the same total measurement rate, $N_q$ pair measurements per unit time, the total rate at which pair weight can be removed is at most $N_q$ for both. It can, however, bring the actual relaxation rate closer to the limit. Since the pair weights follow a random walk with absorption on this space, $\Lambda_0$ is a first-passage rate: the rate at which a walker in configuration space reaches the absorber. A first-passage rate is determined by the space in which the walker moves, by how fast it moves, and by where it is absorbed. The preparation rate can therefore be understood, and changed, by changing each of these three elements. After solving the chain exactly, we change them in turn.

\subsubsection*{Exact solution in one dimension}
For a periodic chain of $N_q$ qubits with nearest-neighbor bond measurements $\{\hat P_{\ell,\ell+1}\}$, we diagonalize the pair dynamics exactly. The slowest decay rate is
\begin{equation}
\Lambda_{\rm pair}
=2\left[1-\cos\left(\frac{\pi}{N_q}\right)\right]
\underset{N_q\to\infty}{\sim}\frac{\pi^2}{N_q^2}.
\tag{\ref{eq:closedgap}}
\end{equation}
The expression is exact for every $N_q$ and gives a relaxation time $\tau_{\rm pair}=1/\Lambda_{\rm pair}\sim N_q^2$. The preparation bound above applies with $\Lambda_0=\Lambda_{\rm pair}$.

\subsubsection*{Preparation laws for general feedback geometries}
The first element is the space in which the pair moves. We consider hypercubic lattices of linear size $L$ in $d$ dimensions, with $N_q=L^d$ qubits, measurements on every nearest-neighbor bond, and feedback on a regular subset of bonds of dimension $d_{\rm fb}$: the bulk ($d_{\rm fb}=d$), a boundary ($d_{\rm fb}=d-1$), or a finite region ($d_{\rm fb}=0$). A pair moves in the $2d$-dimensional space of its two endpoints and is absorbed when both endpoints occupy a feedback bond. The preparation is therefore an absorbing random walk in
\begin{equation}
n=2d-d_{\rm fb}
\tag{\ref{eq:codim}}
\end{equation}
dimensions (Sec.~\ref{sec:localized}). For $n=1,2$ the walk is recurrent: repeated returns to previously visited positions slow the exploration of configuration space. The resulting asymptotic first-passage laws, Eq.~\eqref{eq:capacity}, are $\tau_{\rm pair}\sim L^2$ for $n=1$, $L^2\log L$ for $n=2$, and $L^n$ for $n>2$, up to geometry-dependent prefactors. Regular geometries with the same $n$ therefore share the same scaling with $L$. Their scaling with the number of qubits follows from $N_q=L^d$. For bulk feedback, $n=d$, giving $\tau_{\rm pair}\sim N_q^2$, $N_q\log N_q$, and $N_q$ in $d=1$, $d=2$, and $d\geq3$, respectively. Table~\ref{tab:geometry} collects the laws for bulk, boundary, and finite-region feedback. Figure~\ref{fig:transportgeometry} compares these asymptotic predictions with the exact spectrum at system sizes far beyond many-body simulations.

\subsubsection*{Transport engineering}
The second element is how fast the endpoints move. They can be moved faster by random unitary operations without changing the measurement and feedback protocol. On the chain, random partial-SWAP events on neighboring bonds leave every Dicke state invariant and accelerate endpoint motion with an effective rate $\gamma_{\rm scr}$ set by the event rate and the gate distribution. For $\gamma_{\rm scr}\sim N_q^\alpha$ with $\alpha\geq0$, the exact spectrum gives the asymptotic scaling $\tau_{\rm pair}\sim N_q^{z_{\rm prep}}$, where
\begin{equation}
z_{\rm prep}=\max\{2-\alpha,1\}.
\tag{\ref{eq:zalpha}}
\end{equation}
For $\alpha<1$, relaxation is limited by the time required for the endpoints to reach a feedback bond. For $\alpha>1$, endpoint motion spreads the pair weight over all configurations faster than feedback removes it, so removal sets the relaxation time $\tau_{\rm pair}\sim N_q$. Further acceleration improves neither the exponent nor its limiting prefactor; $\alpha=1$ marks the crossover between the two regimes. Coherent exchange on the measured bonds leaves the decay equation of the total pair weight $\hat W$ and the dynamics in the one-singlet sector unchanged; whether it modifies the relaxation at finite singlet density is left open.

\subsubsection*{Absorber engineering}
The third element is the set of pairs on which feedback acts. Measuring and correcting every pair allows feedback to remove pair weight in every configuration. At the same total measurement rate as on the chain, the slowest rate is exactly
\begin{equation}
\Lambda_{\rm pair}^{\rm all}=\frac{2}{N_q-1},
\tag{\ref{eq:alltoallgap}}
\end{equation}
for every $N_q$. A pair no longer needs to move to a particular feedback bond before it can be removed. This attains the rate that the chain approaches in the limit of fast endpoint motion, without additional unitaries. Restricting feedback instead requires the endpoints to reach a smaller set of bonds, with the preparation laws described above. Extending the range of measured bonds changes both endpoint motion and the absorber when every detected singlet is corrected; the exact generator for any range profile is given in Sec.~\ref{sec:longrange}, where the preparation exponent on the chain is found to interpolate continuously between the absorber-limited and the diffusive value, Eq.~\eqref{eq:longrangescaling}.

\subsubsection*{Global preparation time limit}
For the closed pair dynamics considered here, faster endpoint motion cannot raise the relaxation rate beyond the removal rate averaged over all pair configurations. For an absorber $\mathcal F$, we prove
\begin{equation}
\Lambda_0\leq\frac{\lambda|\mathcal F|}{N_p}.
\tag{\ref{eq:uniformabsorberbound}}
\end{equation}
At other measurement rates, the numerator is replaced by the total measurement rate on feedback bonds. Feedback on all $dN_q$ nearest-neighbor bonds of a periodic hypercubic lattice therefore gives $\tau_{\rm pair}\geq(N_q-1)/(2d)$. The slowest relaxation time is at least linear in the number of qubits in any dimension, however rapidly measurements or additional random unitaries move the endpoints.
The preceding results show when this limit is attained. For regular feedback geometries with $n>2$, measurement-induced endpoint motion attains the scaling of the bound. Repeated returns increase the relaxation time by a factor $\log L$ for $n=2$ and by a factor $L$ for $n=1$. Accelerating endpoint motion on the chain drives the relaxation rate toward the bound from below, producing the plateau $z_{\rm prep}=1$. All-to-all monitoring saturates the same bound exactly at the same total measurement rate. Increasing the total measurement rate on feedback bonds raises this limit; spreading the same rate over more feedback bonds can instead make it attainable.

\subsubsection*{Second closure and unique stationary state}
The pair closure determines how fast the target manifold spanned by $N_q+1$ Dicke states is reached, but does not determine which state is selected within it. We show that each initial state $\hat\rho(0)$ converges to a unique stationary state in this manifold and determine that state explicitly. Let $\lvert D_m\rangle$ be the Dicke state of $N_q$ qubits with total magnetization $J^z=m$. Measurements and feedback leave every operator $\lvert D_m\rangle\langle D_{m'}\rvert$ unchanged. For an initial state confined to a single magnetization sector, the final state is the pure Dicke state of that sector. For a general initial state, the stationary state is fixed by the second exact operator closure. For feedback implemented by single-qubit rotations about the $z$ axis, we construct $(N_q+1)^2$ collective-spin operators  $\hat C_{mm'}$ from powers of the raising and lowering operators $\hat J^\pm$ and projectors $\hat\Pi_m$ onto magnetization $m$, Eq.~\eqref{eq:conservedalgebra}. The operators $\hat C_{mm'}$ are conserved for every feedback strength and every measurement graph. Their expectation values in the initial state are the coefficients of the stationary density matrix,
\begin{equation}
 \hat\rho_\infty
 =\sum_{m,m'}c_{mm'}\lvert D_m\rangle\langle D_{m'}\rvert,
 \qquad
 c_{mm'}=\langle\hat C_{mm'}\rangle_{\hat\rho(0)}.
 \tag{\ref{eq:stationaryform}}
\end{equation}
Hence, the protocol retains precisely the collective-spin information carried by the $\hat C_{mm'}$.

\subsubsection*{Stabilization against physical errors}
In an idealized preparation protocol, the Dicke manifold is absorbing: once it is reached, measurements and feedback act trivially. Realistic systems are subject to physical errors. A local error on a symmetric state creates singlet weight on every pair containing the affected site, so constraint violations are produced continuously and feedback must remove them continuously. This modifies the approach to the target and raises the question how well the target manifold is stabilized against perturbations: whether a steady state with small infidelity is maintained, how small, and how fast the protocol recovers from a disturbance. We consider isotropic single-qubit Pauli errors at total rate $\kappa$ per site, Eq.~\eqref{eq:isotropic}, on the chain. The pair dynamics remains closed and becomes affine, Eq.~\eqref{eq:affine}: the errors add a uniform source of pair weight and shift every noiseless decay rate by $8\kappa/3$, leaving the eigenvectors unchanged. The stationary pair profile is explicit for every $N_q$ and $\kappa$, Eq.~\eqref{eq:coshprofile}, and the slowest recovery rate toward it is $\Lambda_{\rm rec}=\Lambda_{\rm pair}+8\kappa/3$, Eq.~\eqref{eq:recoveryrate}. The stationary infidelity is bounded by the stationary witness, $\epsilon^*_{\rm cert}=2W^*/N_q$. For weak noise and large $N_q$, it is the error rate times the relaxation time, up to the factor $N_q$ by which the witness counts separations,
\begin{equation}
 \epsilon^*_{\rm cert}\sim\kappa N_q^{z_{\rm prep}+1},
 \tag{\ref{eq:fidelityz}}
\end{equation}
in both the transport-limited and the absorber-limited regime; the prefactors are given in Eq.~\eqref{eq:fidelityasymptotic}. Frustration-free control therefore stabilizes the target manifold autonomously whenever $\kappa N_q^{z_{\rm prep}+1}\ll1$.

\subsubsection*{Ensemble fluctuations}
The pair closure gives the mean pair weights and through them the mean witness $W(t)$, but $W(t)$ is an ensemble average: the infidelity bounds hold for the average state, and how strongly the witness fluctuates about its mean is a separate question. The same question arises for the hydrodynamic picture: a transported density has a noise kernel, and for the pair weight this kernel should follow from the measurement and feedback jumps themselves. We therefore determine the symmetrized covariance $\mathsf C^S$ of the pair projectors in the ensemble state, Eq.~\eqref{eq:symcovariance}, which contains the variance of the witness, $\operatorname{Var}(\hat W)=\sum_{a,b}\mathsf C^S_{ab}$. The covariance obeys an exact linear equation driven by the first-moment generator,
\begin{equation}
 \partial_t\mathsf C^S=-\mathsf K\mathsf C^S-\mathsf C^S\mathsf K^T+\mathsf N^S,
 \tag{\ref{eq:coveq}}
\end{equation}
whose source $\mathsf N^S$ is positive and is built directly from the Lindblad jump operators, Eq.~\eqref{eq:Npositive}. Evaluating the source requires products of two pair projectors. The third exact closure states that, for maximal feedback by Pauli rotations, these products lie in an invariant operator space of dimension $O(N_q^4)$, Theorem~\ref{thm:closure2}, so that the covariance is determined exactly by linear equations of polynomial size. The source separates into a measurement and a feedback contribution with different physical roles. The measurement contribution conserves the total pair weight, Eq.~\eqref{eq:noiseconservation}: measurements move pair weight between configurations and generate current noise, but they do not change the total weight. The feedback contribution is the noise of removal, a multiplicative shot noise whose amplitude on a measured bond equals the local removal rate, the pair weight $\langle\hat P_{\ell,\ell+1}\rangle$ on that bond, Eq.~\eqref{eq:reactionnoise}, together with correlations between the pairs attached to that bond. Both contributions vanish on the target manifold, where the dynamics and its fluctuations cease together.

\section{Frustration-free control and operator closure}
\label{sec:framework}

This section introduces the two central objects of our approach: the Lindblad generator of the measurement--feedback protocol of FFC, and the
closed operator spaces on which its adjoint acts as a finite matrix. The latter turns the many-body preparation
problem into the absorbing walker in configuration space of Sec.~\ref{sec:summary}.

\subsection{Measurement--feedback circuit}
The goal of FFC is to prepare a state $\lvert\psi\rangle$ in the common kernel of a
family of local constraint projectors $\{\hat P_\alpha\}$,
\begin{align}
 \hat P_\alpha^2&=\hat P_\alpha,
 \qquad
 \hat P_\alpha\lvert\psi\rangle=0
 \quad\text{for all }\alpha .
 \label{eq:darkcondition}
\end{align}
Each constraint $\alpha$ is measured independently according to a Poisson
process with rate $\gamma_\alpha$. At each measurement event, the binary
projective measurement
$\{\mathds{1}-\hat P_\alpha,\hat P_\alpha\}$ is performed. An outcome in the
subspace projected onto by $\hat P_\alpha$ identifies a violated constraint
and is followed by a unitary feedback operation $\hat V_\alpha$. No feedback
is applied after the complementary outcome.
A single measurement--feedback
event therefore implements the completely positive, trace-preserving channel
\begin{align}
 \Phi_\alpha[\hat\rho]
 &=\sum_{s=0,1}\hat K_{\alpha s}\hat\rho
       \hat K_{\alpha s}^\dagger,
 \label{eq:finitechannel}\\
 \hat K_{\alpha0}&=\mathds{1}-\hat P_\alpha,
 \quad
 \hat K_{\alpha1}=\hat V_\alpha\hat P_\alpha,\quad
 \mathds{1}=\sum_{s=0,1}\hat K_{\alpha s}^\dagger\hat K_{\alpha s}.\nonumber
\end{align}

Events on different constraints are independent, and two events on the
same constraint within an interval $dt$ have probability $O(dt^2)$. The
ensemble state therefore obeys the Lindblad equation
\begin{align}
 \partial_t\hat\rho
 =\mathcal L\hat\rho
 &=\sum_\alpha\gamma_\alpha
 \left(\Phi_\alpha[\hat\rho]-\hat\rho\right)
 =\sum_{\alpha,s}\gamma_\alpha
 \mathcal D[\hat K_{\alpha s}]\hat\rho,
 \label{eq:master}
 \\
 \mathcal D[\hat K]\hat\rho
 &\equiv\hat K\hat\rho\hat K^\dagger
 -\tfrac12\{\hat K^\dagger\hat K,\hat\rho\},
\end{align}
whose jump operators are the Kraus operators of a single
measurement--feedback event. With no feedback~\cite{Leung2025,Mandal2026},
$\hat V_\alpha=\mathds{1}$, both Kraus operators generate the same
dissipator,
\begin{align}
 \left.\mathcal L\right|_{\hat V_\alpha=\mathds{1}}\hat\rho
 &=2\sum_\alpha\gamma_\alpha\mathcal D[\hat P_\alpha]\hat\rho .
 \label{eq:measurementrateconvention}
\end{align}

Let $\mathcal H_D$ denote the common kernel of the constraint projectors.
On any two states $\lvert\psi_\nu\rangle,\lvert\psi_\mu\rangle\in\mathcal H_D$,
$\hat K_{\alpha0}$ acts as the identity and $\hat K_{\alpha1}$ vanishes, so
\begin{align}
 \Phi_\alpha[
 \lvert\psi_\nu\rangle\langle\psi_\mu\rvert]
 &=
 \lvert\psi_\nu\rangle\langle\psi_\mu\rvert,
 &
 \mathcal L[
 \lvert\psi_\nu\rangle\langle\psi_\mu\rvert]
 &=
 0.
 \label{eq:darkoperators}
\end{align}
Every operator supported on $\mathcal H_D$ is stationary: $\mathcal H_D$ is
a decoherence-free subspace. Throughout, we consider protocols for which
the stationary space is exactly $\mathcal B(\mathcal H_D)$, so that no
stationary operator has support outside $\mathcal H_D$; for the Dicke
protocol this is proven in Lemma~\ref{lem:stationaryspace}. When exact
symmetries split the Hilbert space into invariant sectors, $\mathcal H_D$
contains one dark state per sector and the stationary space consists of
their populations and coherences. Additional operations that act trivially
on $\mathcal H_D$ change the relaxation toward the target manifold but not
the manifold itself; this is the freedom used in Sec.~\ref{sec:scrambling}.

\subsection{Adjoint dynamics and linear closure}

The approach to the target manifold is governed by the dynamics of the projectors $\hat P_\alpha$ under the adjoint generator $\mathcal L^\dagger$. For an arbitrary operator $\hat O$ it is defined by duality with respect to the trace,
\begin{align}
 \Tr\!\left[\hat O\,\mathcal L\hat\rho\right]
 &=\Tr\!\left[(\mathcal L^\dagger\hat O)\hat\rho\right],\\
 \partial_t\avg{\hat O}_t
 &=\avg{\mathcal L^\dagger\hat O}_t .
 \label{eq:adjointgeneral}
\end{align}
For the generator of Eq.~\eqref{eq:master},
\begin{align}
 \mathcal L^\dagger\hat O
 &=\sum_\alpha\gamma_\alpha
 \left(\sum_{s=0,1}\hat K_{\alpha s}^\dagger
 \hat O\hat K_{\alpha s}-\hat O\right)\label{eq:adjsplit}\\
 &=\sum_\alpha\gamma_\alpha\left(
 \underbrace{-[\hat P_\alpha,
 [\hat P_\alpha,\hat O]]}_{\text{measurement}}
 +\underbrace{\hat P_\alpha
 (\hat V_\alpha^\dagger\hat O\hat V_\alpha-\hat O)
 \hat P_\alpha}_{\text{feedback}}
 \right).
 \nonumber
\end{align}
The first term is the unconditioned projective measurement. It dephases
coherences between the two eigenspaces of $\hat P_\alpha$. The second term
is the conditional feedback, vanishing for
$\hat V_\alpha=\mathds{1}$.

Equation~\eqref{eq:adjsplit} is a linear map on operator space. In general it
mixes the $\hat P_\alpha$ with the full $4^{N_q}$-dimensional operator space of $N_q$ qubits. The structure we exploit is that, for suitable protocols, there is
a space
$\mathsf Q=\operatorname{span}\{\hat Q_a\}$ of polynomial dimension that
contains the measured projectors, $\{\hat P_\alpha\}\subseteq\mathsf Q$,
and is mapped into itself,
\begin{align}
 \mathcal L^\dagger\hat Q_a
 &=-\sum_b\mathsf K_{ab}\hat Q_b,
 \qquad
 \mathsf K=\mathsf M+\mathsf A .
 \label{eq:closure}
\end{align}
We call $\mathsf Q$ a closed operator space and $\mathsf K$ its generator.
The transport matrix $\mathsf M$ collects the measurement term and the
absorption matrix $\mathsf A$ the feedback term of Eq.~\eqref{eq:adjsplit}.

Equation~\eqref{eq:closure} is an operator identity and holds for any state. It has three consequences, which together
reduce the preparation problem to linear algebra of polynomial size. First,
the Heisenberg evolution of the closed space is the finite matrix
exponential, $e^{t\mathcal L^\dagger}\hat Q_a=\sum_b(e^{-\mathsf Kt})_{ab}\hat Q_b$. Second, since the measured
projectors belong to $\mathsf Q$, every constraint violation decays inside
$\mathsf Q$: whether, and how fast, the target manifold is reached is
decided by $\mathsf K$ alone. Third, the eigenvalues of $\mathsf K$ are
exact Liouvillian eigenvalues. If $\bm v^T\mathsf K=\Lambda\bm v^T$, then
\begin{align}
\mathcal L^\dagger\hat O_{\bm v}
 =
 -\Lambda\hat O_{\bm v}\quad \text{ for } \hat O_{\bm v}
 \equiv
 \sum_a v_a\hat Q_a,
 \label{eq:closedeigenoperator}
\end{align}
so $-\Lambda$ is an eigenvalue of $\mathcal L^\dagger$, and since
$\operatorname{spec}(\mathcal L^\dagger)=\operatorname{spec}(\mathcal L)^*$,
the rate $\operatorname{Re}\Lambda$ is an exact decay rate of the full
dynamics. Expectation values in any state follow by taking the trace
against $\hat\rho(0)$, $\partial_t\langle\hat Q_a\rangle_t=-\sum_b\mathsf K_{ab}\langle\hat Q_b\rangle_t$. When the $\hat Q_a$ are spatially resolved few-body observables
and the decay rates of $\mathsf K$ vanish with increasing system size, we
speak of a hydrodynamic closure.

Whether a closed space of polynomial dimension exists depends on the
algebra of the constraint projectors and feedback operations. In
Sec.~\ref{sec:benchmark} we show that for the SWAP protocol $\mathsf Q$ is
spanned by the $N_p$ pair projectors, that $\mathsf M$ moves one endpoint
of a pair along a measured bond, and that $\mathsf A$ removes the pairs on
the feedback bonds. This is the walker in configuration space of
Sec.~\ref{sec:summary}, as an exact operator identity.

Two remarks delimit what the closure provides. The closed space fixes the
approach to the target manifold but not the state that is reached within
it; that state is fixed by the zero modes of $\mathcal L^\dagger$, which lie
outside $\mathsf Q$ and are constructed in
Sec.~\ref{sec:stationaryensemble}. Closure of the $\hat Q_a$ also does not
imply closure of their products. Although $e^{t\mathcal L^\dagger}$
generates the Heisenberg-picture evolution, it is not an algebra
homomorphism: for the Lindblad generator in Eq.~\eqref{eq:master}, the
exact deviation from the Leibniz rule is
\begin{align}
 \mathcal L^\dagger(\hat A\hat B)
 -(\mathcal L^\dagger\hat A)\hat B
 -\hat A(\mathcal L^\dagger\hat B)
 &=
 \sum_{\alpha,s}\gamma_\alpha
 [\hat K_{\alpha s}^\dagger,\hat A]
 [\hat B,\hat K_{\alpha s}].
 \label{eq:lindbladleibniz}
\end{align}
Invariance of $\mathsf Q$ under $\mathcal L^\dagger$ therefore does not
imply invariance of the operator products required for ensemble second
moments; their closure is established independently in
Sec.~\ref{sec:fluctuations}.

\section{Dicke-state preparation and exact pair hydrodynamics}
\label{sec:benchmark}

We now apply the framework to the preparation of Dicke states by SWAP
measurements and single-qubit feedback. Sec.~\ref{sec:protocol} defines the
protocol. Sec.~\ref{sec:endpoint} derives the pair closure,
Eq.~\eqref{eq:exactpair}: the walker in configuration space, with the
transport matrix $\mathsf M$ and the absorption matrix $\mathsf A$. Secs.~\ref{sec:separation} and \ref{sec:spectrum} solve the walker on
the periodic chain exactly, and Sec.~\ref{sec:absorberbound} proves a global
absorber bound that applies to all such protocols.

\subsection{Dicke target and local measurement--feedback protocol}
\label{sec:protocol}

We consider $N_q$ spin-$1/2$ degrees of freedom on a connected periodic lattice in $d$ spatial dimensions. The collective spin is
$\bm J=\frac12\sum_{\ell}\bm\sigma_\ell$. In the sector
$J^z=m$, our target is the Dicke state
\begin{align}
 \lvert D_m\rangle
 &=\binom{N_q}{N_q/2+m}^{-1/2}
 \sum_{\substack{\bm s\in\{\uparrow,\downarrow\}^{N_q}\\
 N_\uparrow(\bm s)=N_q/2+m}}
 \lvert\bm s\rangle.
 \label{eq:dickestate}
\end{align}
It is the unique state with $J=N_q/2$ and $J^z=m$, and it is invariant
under every permutation of the spins \cite{Dicke1954}. We impose this
permutation symmetry locally through two-site SWAP constraints. The
resulting spin-$1/2$ protocol was introduced in
Ref.~\cite{Doerstel2026} and studied numerically in
Ref.~\cite{Hauser2024}.

For two sites $\ell$ and $m$, the SWAP operator and the corresponding
constraint projector are
\begin{align}
 \hat S_{\ell m}
 &=\sum_{s,s'\in\{\uparrow,\downarrow\}}
 \lvert ss'\rangle\langle s's\rvert_{\ell m}
 =\frac12\left(\mathds{1}
 +\bm\sigma_\ell\cdot\bm\sigma_m\right),
 \label{eq:swapqubit}\\
 \hat P_{\ell m}
 &=\frac12\left(\mathds{1}-\hat S_{\ell m}\right)
 =\frac14\left(\mathds{1}
 -\bm\sigma_\ell\cdot\bm\sigma_m\right)
 =\lvert s_{\ell m}\rangle\langle s_{\ell m}\rvert,
 \label{eq:swapproj}
\end{align}
with $\lvert s_{\ell m}\rangle=(\lvert\uparrow_\ell\downarrow_m\rangle
-\lvert\downarrow_\ell\uparrow_m\rangle)/\sqrt2$. Thus
$\hat P_{\ell m}$ projects onto the two-spin singlet, while its kernel is the
symmetric triplet subspace. We identify the constraint label of
Sec.~\ref{sec:framework} as $\alpha=(\ell,m)$ and hence
$\hat P_\alpha=\hat P_{\ell m}$.

Unless stated otherwise, only nearest-neighbor bonds are measured; on a
hypercubic lattice their set is
\begin{align}
 E_d=\left\{(\ell,\ell+\bm e_\mu):
 \ell \text{ on the lattice},\ \mu=1,\ldots,d\right\},
 \label{eq:measuredbonds}
\end{align}
and, unless stated otherwise, we consider the homogeneous protocol
\begin{align}
 \gamma_{\ell m}=\gamma\equiv1 .
 \label{eq:homogeneousrates}
\end{align}
The measurement rate therefore sets the unit of time. Both restrictions are
relaxed later: Sec.~\ref{sec:longrange} extends the monitored bonds to
arbitrary range and confines feedback to a lower-dimensional subset, and
Sec.~\ref{sec:broader} monitors every pair at fixed total measurement effort.
Because nearest-neighbor transpositions generate all permutations on a
connected lattice, the common kernel of
$\{\hat P_{\ell m}:(\ell,m)\in E_d\}$ is the fully symmetric
$J=N_q/2$ manifold. Restricting the dynamics to a fixed-$J^z$ sector
makes $\lvert D_m\rangle$ the unique dark state.

For each measured bond, the feedback unitary must satisfy
\begin{align}
 [\hat V_{\ell m},\hat P_{\ell m}]&\ne0,
 &
 \supp(\hat V_{\ell m})
 &\subseteq\supp(\hat P_{\ell m})=\{\ell,m\}.
 \label{eq:feedbackconditions}
\end{align}
The first condition allows the feedback to transfer a detected singlet out
of the violated subspace, while the second confines the correction to the
measured bond. Closure of the pair-projector dynamics imposes the additional
requirement that the feedback contribution preserve the linear span of the
pair projectors. This is achieved by a single-qubit rotation on either
endpoint,
\begin{align}
 \hat V_{\ell m}
 &=e^{i\varphi\sigma_\ell^\nu}
 =\mathds{1}\cos\varphi
 +i\sigma_\ell^\nu\sin\varphi,
 \quad
 \nu\in\{x,y,z\}.
 \label{eq:paulifeedback}
\end{align}
Since the singlet has vanishing one-spin polarization,
\begin{align}
 \hat P_{\ell m}\sigma_\ell^\nu\hat P_{\ell m}&=0,
 &
 \hat P_{\ell m}\hat V_{\ell m}\hat P_{\ell m}
 &=\cos\varphi\,\hat P_{\ell m}.
 \label{eq:feedbackcompression}
\end{align}
The feedback contribution to the adjoint evolution of a pair projector is
therefore
\begin{align}
 \hat P_{\ell m}\hat V_{\ell m}^\dagger
 [\hat P_{\ell' m'},\hat V_{\ell m}]\hat P_{\ell m}
 &=-\lambda\,\hat P_{\ell' m'}
 \delta_{\{\ell,m\},\{\ell',m'\}},
 \nonumber\\
 \lambda&=\sin^2\varphi .
 \label{eq:feedbacklocal}
\end{align}
Thus the feedback removes weight only from the detected pair projector and
does not act on any other pair projector, including pairs sharing one
endpoint with the measured bond. The angle $\varphi$ tunes the dimensionless
absorption strength $\lambda$ independently of the measurement-induced
transport. To preserve $J^z$, we choose $\nu=z$; unless stated otherwise,
we use the maximal correction $\varphi=\pi/2$, for which
$\hat V_{\ell m}=i\sigma_\ell^z$ and $\lambda=1$.

Feedback is applied on a subset $\mathcal F$ of the measured bonds, the
feedback bonds. We call $\mathcal F$ the absorber: a pair projector
$\hat P_{nm}$ is absorbed if and only if $\{n,m\}\in\mathcal F$. We write
$\mathbf 1_{\mathcal F}(nm)$ for the indicator of this set,
$\mathbf 1_{\mathcal F}(nm)=1$ if $\{n,m\}\in\mathcal F$ and
$\mathbf 1_{\mathcal F}(nm)=0$ otherwise. Unless stated otherwise, every
measured bond is a feedback bond, $\mathcal F=E_d$, so that
$|\mathcal F|=dN_q$.

\subsection{Exact absorbing pair dynamics}
\label{sec:endpoint}

The result of this subsection is Eq.~\eqref{eq:exactpair}: the adjoint
generator maps every pair projector onto pair projectors, measurement
moving one endpoint of the pair along a measured bond and feedback removing
the pair when its endpoints form a feedback bond. We first show that
measurement alone preserves the span of the pair projectors on any
measurement graph, and then add the feedback.

For every projector $\hat P_\alpha$, define the Hermitian involution
\begin{align}
 \hat U_\alpha&=\mathds{1}-2\hat P_\alpha,
 \qquad
 \hat U_\alpha^2=\mathds{1}.
\end{align}
Let the measurement contribution $\mathcal{L}_M\equiv \left.\mathcal{L}\right|_{\hat V_\alpha=\mathds{1}}$ in Eq.~\eqref{eq:adjsplit}. It can then be written
as
\begin{align}
 -[\hat P_\alpha,[\hat P_\alpha,\hat O]]
 &=\frac12\left(
 \hat U_\alpha\hat O\hat U_\alpha-\hat O
 \right),
 \label{eq:conjwalk}\\
 \mathcal L^\dagger_M\hat O
 &\equiv\sum_\alpha\frac{\gamma_\alpha}{2}
 \left(
 \hat U_\alpha\hat O\hat U_\alpha-\hat O
 \right).
 \label{eq:measwalk}
\end{align}
By Eq.~\eqref{eq:measwalk}, an operator space $\mathsf Q$ is invariant under
the measurement contribution, for arbitrary rates $\gamma_\alpha$, whenever
$\hat U_\alpha\mathsf Q\hat U_\alpha\subseteq\mathsf Q$ for every $\alpha$.

For the SWAP constraints,
$\hat U_{\ell m}=\hat S_{\ell m}$, so $\hat U_{\ell m}\hat O \hat U_{\ell m}$ simply exchanges the site
labels $\ell$ and $m$: a measurement maps each pair projector onto a pair
projector with one endpoint moved, so the span of the pair projectors is
invariant under the measurement dynamics, independently of the measurement
geometry and local Hilbert-space dimension. Equation~\eqref{eq:feedbacklocal} shows that the
feedback contribution preserves the same space and completes the closure
under the full generator. Since $\hat U_{\ell m}\hat O \hat U_{\ell m}$ is an algebra automorphism,
products of any number of pair operators also remain closed under
the measurement, see Sec.~\ref{sec:fluctuations}.

The endpoint motion follows from the permutation-group identity
$\hat S_{ij}\hat S_{ik}\hat S_{ij}=\hat S_{jk}$. Together with
$\hat S_{ij}^2=\mathds{1}$, it gives, for three distinct sites $i,j,k$,
\begin{align}
 -[\hat P_{ij},[\hat P_{ij},\hat P_{ik}]]
 &=\frac12\left(\hat P_{jk}-\hat P_{ik}\right).
 \label{eq:endpointmove}
\end{align}
A measurement on $(i,j)$ therefore replaces the endpoint $i$ of
$\hat P_{ik}$ by $j$ at half the measurement rate. A measurement on a
disjoint bond has no effect, while the contribution from a coincident bond
vanishes because
$[\hat P_{ij},\hat P_{ij}]=0$.

We define the pair-graph Laplacian on unordered noncoincident pairs
$\{n,m\}$ by
\begin{align}
 (\Delta_{\rm pair}\hat P)_{nm}
 &=
 \sum_{\mu=1}^{d}
 \sum_{\substack{\eta=\pm1\\
 n+\eta\bm e_\mu\ne m}}
 \left(
 \hat P_{n+\eta\bm e_\mu,m}-\hat P_{nm}
 \right)
 \nonumber\\
 &\quad+
 \sum_{\mu=1}^{d}
 \sum_{\substack{\eta=\pm1\\
 m+\eta\bm e_\mu\ne n}}
 \left(
 \hat P_{n,m+\eta\bm e_\mu}-\hat P_{nm}
 \right).
 \label{eq:pairlap}
\end{align}
Combining Eq.~\eqref{eq:endpointmove} with the feedback
identity~\eqref{eq:feedbacklocal}, applied on every bond of the absorber
$\mathcal F$, gives
\begin{align}
 \boxed{
 \mathcal L^\dagger\hat P_{nm}
 =
 \frac12(\Delta_{\rm pair}\hat P)_{nm}
 -\lambda\,\mathbf 1_{\mathcal F}(nm)\hat P_{nm}.}
 \label{eq:exactpair}
\end{align}
This is an exact operator identity on the full Hilbert space. Measurement moves either endpoint on
the graph of unordered pairs, while feedback absorbs a pair whose endpoints
form a feedback bond. The absorbing pair description was introduced in Ref.~\cite{Doerstel2026}
for the bulk; Eq.~\eqref{eq:exactpair} adds the exclusion of coincident
endpoints and the exact treatment of the absorbing and outer boundaries,
which is what makes the spectrum below exact for every $N_q$.

Equation~\eqref{eq:exactpair} is a closed linear equation for the
$N_p=N_q(N_q-1)/2$ pair projectors themselves. Collecting them in the vector
$\hat{\bm P}$, it takes the form of Eq.~\eqref{eq:closure},
\begin{align}
 \mathcal L^\dagger\hat{\bm P}&=-\mathsf K\hat{\bm P},
 &
 \mathsf K&=\mathsf M+\mathsf A,
 \nonumber\\
 \mathsf M&=-\tfrac12\Delta_{\rm pair},
 &
 \mathsf A&=\lambda\operatorname{diag}(\mathbf 1_{\mathcal F}),
 \label{eq:Kdef}
\end{align}
with the transport matrix $\mathsf M$, which moves one endpoint of a pair
along a measured bond, and the absorption matrix $\mathsf A$, which removes
the pairs whose endpoints form a feedback bond. The pair graph is connected and
undirected, so $\mathsf M$ is symmetric and positive semidefinite with
$\mathsf M\bm 1=0$. Since the absorber is nonempty,
\begin{align}
 \mathsf K&=\mathsf K^T>0
 \qquad\text{for }\lambda>0 ,
 \label{eq:Kspd}
\end{align}
and the pair sector has a strictly positive relaxation rate. Equation
\eqref{eq:Kdef} is the generator of a random walk on the pair graph with
absorption on the absorber $\mathcal F$, performed by the projectors $\hat P_{nm}$
as an operator identity. Their expectation values $q_{nm}=\langle\hat P_{nm}\rangle$,
collected in the vector $\bm q$, obey the same equation,
$\partial_t\bm q=-\mathsf K\bm q$, and are the quantities compared with
experiment below.

\subsection{Reduction to pair separation coordinates}
\label{sec:separation}

On a periodic chain of $N_q$ sites the absorbing walk in
Eq.~\eqref{eq:Kdef} can be solved in closed form. Since the generator is
translation invariant, it closes on the $R=\lfloor N_q/2\rfloor$ operators
of the pair separation coordinate $r$,
\begin{align}
 \hat P_r
 &\equiv\sum_{n=1}^{N_q}\hat P_{n,n+r},
 &
 \hat P_r&=\hat P_{N_q-r},
 \qquad
 1\le r\le R,
 \label{eq:Pr}
\end{align}
with site labels modulo $N_q$. For $r<N_q/2$, the sum contains every pair
at separation $r$ once. For even $N_q$, $\hat P_{N_q/2}$ contains each
diametrically opposite pair twice.

Equation~\eqref{eq:exactpair} then yields
\begin{align}
 \mathcal L^\dagger\hat P_r
 &=
 \hat P_{r+1}+\hat P_{r-1}-2\hat P_r,
 \qquad
 1<r<R,
 \nonumber\\
 \mathcal L^\dagger\hat P_1
 &=
 \hat P_2-\hat P_1-\lambda\hat P_1,
 \nonumber\\
 \mathcal L^\dagger\hat P_R
 &=c_R(\hat P_{R-1}-\hat P_R),
 \qquad
 c_R=1+\delta_{N_q,2R}.
 \label{eq:folded}
\end{align}

We define the witness $\hat W$ for the Dicke state as
\begin{align}
 \hat W
 &\equiv\sum_{n<m}\hat P_{nm}
 =
 \sum_{r=1}^{R-1}\hat P_r
 +\frac{\hat P_R}{1+\delta_{N_q,2R}},
 \label{eq:Wfolded}
\end{align}
where the last term has weight $1/2$ for even $N_q$ because $\hat P_R$
then counts each pair twice. Summing Eq.~\eqref{eq:folded} gives the exact
loss identity
\begin{align}
 \mathcal L^\dagger\hat W
 &=-\lambda\hat P_1 .
 \label{eq:Wloss}
\end{align}
Measurement conserves the total pair weight and moves it between
separations; feedback removes it at $r=1$.

Taking the trace against the state and writing
$p_r(t)=\langle\hat P_r\rangle_t/N_q$ for the pair densities,
Eq.~\eqref{eq:folded} becomes
\begin{align}
 \partial_tp_1
 &=p_2-p_1-\lambda p_1,
 \label{eq:foldleft}\\
 \partial_tp_r
 &=p_{r+1}+p_{r-1}-2p_r,
 \qquad
 1<r<R,
 \label{eq:foldbulk}\\
 \partial_tp_R
 &=c_R(p_{R-1}-p_R),
 \label{eq:foldright}
\end{align}
and $W(t)=\langle\hat W\rangle_t$ obeys $\partial_tW=-\lambda N_qp_1$.

\subsection{Spectrum of the reduced generator}
\label{sec:spectrum}

Equations~\eqref{eq:foldleft}--\eqref{eq:foldright} are solved by
$p_r(t)=e^{-\Lambda t}f_r$ with
\begin{align}
 \Lambda(k)&=2(1-\cos k),
 &
 f_r&=\cos\left[k\left(\tfrac{N_q}{2}-r\right)\right],
 \label{eq:mode}
\end{align}
for both even and odd $N_q$: the boundary condition at $r=R$ in
Eq.~\eqref{eq:foldright} is satisfied identically for both parities
(Appendix~\ref{app:spectrum}). The boundary condition at $r=1$ fixes the
allowed momenta,
\begin{align}
 \cos\left(\frac{kN_q}{2}\right)
 =(1-\lambda)
 \cos\left[k\left(\frac{N_q}{2}-1\right)\right].
 \label{eq:quantization}
\end{align}
For $0<\lambda\le1$, Eq.~\eqref{eq:quantization} has $R$ roots, which give
the complete spectrum of Eqs.~\eqref{eq:foldleft}--\eqref{eq:foldright}.
The smallest root $k_0$ gives the slowest rate in the pair sector,
$\Lambda_{\rm pair}=2(1-\cos k_0)$. This is also the lowest eigenvalue of
the full $N_p$-dimensional generator $\mathsf K$: $\mathsf K$ is
irreducible with nonpositive off-diagonal elements, so by the
Perron--Frobenius theorem its lowest eigenvector is unique and positive,
and since $\mathsf K$ commutes with translations this eigenvector is
translation invariant.

For any fixed $\lambda>0$, Eq.~\eqref{eq:quantization} gives
$k_0=\pi/N_q+O(N_q^{-2})$ and $\Lambda_{\rm pair}=\pi^2/N_q^2+O(N_q^{-3})$:
the pair relaxation has dynamical exponent $z=2$. For maximal feedback,
$\lambda=1$, the quantization condition reduces to $\cos(kN_q/2)=0$, with
$k_0=\pi/N_q$ exactly and
\begin{align}
 \boxed{
 \Lambda_{\rm pair}(N_q)
 =2\left(1-\cos\frac{\pi}{N_q}\right)
 \xrightarrow[N_q\to\infty]{}
 \frac{\pi^2}{N_q^2}.
 }
 \label{eq:closedgap}
\end{align}
This is the exact form of the diffusive scaling
$\tau_{\rm pair}=\Lambda_{\rm pair}^{-1}\sim N_q^2$ found in
Ref.~\cite{Doerstel2026}.

At $\lambda=1$, the complete set of modes is
\begin{align}
 k_j&=\frac{(2j+1)\pi}{N_q},
 \quad \Lambda_j=2\left[
 1-\cos\frac{(2j+1)\pi}{N_q}
 \right],\nonumber\\
 f_r^{(j)}&=\sin\frac{(2j+1)\pi r}{N_q},
  \quad j=0,\ldots,R-1,
 \label{eq:fullspectrum}
\end{align}
and the modes are orthogonal with the same weight of $\hat P_R$ as in
Eq.~\eqref{eq:Wfolded},
\begin{align}
 \sum_{r=1}^{R-1}
 f_r^{(j)}f_r^{(j')}
 +\frac{f_R^{(j)}f_R^{(j')}}{1+\delta_{N_q,2R}}
 =\frac{N_q}{4}\delta_{jj'}.
 \label{eq:modeorthogonality}
\end{align}
Any initial condition is therefore solved by a finite sine transform. For
even $N_q$ and the N\'eel initial state,
\begin{align}
 p_r(0)=\frac{1-(-1)^r}{4},
 \label{eq:neelinitial}
\end{align}
this gives
\begin{align}
 p_r(t)
 =\frac{1}{N_q}
 \sum_{j=0}^{R-1}
 \frac{\sin[(2j+1)\pi r/N_q]}
 {\sin[(2j+1)\pi/N_q]}
 e^{-\Lambda_j t},
 \label{eq:pclosed}
\end{align}
the exact time dependence of every pair density for this initial state.
At $t=0$ it gives $W(0)=N_q^2/8$; at long times only the $j=0$ mode
survives and the decay rate is Eq.~\eqref{eq:closedgap}.

\subsection{Absorber bound: a transport-independent limit on the rate}
\label{sec:absorberbound}

On the chain, the relaxation time $\Lambda_{\rm pair}^{-1}\sim N_q^{2}$ is
the first-passage time of a diffusive random walker. Here we derive a
general lower bound for the relaxation time that is set by the absorber
geometry.

In Eq.~\eqref{eq:Kdef}, the transport matrix conserves the total pair
weight, $\mathsf M\bm 1=0$, and is symmetric. The absorption matrix
$\mathsf A=\lambda\operatorname{diag}(\mathbf 1_{\mathcal F})$ is
diagonal. Here we write $\mathsf A=\operatorname{diag}(a_1,\ldots,a_{N_p})$,
with $a_i$ the rate at which weight is removed from the pair $i$: for the
protocol, $a_i=\lambda$ if the pair $i$ is in the absorber $\mathcal F$ and
$a_i=0$ otherwise, so that $\sum_ia_i=\lambda|\mathcal F|$. We derive a
bound for general rates $a_i\geq0$ that includes the long-range protocols
of Sec.~\ref{sec:longrange}.

\begin{thmbox}{Proposition}\label{prop:absorber}
Let $\mathsf K=\mathsf M+\mathsf A$ with $\mathsf M=\mathsf M^T\geq0$,
$\mathsf M\bm 1=0$, and $\mathsf A=\operatorname{diag}(a_i)$, $a_i\geq0$,
and let no vector be annihilated by both,
\begin{align}
 \ker\mathsf M\cap\ker\mathsf A=\{\bm 0\}.
 \label{eq:absorberreachability}
\end{align}
Then $\mathsf K>0$ and its smallest eigenvalue obeys
\begin{align}
 \Lambda_0
 \leq
 \frac{1}{N_p}
 \sum_{i=1}^{N_p}a_i .
 \label{eq:absorberbound}
\end{align}
For feedback of strength $\lambda$ on the absorber $\mathcal F$,
\begin{align}
 \Lambda_0\leq
 \lambda\frac{|\mathcal F|}{N_p}.
 \label{eq:uniformabsorberbound}
\end{align}
\end{thmbox}

\noindent\emph{Proof.}
For any vector $\bm x$, $\bm x^T\mathsf K\bm x=\bm x^T\mathsf M\bm x
+\sum_ia_ix_i^2\geq0$, with equality only if
$\bm x\in\ker\mathsf M\cap\ker\mathsf A$; hence $\mathsf K>0$. The Rayleigh
quotient on the uniform vector gives
$\Lambda_0\leq\bm 1^T\mathsf K\bm 1/\bm 1^T\bm 1=\sum_ia_i/N_p$, since
$\mathsf M\bm 1=0$. \hfill$\square$

When $\mathsf M$ is the Laplacian of a connected transport graph,
Eq.~\eqref{eq:absorberreachability} holds as soon as the absorber is
nonempty; for the pair sector this is Eq.~\eqref{eq:Kspd}. The symmetry of
$\mathsf M$ is essential: for a non-symmetric generator the Rayleigh
quotient does not bound the slowest rate.

The bound says that transport, however fast, cannot remove pair weight
faster than the removal rate averaged over all $N_p$ pairs: the relaxation
time $\tau_0=\Lambda_0^{-1}$ is at least $N_p/\sum_ia_i$. The bound is
attained in the limit of fast transport, $\mathsf M\to s\mathsf M$,
$s\to\infty$, where the lowest mode becomes uniform over the pairs and
$\Lambda_0\to\sum_ia_i/N_p$. A faster rate requires a larger absorber or a
larger removal rate on it.

For feedback on all $dN_q$ nearest-neighbor bonds of a periodic hypercubic
lattice, $|\mathcal F|=dN_q$ and $N_p=N_q(N_q-1)/2$, so
\begin{align}
 \Lambda_0
 \leq
 \frac{2d\lambda}{N_q-1},
 \qquad
 \tau_0
 \geq
 \frac{N_q-1}{2d\lambda}.
 \label{eq:localfloor}
\end{align}
No transport can bring the pair relaxation time below $O(N_q/\lambda)$
while feedback acts on $O(N_q)$ bonds, in any dimension.

We note that enlarging the absorber does not raise this bound unless
the total measurement rate on feedback bonds is raised with it. Because
overlapping pairs do not commute, one can at most measure $N_q/2$ of the
$N_p=N_q(N_q-1)/2$ pairs simultaneously, and it requires at least $N_q-1$
layers of measurement and feedback to measure all possible pairs. Hence,
enlarging the absorber to the maximum set of pairs $|\mathcal F|\to N_p$
at a fixed total rate of $N_q$ pair measurements per unit time reduces
the measurement rate per pair from $1$ to $2/(N_q-1)$, and with it the
removal rate $a_i$ on each pair, while the feedback strength $\lambda$ is
unchanged. The chain, which measures its $N_q$ bonds at unit rate, and
any protocol that monitors other pairs at the same total rate
(Sec.~\ref{sec:resourcecomparison}) therefore have
$\sum_ia_i\leq\lambda N_q$ and $\tau_0\geq(N_q-1)/2\lambda$,
whichever pairs are monitored. What a larger absorber changes is whether
the bound is attained: the chain reaches it only in the limit of fast
transport, whereas all-to-all monitoring attains it exactly,
Sec.~\ref{sec:resourcecomparison}. Equation~\eqref{eq:localfloor} bounds
the slowest rate; the preparation time of a given initial state also
depends on its overlap with the slowest mode, Sec.~\ref{sec:certificate}.

For a lattice of linear size $L$, $N_q=L^d$, and feedback bonds forming a
regular $d_{\rm fb}$-dimensional set, $|\mathcal F|\sim L^{d_{\rm fb}}$
while $N_p\sim L^{2d}$, so
\begin{align}
 \boxed{
 \Lambda_0
 \lesssim
 \lambda L^{-n},
 \qquad
 \tau_0
 \gtrsim
 \lambda^{-1} L^n,
 \qquad
 n
 =
 2d-d_{\rm fb}.}
 \label{eq:codimensionbound}
\end{align}
Bulk feedback has $d_{\rm fb}=d$ and $\tau_0\gtrsim N_q/\lambda$ in any
dimension. Whether the bound is attained depends on the transverse
dimension $n$, Sec.~\ref{sec:localized}.

\begin{figure*}[t]
 \centering
 \includegraphics[width=0.92\textwidth]{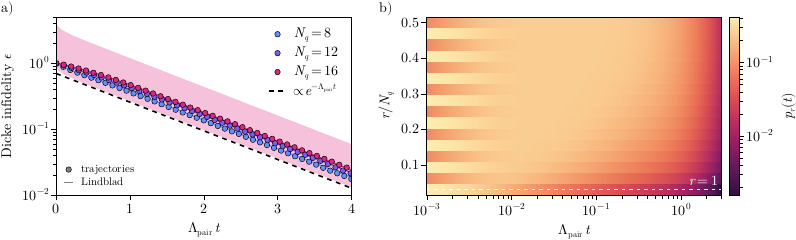}
 \caption{\textbf{Exact pair dynamics and fidelity bound.}
(a) Exact  Dicke-state infidelity following the N\'eel initial state for
$N_q=8,12,16$: lines are the exact Lindblad evolution in the $J^z=0$ sector,
dots the corresponding averages over $10^5$ quantum trajectories. The shaded band is the explicit
time-dependent two-sided bound of Eq.~\eqref{eq:twosided} at
$N_q=16$, evaluated from the closed pair equation alone. The many-body data lie
inside the band at every time and every size. The upper and lower
envelopes are the two sides of Eq.~\eqref{eq:twosided}. The dashed line decays
with the exact pair-sector rate
$\Lambda_{\rm pair}=2[1-\cos(\pi/N_q)]$, Eq.~\eqref{eq:closedgap}.
(b) The exact closed-form relaxation $p_r(t)$ of Eq.~\eqref{eq:pclosed} at
$N_q=32$ (finite sine transform of the N\'eel initial condition,
Eq.~\eqref{eq:neelinitial}; logarithmic time and color scales). The three
regimes of the pair dynamics are visible: the even--odd structure of the
initial state at early times, diffusive redistribution with depletion at the
absorbing boundary $r=1$ (white line), and the final single-mode decay at
rate $\Lambda_{\rm pair}$.}
 \label{fig:certificate}
\end{figure*}

\subsection{Two-sided fidelity bound}
\label{sec:certificate}

In an experimental implementation of the state preparation protocol, the
relevant criterion is the fidelity with the target state. Here we derive
upper and lower bounds for the infidelity in terms of the pair weight
$W=\langle\hat W\rangle$, and from them bounds on the time needed to
reach a given infidelity.

For spin $1/2$, the total pair operator of
Eq.~\eqref{eq:Wfolded} can be expressed in terms of the collective spin as
\begin{align}
 \hat W
 &=
 \frac12
 \left[
 J_{\max}(J_{\max}+1)\mathds{1}
 -\bm J^2
 \right],
 &
 J_{\max}&=\frac{N_q}{2}.
 \label{eq:Jsum}
\end{align}
The identity is derived in Appendix~\ref{app:imperfections}. Let
\begin{align}
 \mathcal H_{\rm sym}
 =\operatorname{span}
 \left\{
 \lvert D_m\rangle:
 m=-\frac{N_q}{2},\ldots,\frac{N_q}{2}
 \right\}
 \label{eq:symmetricmanifold}
\end{align}
be the fully symmetric manifold spanned by all Dicke states, i.e., the
common kernel of all pair projectors. We define the projector onto this
kernel and its population, i.e., the Dicke space fidelity, by
\begin{align}
 \hat\Pi_{\rm D}
 &\equiv
 \sum_{m=-J_{\max}}^{J_{\max}}
 \lvert D_m\rangle\langle D_m\rvert,
 &
 F_{\rm D}(t)
 &\equiv
 \operatorname{Tr}
 \left[
 \hat\rho(t)\hat\Pi_{\rm D}
 \right].
 \label{eq:symmetricfidelity}
\end{align}
The Dicke space infidelity
\begin{align}
 \epsilon(t)
 \equiv
 1-F_{\rm D}(t)
 \label{eq:infidelitydefinition}
\end{align}
measures the leakage from the decoherence-free symmetric manifold.

On a state of total spin $J$, the eigenvalue of $\hat W$ is
\begin{align}
 W_J
 =
 \frac12
 \left[
 J_{\max}(J_{\max}+1)-J(J+1)
 \right].
 \label{eq:WJeigenvalue}
\end{align}
It vanishes for $J=J_{\max}$, has the smallest nonzero value $N_q/2$ at
$J=J_{\max}-1$, and is bounded above by $N_q(N_q+2)/8$. Consequently,
$\hat W$ bounds the projector onto the complement of the Dicke manifold
from above and below,
\begin{align}
 \frac{8}{N_q(N_q+2)}\hat W
 \leq
 \mathds{1}-\hat\Pi_{\rm D}
 \leq
 \frac{2}{N_q}\hat W.
 \label{eq:fidelityoperator}
\end{align}
This yields an upper and lower bound for the Dicke infidelity,
\begin{align}
 \frac{8W(t)}{N_q(N_q+2)}
 \leq
 \epsilon(t)
 \leq
 \frac{2W(t)}{N_q}.
 \label{eq:fidelitybound}
\end{align}
Since $\hat W\geq0$ and $\ker\hat W=\mathcal H_{\rm sym}$, $W(t)=0$
implies that $\hat\rho(t)$ is supported entirely on $\mathcal H_{\rm sym}$,
where it remains stationary.

The bounds that follow use only $\mathsf K=\mathsf K^{T}>0$ and therefore
hold on any connected measurement graph. With the pair weights
$q_{nm}(t)$ of Eq.~\eqref{eq:Kdef} collected in the vector $\bm q(t)$,
$W(t)=\bm 1^{T}e^{-\mathsf Kt}\bm q(0)$, and
$\mathsf K=\mathsf K^{T}>0$ gives
\begin{align}
 \left\lVert e^{-\mathsf Kt}\right\rVert_2
 =
 e^{-\Lambda_0 t}.
 \label{eq:semigroupnorm}
\end{align}
Moreover, $0\leq q_{nm}(0)\leq1$ implies
$\lVert\bm q(0)\rVert_2^2\leq W(0)$, while
$\lVert\bm 1\rVert_2=\sqrt{N_p}$. Hence
\begin{align}
 W(t)
 \leq
 \sqrt{N_pW(0)}
 e^{-\Lambda_0 t}.
 \label{eq:Wupper}
\end{align}
Using $N_p=N_q(N_q-1)/2$ and the state-independent bound
$W(0)\leq N_q(N_q+2)/8$, Eq.~\eqref{eq:fidelitybound} yields
\begin{align}
 \epsilon(t)
 \leq
 \frac{\sqrt{(N_q-1)(N_q+2)}}{2}
 e^{-\Lambda_0 t}.
 \label{eq:timecertificate}
\end{align}
Let $T_\epsilon$ be the time after which $\epsilon(t)\leq\epsilon$ for all
$t\geq T_\epsilon$. Equation~\eqref{eq:timecertificate} gives, for every
initial state,
\begin{align}
 T_\epsilon
 &\leq
 \frac{1}{\Lambda_0}
 \log
 \left[
 \frac{\sqrt{(N_q-1)(N_q+2)}}{2\epsilon}
 \right].
 \label{eq:Teps}
\end{align}

The asymptotic infidelity can be obtained from a decomposition into the
eigenmodes of $\mathsf K$ and their occupations. This yields
\begin{align}
 W(t)
 &=
 \sum_{j=0}^{R-1}W_je^{-\Lambda_jt} ,
 \label{eq:Wmodes}
\end{align}
where $W_j$ is the pair weight of the initial state in the mode $j$ and
$\sum_jW_j=W(0)$. Equation~\eqref{eq:fidelitybound} then bounds the
Dicke infidelity at every time,
\begin{align}
 \boxed{
 \frac{8}{N_q(N_q+2)}
 \sum_{j=0}^{R-1}
 W_je^{-\Lambda_jt}
 \leq
 \epsilon(t)
 \leq
 \frac{2}{N_q}
 \sum_{j=0}^{R-1}
 W_je^{-\Lambda_jt}.
 }
 \label{eq:twosided}
\end{align}
At long times only the slowest mode survives. Its weight $W_0$ is
positive for every initial state with nonzero pair weight, because the
slowest eigenvector of $\mathsf K$ is positive (Sec.~\ref{sec:spectrum}).
The infidelity therefore decays exactly at the rate $\Lambda_{\rm pair}$,
the two sides of Eq.~\eqref{eq:twosided} differ by a factor polynomial in
$N_q$, and the preparation time is fixed from both sides,
\begin{align}
 T_\epsilon
 =
 \frac{1}{\Lambda_{\rm pair}}
 \left[
 \log\frac{W_0}{\epsilon}
 +O(\log N_q)
 \right].
 \label{eq:Tepstwo}
\end{align}
Equation~\eqref{eq:Teps} is the guarantee for every initial state;
Eq.~\eqref{eq:Tepstwo} is two-sided and uses the initial pair weights.

An important case is the one-dimensional chain with an initial N\'eel
state: it is the product state in the sector $J^z=0$, whose Dicke state
is the most entangled one. For even $N_q$ its mode weights are
$W_j=1/\Lambda_j$, so that $W(0)=\sum_j\Lambda_j^{-1}=N_q^2/8$ and
$W_0=\Lambda_{\rm pair}^{-1}$.

Figure~\ref{fig:certificate}(a) tests the fidelity bound against the
dynamics it is meant to bound. The dots are the Dicke infidelity of the
full measurement--feedback dynamics for $N_q=8,12,16$ starting from the
N\'eel state, averaged over $10^5$ quantum trajectories, and the lines the
exact Lindblad evolution in the $J^z=0$ sector. The shaded band is
Eq.~\eqref{eq:twosided} at $N_q=16$, evaluated from the $R$ eigenvalues of
the reduced generator alone. The many-body infidelity lies inside the band
at all times, and its late-time slope is the pair-sector rate
$\Lambda_{\rm pair}$. Figure~\ref{fig:certificate}(b) shows the
corresponding closed-form relaxation $p_r(t)$ of Eq.~\eqref{eq:pclosed} at
$N_q=32$, beyond the reach of the many-body simulation.

\subsection{The Dicke manifold forms the complete decoherence-free subspace}
\label{sec:dfs}

The evolution of the pair projectors $\hat P_{nm}$ determines the
relaxation to the fully symmetric Dicke subspace and sets time-dependent
fidelity bounds. In this section, we show that the Dicke manifold is the
only stationary manifold, i.e., the complete decoherence-free subspace. In
Sec.~\ref{sec:stationaryensemble}, we use this result to construct the
unique stationary state for each initial state of the preparation
protocol.

For any magnetization $J^z=m$, the Dicke state $\lvert D_m\rangle$ is
fully symmetric and hence annihilated by every pair projector,
$\hat P_\alpha\lvert D_m\rangle=0$. That means, for any pair $\alpha$,
$\hat K_{\alpha0}=\mathds{1}-\hat P_\alpha$ acts as the identity on the
Dicke states, whereas $\hat K_{\alpha1}=\hat V_\alpha\hat P_\alpha$
annihilates them. Hence
\begin{align}
 \Phi_\alpha
 \left[
 \lvert D_m\rangle\langle D_{m'}\rvert
 \right]
 &=
 \lvert D_m\rangle\langle D_{m'}\rvert,
 &
 \mathcal L
 \left[
 \lvert D_m\rangle\langle D_{m'}\rvert
 \right]
 &=0.
 \label{eq:dfs}
\end{align}
The symmetric manifold $\mathcal H_{\rm sym}$ of
Eq.~\eqref{eq:symmetricmanifold} is therefore a decoherence-free subspace
\cite{Zanardi1997,Lidar1998}, and
\begin{align}
 \mathcal B(\mathcal H_{\rm sym})
 \subseteq
 \ker\mathcal L.
 \label{eq:dfsalgebra}
\end{align}
In particular, the stationary space contains all $(N_q+1)^2$ operators
$\lvert D_m\rangle\langle D_{m'}\rvert$. We show below that for this protocol they exhaust the kernel.

\begin{thmbox}{Lemma}\label{lem:stationaryspace}
(Stationary space.)\;
Let feedback of strength $\lambda=\sin^2\varphi>0$ act on every measured bond,
and let
\begin{align}
 \mathcal K
 &\equiv
 \bigcap_\alpha\ker\hat P_\alpha
 \label{eq:commonkernel}
\end{align}
be the common kernel of the measured projectors. Then
\begin{align}
 \ker\mathcal L
 &=
 \mathcal B(\mathcal K).
 \label{eq:kernelequality}
\end{align}
If the measured bonds form a connected graph, then
$\mathcal K=\mathcal H_{\rm sym}$ and $\dim\ker\mathcal L=(N_q+1)^2$.
\end{thmbox}

The proof is given in Appendix~\ref{app:stationaryproof}; its physical
content is the following. By the loss identity
$\mathcal L^\dagger\hat W=-\lambda\sum_\alpha\hat P_\alpha$, a stationary
state carries no pair weight on the measured bonds. Measurement moves pair
weight along the connected measurement graph, so a state with no pair
weight on the measured bonds can only be stationary if it has no pair
weight at all, $W=0$, that is, if it lies in $\mathcal H_{\rm sym}$. The
same argument applies to the coherences, which gives
Eq.~\eqref{eq:kernelequality}. The loss identity also excludes modes that
oscillate without decaying, so every initial state converges to a
stationary state. Both conditions of the Lemma are necessary: without
feedback, $\lambda=0$, every collective-spin operator is stationary, and
for a disconnected measurement graph each component of the graph relaxes to its own
Dicke manifold.

For any initial state, the asymptotic density matrix therefore has the form
\begin{align}
 \hat\rho_\infty
 =
 \sum_{m,m'}
 c_{mm'}^{(\infty)}
 \lvert D_m\rangle\langle D_{m'}\rvert.
 \label{eq:dfsasymptotic}
\end{align}
The coefficients $c^{(\infty)}_{mm'}$ are determined in Sec.~\ref{sec:stationaryensemble}.

\subsection{The kernel of $\mathcal L^\dagger$ and the stationary
ensemble}
\label{sec:stationaryensemble}

Lemma~\ref{lem:stationaryspace} identifies the kernel of $\mathcal L$: it
is spanned by the $(N_q+1)^2$ operators
$\lvert D_m\rangle\langle D_{m'}\rvert$ on the Dicke manifold. Here we
determine the stationary state a given initial state relaxes to.

Since the generator $\mathcal L$ is not normal, the projection onto its
kernel is not the orthogonal one. It is fixed by the zero modes of the
adjoint generator, normalized against the operators
$\lvert D_m\rangle\langle D_{m'}\rvert$,
\begin{align}
 \mathcal L^\dagger\hat C_{mm'}
 &=
 0,
 &
 \operatorname{Tr}
 \!\left[
 \hat C_{mm'}
 \lvert D_n\rangle\langle D_{n'}\rvert
 \right]
 &=
 \delta_{mn}\delta_{m'n'} .
 \label{eq:leftdef}
\end{align}
The second condition assigns to each stationary operator its own
coefficient, $c_{mm'}(t)=\operatorname{Tr}[\hat C_{mm'}\hat\rho(t)]$, and
the first condition makes this coefficient a conserved quantity,
$\partial_tc_{mm'}(t)=0$. The asymptotic coefficients of
Eq.~\eqref{eq:dfsasymptotic} are therefore determined by the initial
state,
\begin{align}
 c^{(\infty)}_{mm'}
 &=
 c_{mm'}(t)
 =
 \operatorname{Tr}
 \!\left[
 \hat C_{mm'}\hat\rho(0)
 \right].
 \label{eq:stationaryform}
\end{align}
With the convention below, $\hat C_{mm'}^\dagger=\hat C_{m'm}$, so
$c^{(\infty)}_{m'm}=c^{(\infty)*}_{mm'}$ and $\hat\rho_\infty$ is
Hermitian. We now construct the operators $\hat C_{mm'}$.

\begin{thmbox}{Proposition}\label{prop:conservedalgebra}
Let the feedback be $\hat V_{\ell m}=e^{i\varphi\sigma_\ell^z}$ and let
$\hat\Pi_m$ be the projector onto the sector $J^z=m$,
\begin{align}
 \hat\Pi_m
 &=
 \prod_{m''\neq m}
 \frac{\hat J^z-m''}{m-m''} .
 \label{eq:magprojector}
\end{align}
Define, for $J=N_q/2$ and $k=m'-m\geq0$,
\begin{align}
 \hat C_{mm'}
 &=
 \frac{(\hat J^{+})^{k}\hat\Pi_m}
 {\prod_{j=0}^{k-1}\sqrt{(J-m-j)(J+m+j+1)}} ,
 \label{eq:conservedalgebra}
\end{align}
and $\hat C_{mm'}=\hat C_{m'm}^\dagger$ for $k<0$. Then:

(i) For every angle $\varphi$ and every set of measured bonds,
Eq.~\eqref{eq:leftdef} holds.

(ii) If $\lambda=\sin^2\varphi>0$ and the measured bonds form a connected
graph, these $(N_q+1)^2$ operators form a basis of
$\ker\mathcal L^\dagger$.
\end{thmbox}

\noindent\emph{Proof.}
Every collective operator commutes with the constraint projectors, since
$\hat P_\alpha$ is built from a SWAP and is therefore $\mathrm{SU}(2)$
invariant: $[\hat J^\mu,\hat P_\alpha]=0$. Hence, for any operator
$\hat O$ constructed from the collective operators, Eq.~\eqref{eq:adjsplit}
reduces to
\begin{align}
 \mathcal L^\dagger\hat O
 &=
 \sum_\alpha\gamma_\alpha
 \left[
 \hat P_\alpha
 \hat V_\alpha^\dagger\hat O\hat V_\alpha
 \hat P_\alpha
 -\hat O\hat P_\alpha
 \right].
 \label{eq:collectiveadjoint}
\end{align}
The unitary transformation by
$\hat V_\alpha=\cos\varphi+i\sin\varphi\,\sigma_\ell^z$, where $\ell$ is
the site on which the feedback of bond $\alpha$ acts, sends a single
raising operator to a pure phase,
$\hat V_\alpha^\dagger\sigma_\ell^+\hat V_\alpha
=e^{-2i\varphi}\sigma_\ell^+$, so that
$\hat V_\alpha^\dagger\hat J^+\hat V_\alpha
=\hat J^++\xi\,\sigma_\ell^+$ with
$\xi=e^{-2i\varphi}-1$. Using $(\sigma_\ell^+)^2=0$, one finds
\begin{align}
 \hat V_\alpha^\dagger(\hat J^+)^k\hat V_\alpha
 &=
 (\hat J^+)^k
 +k\,\xi\,\sigma_\ell^+(\hat J^+)^{k-1}.
 \label{eq:ladderconjugation}
\end{align}
Since $\hat\Pi_m$ is a polynomial in $\hat J^z$, it commutes with
$\sigma_\ell^z$ and with $\hat P_\alpha$. Inserting
Eq.~\eqref{eq:ladderconjugation} into Eq.~\eqref{eq:collectiveadjoint} and
using $[\hat J^+,\hat P_\alpha]=0$ leaves the single term
$k\xi(\hat J^+)^{k-1}\hat\Pi_m\,
\hat P_\alpha\sigma_\ell^+\hat P_\alpha$, which vanishes by
Eq.~\eqref{eq:feedbackcompression}: the singlet has no one-spin
polarization, so $\hat P_\alpha\sigma^\pm_\ell\hat P_\alpha=0$. Hence
$\mathcal L^\dagger\hat C_{mm'}=0$. Acting on the Dicke manifold,
$\hat\Pi_m\lvert D_n\rangle=\delta_{mn}\lvert D_m\rangle$ and
$(\hat J^+)^k\lvert D_m\rangle
=\lvert D_{m+k}\rangle\prod_{j=0}^{k-1}\sqrt{(J-m-j)(J+m+j+1)}$, so the
normalization in Eq.~\eqref{eq:conservedalgebra} is the one that enforces
the second condition in Eq.~\eqref{eq:leftdef}. The $(N_q+1)^2$ operators
are therefore linearly independent. Under the hypotheses of part (ii),
Lemma~\ref{lem:stationaryspace} gives
$\dim\ker\mathcal L^\dagger=\dim\ker\mathcal L=(N_q+1)^2$, so they exhaust
the kernel. \hfill$\square$

Equations~\eqref{eq:stationaryform} and \eqref{eq:conservedalgebra}
together give the stationary ensemble explicitly. With $k=m'-m\geq0$,
\begin{align}
 \boxed{
 c^{(\infty)}_{mm'}
 =
 \frac{
 \big\langle
 (\hat J^+)^{k}\hat\Pi_m
 \big\rangle_{\hat\rho(0)}
 }{
 \prod_{j=0}^{k-1}
 \sqrt{(J-m-j)(J+m+j+1)}
 } ,
 }
 \label{eq:coherencerule}
\end{align}
and $c^{(\infty)}_{mm'}=c^{(\infty)*}_{m'm}$ for $m'<m$. For $k=0$ one has
$\hat C_{mm}=\hat\Pi_m$, and $c^{(\infty)}_{mm}$ is the conserved
population of the sector $J^z=m$. Occupations and coherences alike are
collective-spin moments of the initial state, and the protocol retains
this information.

Two consequences follow. First, the mixedness of the stationary ensemble
is explicit,
\begin{align}
 \operatorname{Tr}\hat\rho_\infty^2
 &=
 \sum_{m,m'}
 \big\lvert c^{(\infty)}_{mm'}\big\rvert^2 .
 \label{eq:stationarypurity}
\end{align}
The stationary state is pure if and only if the conserved coefficient
matrix $c^{(\infty)}$ has rank one. A pure initial state need not satisfy
this condition and can therefore relax to a mixed stationary ensemble.
Conversely, support in several magnetization sectors does not by itself
imply mixedness; what matters is whether the conserved inter-sector
coefficients form a rank-one matrix.

Second, if the initial state is diagonal in the $\sigma^z$ product basis,
including classical mixtures and, more generally, states without
magnetization coherence, all moments of the ladder operators $\hat J^\pm$
vanish and
\begin{align}
 \hat\rho_\infty
 &=
 \sum_m p_m\lvert D_m\rangle\langle D_m\rvert,
 &
 S(\hat\rho_\infty)
 &=
 H(\{p_m\}),
 \label{eq:stationaryentropy}
\end{align}
with $p_m=\operatorname{Tr}[\hat\Pi_m\hat\rho(0)]$ the initial
magnetization distribution, $S$ the von Neumann entropy and $H$ the
Shannon entropy. In this case the protocol transfers the initial classical
uncertainty in the conserved magnetization directly to the stationary
Dicke ensemble without generating additional mixing.

The stationary manifold, i.e., the kernel of $\mathcal L$, is the same for
every feedback direction. However, the dual basis is not: for feedback
about an axis $\hat n$, i.e., for
$\hat V_\alpha=\cos\varphi+i\sin\varphi\,\hat n\cdot\bm\sigma_\ell$, the
conserved operators are the collective-spin moments in the frame rotated
to that axis. The same initial state therefore relaxes to different points
of the same manifold under $\sigma^z$ and $\sigma^x$ feedback.

\subsection{Liouvillian gap and relation to the parent Hamiltonian}
\label{sec:liouvilliangap}

The Liouvillian gap is the slowest nonzero decay rate,
\begin{equation}
 \Delta_{\rm Liou}
 \equiv \min_{\substack{\zeta\in\operatorname{spec}(\mathcal L)\\
                         \zeta\neq0}}(-\operatorname{Re}\zeta).
 \label{eq:liouvilliangapdefinition}
\end{equation}
Let $D=\sum_m\lvert D_m\rangle\langle D_m\rvert$ project onto the Dicke
manifold and $R=\mathds{1}-D$. The four blocks of $\rho$ are
$\rho_D=D\rho D$, $\rho_R=R\rho R$, and the coherences $X=D\rho R$
and $X^\dagger$. Since $\hat P_\alpha D=0$, the feedback term
vanishes on $X$, and
$\mathcal L(X)=-X\hat G$ and $\mathcal L(X^\dagger)=-\hat G X^\dagger$,
where $\hat G=\sum_\alpha\gamma_\alpha\hat P_\alpha$.
The slowest homogeneous coherence decay rate $\Lambda_{\rm coh}$ is
therefore the smallest positive eigenvalue of $\hat G$.
Feedback can populate $\rho_D$ and generate coherences from $\rho_R$;
the coherences feed neither block, and states supported on $D$ are
stationary. The Liouvillian is thus block triangular, and its spectrum
is the union of the spectra of its diagonal blocks.

The decay of the autonomous $\rho_R$ block follows directly from
Eqs.~\eqref{eq:fidelitybound} and \eqref{eq:Wupper}:
\[
 \|\rho_R(t)\|_1=\Tr\rho_R(t)=\epsilon(t)
 \leq \frac{2W(t)}{N_q}=O(e^{-\Lambda_0t}).
\]
Positive operators span this block, so no mode there decays more slowly.
The pair eigenoperator at $-\Lambda_0$ is supported entirely on $R$,
so this rate is attained: $\Lambda_R=\Lambda_0$. Together the blocks
give $\Delta_{\rm Liou}=\min\{\Lambda_0,\Lambda_{\rm coh}\}$.
For the SWAP measurement and feedback protocol on a connected weighted
graph with at least one active feedback bond,
Appendix~\ref{app:coherencegap} shows that
$\Lambda_{\rm coh}\geq\Lambda_0$~\cite{Zhou2026private}. Thus
\begin{equation}
 \Delta_{\rm Liou}=\Lambda_0.
 \label{eq:gapidentity}
\end{equation}

On the periodic chain at unit measurement rate, $\hat G$ is the parent
Hamiltonian
\begin{equation}
 \hat H=\sum_\ell\hat P_{\ell,\ell+1}
 =\frac14\sum_\ell
 \left(\mathds{1}-\bm\sigma_\ell\cdot\bm\sigma_{\ell+1}\right).
 \label{eq:parenthamiltonian}
\end{equation}
Its gap is $\Lambda_{\rm coh}=\Delta_H=1-\cos(2\pi/N_q)$.
At maximal feedback, Eq.~\eqref{eq:closedgap} gives
$\Lambda_{\rm pair}<\Lambda_{\rm coh}$, so
\begin{equation}
 \Delta_{\rm Liou}=\Lambda_{\rm pair}
 =\frac{\Delta_H}{2\cos^2\!\left(\frac{\pi}{2N_q}\right)}.
 \label{eq:gapratio}
\end{equation}
Both rates scale as $N_q^{-2}$, with momenta $2\pi/N_q$ for the
lowest magnon and $\pi/N_q$ for the absorbing pair dynamics.

\section{Controlling hydrodynamics: random and coherent transport}
\label{sec:scrambling}

The exact pair closure identifies transport to the absorber as the
dynamical bottleneck of the FFC protocol with local operations.
Accelerating transport by other means than measurement and feedback may
hence be a suitable strategy to accelerate preparation with alternative
resources. This makes target-preserving unitary dynamics a natural control
resource: it can modify the redistribution of pair weight while leaving
the Dicke manifold and the absorber unchanged, and it may be applied
concurrently with the measurement--feedback process. We consider two
distinct realizations. Stochastic partial-SWAP events preserve the pair
closure and renormalize its transport coefficient, allowing the resulting
acceleration to be solved exactly. Coherent exchange via the exact parent
Hamiltonian acts through a commutator and does not preserve the pair
sector on the full Hilbert space. Single-qubit feedback removes at most
one singlet at a time, so the final step into the Dicke manifold removes
the last singlet. We determine the effect of coherent exchange on this
decay exactly within the one-singlet restriction. In the magnetization
sectors $m=\pm(N_q/2-1)$, which allow at most one singlet, the result
applies throughout preparation.

\subsection{Random partial-SWAP events}

We supplement the measurement--feedback dynamics by independent unitary
events that (i) preserve the Dicke state manifold and (ii) require local
resources. The simplest such realization is random partial-SWAP
unitaries
\begin{align}
 \hat U_\ell(\varphi)
 &=
 e^{i\varphi\hat S_{\ell,\ell+1}}
 =
 \mathds{1}\cos\varphi
 +i\hat S_{\ell,\ell+1}\sin\varphi,
 \label{eq:partialswap}
\end{align}
applied on bond $\ell$ at random times, with the angle $\varphi$ drawn
from a distribution symmetric about zero. Since
$\hat S_{\ell,\ell+1}\lvert D_m\rangle=\lvert D_m\rangle$, every Dicke
state acquires only a phase, so the events act trivially on the Dicke
manifold. Averaged over the angle, an event applies the full SWAP
$\hat S_{\ell,\ell+1}$ with probability $\overline{\sin^2\varphi}$ and
does nothing otherwise; the cross term $\propto\sin\varphi\cos\varphi$
averages to zero for a symmetric distribution. Random partial-SWAP events
are therefore equivalent to full SWAPs applied at the rate
$\gamma_{\rm scr}$, the event rate times $\overline{\sin^2\varphi}$, and
contribute to the adjoint generator
\begin{align}
 \left.\mathcal L^\dagger\hat O\right|_{\rm scr}
 &=
 \gamma_{\rm scr}
 \sum_\ell
 \left(
 \hat S_{\ell,\ell+1}\hat O\hat S_{\ell,\ell+1}-\hat O
 \right).
 \label{eq:scradjoint}
\end{align}
This is the measurement term of Eq.~\eqref{eq:measwalk} with
$\gamma_\alpha/2$ replaced by $\gamma_{\rm scr}$: the events move the
endpoints of a pair exactly as the measurements do, and by
Sec.~\ref{sec:endpoint} they preserve the pair closure and the products of
pair projectors. Adding Eq.~\eqref{eq:scradjoint} to
Eq.~\eqref{eq:exactpair},
\begin{align}
 \mathcal L^\dagger\hat P_{nm}
 &=
 \tfrac12\underbrace{\left(
 1+2\gamma_{\rm scr}
 \right)}_{\mathbb D_{N_q}}
 (\Delta_{\rm pair}\hat P)_{nm}
 -\lambda\,\mathbf 1_{\mathcal F}(nm)\hat P_{nm}.
 \label{eq:scrfullpair}
\end{align}
The events change only the transport matrix; the absorption matrix remains
$\lambda\operatorname{diag}(\mathbf 1_{\mathcal F})$. For an asymmetric
angle distribution the cross term survives and adds a coherent exchange
$\propto\overline{\sin\varphi\cos\varphi}\sum_\ell[\hat S_{\ell,\ell+1},\cdot]$,
which is the subject of Sec.~\ref{sec:coherent}.

On the periodic chain, Eqs.~\eqref{eq:foldleft}--\eqref{eq:foldright}
hold with every transport term multiplied by
\begin{align}
 \mathbb D_{N_q}
 =
 1+2\gamma_{\rm scr}(N_q),
 \label{eq:Dtotal}
\end{align}
the diffusion coefficient of the pair separation. We indicate that the rate $\gamma_{\rm scr}(N_q)$ may be chosen to scale with the number of qubits.  The factor of two is
the ratio of the two coefficients in Eq.~\eqref{eq:scrfullpair}. Equation
\eqref{eq:Dtotal} is the exact form of the diffusion
enhancement identified in Ref.~\cite{Doerstel2026}. For a size-independent
rate, $\mathbb D_{N_q}=O(1)$, the diffusive prefactor changes while the
dynamical exponent remains $z=2$. The consequences of a size-dependent
transport coefficient for the spectrum are derived next.

\subsection{Exact crossover from transport-limited to absorber-limited relaxation}
\label{sec:crossover}

\begin{figure*}[t]
\centering
\includegraphics[width=\textwidth]{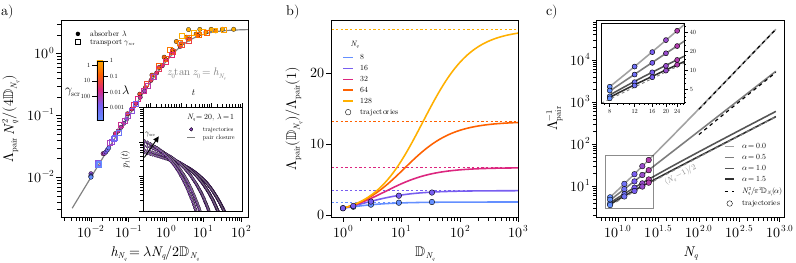}
\caption{\textbf{Transport-to-absorber crossover of the preparation rate.}
Rates are quoted in units of the measurement rate.
(a) The compensated rate $\Lambda_{\rm pair}N_q^{2}/4\mathbb{D}_{N_q}$ depends on
the microscopic protocol only through $h_{N_q}=\lambda N_q/2\mathbb{D}_{N_q}$,
Eq.~\eqref{eq:hLscr}: filled symbols vary the removal rate $\lambda$ at
fixed transport, open symbols vary the partial-SWAP rate $\gamma_{\rm scr}$
(hence $\mathbb{D}_{N_q}$) at fixed $\lambda=1$. Both collapse onto the root of
$z_0\tan z_0=h_{N_q}$, Eq.~\eqref{eq:crossoverfunction}. Inset: the pair weight on the measured bonds,
$p_1(t)$ at $N_q=20$, $\lambda=1$ for $\gamma_{\rm scr} =0,0.5, 2, 8, 32$, from the full measurement--feedback quantum simulation (dots, ``trajectories'')
and from the reduced pair generator (lines, ``pair closure''). The Lindbladian and the reduced pair generator 
coincide, Eq.~\eqref{eq:exactpair}.
(b) Both axes are normalized by the bare protocol, $\mathbb{D}_{N_q}=1$:
at fixed $N_q$, increasing $\mathbb{D}_{N_q}$ drives the rate toward the
absorber bound $\Lambda_{\rm max}=2\lambda/(N_q-1)$,
Eq.~\eqref{eq:reactionlimitedscr}, which in this normalization is the dashed
plateau $\Lambda_{\rm max}/\Lambda_{\rm pair}(1)$. Lines are the
reduced generator for $N_q=8,\ldots,128$; the circles, at the two smallest
sizes, are least-squares slopes of $\log p_1(t)$ from quantum-trajectory
averages, each divided by its own $\gamma_{\rm scr}=0$ run. 
(c) $\Lambda^{-1}_{\rm pair}$ for the size-dependent partial-SWAP rate
$\gamma_{\rm scr}=N_q^{\alpha}/2$, i.e.\ Eq.~\eqref{eq:scrscaling} with
$\gamma_0=1/2$, exhibit $z_{\rm prep}=\max\{2-\alpha,1\}$,
Eq.~\eqref{eq:zalpha}; the dashed lines are the transport-limited asymptote
$N_q^2/\pi^2\mathbb{D}_{N_q}(\alpha)$ and the absorber-limited floor
$(N_q-1)/2$. Circles are again quantum-trajectory averages at the
accessible sizes. The inset enlarges the range of sizes accessible to the many-body simulation. For all quantum data, the error bars are smaller than the symbols and are omitted.}
\label{fig:scrambling}
\end{figure*}

Equation~\eqref{eq:Dtotal} shows that random partial-SWAP events modify
the pair dynamics only through the transport coefficient
$\mathbb D_{N_q}$, while the removal and the boundary conditions are
those of Sec.~\ref{sec:spectrum}. Repeating this calculation gives the
exact spectrum and quantization condition
\begin{align}
 \Lambda(k)
 &=
 2\mathbb D_{N_q}(1-\cos k),
 \label{eq:scrdispersion}\\
 \cos\left(\frac{kN_q}{2}\right)
 &=
 \left(
 1-\frac{\lambda}{\mathbb D_{N_q}}
 \right)
 \cos\left[
 k\left(\frac{N_q}{2}-1\right)
 \right].
 \label{eq:scrquantization}
\end{align}
The smallest positive solution $k_0$ of Eq.~\eqref{eq:scrquantization}
determines the exact pair-sector gap,
\begin{align}
 \Lambda_{\rm pair}(N_q)
 =
 2\mathbb D_{N_q}
 \left(
 1-\cos k_0
 \right).
 \label{eq:scrgap}
\end{align}

This controls the crossover from transport-limited to
absorber-limited relaxation through the ratio
\begin{align}
 h_{N_q}
 &\equiv
 \frac{\lambda N_q}{2\mathbb D_{N_q}}.
 \label{eq:hLscr}
\end{align}
Large $h_{N_q}$ corresponds to transport-limited relaxation, while small
$h_{N_q}$ corresponds to rapid mixing followed by absorption from an
approximately uniform pair distribution.

We determine the crossover scaling function by taking
$N_q\rightarrow\infty$ at fixed $h_{N_q}$ and writing
$k_0=2z_0/N_q+O(N_q^{-2})$. Expanding the exact quantization condition
\eqref{eq:scrquantization} gives
\begin{align}
 z_0\tan z_0
 &=
 h_{N_q},
 &
 \Lambda_{\rm pair}
 &=
 \frac{4\mathbb D_{N_q}}{N_q^2}
 z_0(h_{N_q})^2
 \left[
 1+O(N_q^{-1})
 \right].
 \label{eq:crossoverfunction}
\end{align}

We now take
\begin{align}
 \gamma_{\rm scr}(N_q)
 &=
 \gamma_0N_q^\alpha,
 &
 \mathbb D_{N_q}
 &=
 1+2\gamma_0N_q^\alpha,
 \qquad
 \alpha\geq0,
 \label{eq:scrscaling}
\end{align}
and consider three cases.
(i) For $0\leq\alpha<1$, $h_{N_q}\rightarrow\infty$. The smallest root of
Eq.~\eqref{eq:scrquantization} then satisfies
$k_0=(\pi/N_q)[1+o(1)]$ and therefore
\begin{align}
 \Lambda_{\rm pair}
 &=
 \frac{\pi^2\mathbb D_{N_q}}{N_q^2}
 \left[
 1+o(1)
 \right]
 \sim
 N_q^{\alpha-2},
 \qquad
 0\leq\alpha<1.
 \label{eq:diffusionlimitedscr}
\end{align}
(ii) At $\alpha=1$, $\mathbb D_{N_q}=2\gamma_0N_q[1+o(1)]$ and
$h_{N_q}\rightarrow\lambda/4\gamma_0$. Hence $z_0$ approaches the root of
$z_0\tan z_0=\lambda/4\gamma_0$, a constant, and
Eq.~\eqref{eq:crossoverfunction} gives
\begin{align}
 \Lambda_{\rm pair}&=
 \frac{8\gamma_0z_0^2}{N_q}
 \left[
 1+o(1)
 \right]\sim N_q^{-1}.
 \label{eq:criticalrate}
\end{align}
(iii) For $\alpha>1$, $h_{N_q}\rightarrow0$. Transport equilibrates the
pair weight before absorption, and the lowest mode becomes uniform over
the pair graph. Its decay rate approaches the removal rate averaged over
all pair configurations,
\begin{align}
 \Lambda_{\rm pair}
 &=
 \frac{2\lambda}{N_q-1}
 \left[
 1+o(1)
 \right]
 \sim
 \frac{2\lambda}{N_q},
 \qquad
 \alpha>1.
 \label{eq:reactionlimitedscr}
\end{align}
This is the saturation of the absorber bound \eqref{eq:localfloor}.
Increasing the transport beyond this regime cannot improve either the
leading exponent or its asymptotic prefactor.

Combining the three regimes gives the preparation exponent,
$\Lambda_{\rm pair}\sim N_q^{-z_{\rm prep}}$ with
\begin{align}
 \boxed{
 z_{\rm prep}
 =
 \max\{2-\alpha,1\}.
 }
 \label{eq:zalpha}
\end{align}
For $0\leq\alpha<1$, preparation is transport limited: the walker needs
the time $\sim N_q^{2-\alpha}$ to reach the absorber. At $\alpha=1$,
transport and absorption compete at the same order and
Eq.~\eqref{eq:criticalrate} fixes the prefactor. For $\alpha>1$,
preparation is absorber limited: the size of the absorber relative to the
configuration space sets the rate. The $z_{\rm prep}=1$ plateau therefore
follows from the transport-independent absorber bound and applies to any
symmetric, weight-conserving transport within the closed pair sector.

Figure~\ref{fig:scrambling} collects these results and, at the sizes
where both are available, compares them with the full
measurement--feedback Lindbladian. Panel (a) shows the collapse of the
compensated rate onto the single variable $h_{N_q}$, with an inset
comparing the pair weight on the measured bonds, $p_1(t)$, obtained
from the many-body dynamics with the same quantity obtained from the
reduced generator. Panels (b) and (c) repeat the comparison for the rates
themselves.

\emph{Resource scaling.} Acceleration below the bare $O(N_q^2)$ scale uses
a size-dependent additional resource. Let $T_{\rm prep}$ be the time to
reach a fixed target infidelity, $N_{\rm M}=N_qT_{\rm prep}$ the number of
measurement attempts at the total rate $N_q$, and
$N_{\rm scr}=N_q\gamma_{\rm scr}T_{\rm prep}\propto N_q^{1+\alpha}T_{\rm prep}$
the number of partial-SWAP events. Ignoring logarithmic accuracy factors,
for $0\leq\alpha\leq1$ one has $T_{\rm prep}\sim N_q^{2-\alpha}$,
$N_{\rm M}\sim N_q^{3-\alpha}$, and $N_{\rm scr}\sim N_q^3$. For
$\alpha>1$ the preparation time remains $O(N_q)$ while
$N_{\rm scr}\sim N_q^{\alpha+2}$. Thus additional partial-SWAP events
beyond $\alpha=1$ increase cost without improving elapsed time. Complete
resource accounting, including accuracy dependence, is given in
Sec.~\ref{sec:resourcecomparison} and Appendix~\ref{app:resources}.

\subsection{Coherent exchange and one-singlet relaxation}
\label{sec:coherent}

As a second option to enhance transport, we consider coherent exchange,
as induced, e.g., by the parent Hamiltonian. It differs from
random partial-SWAP events in two ways: (i) it preserves the closure of the summed pair weights only in the one-singlet sector; for more than one singlet it yields a
many-body problem which no longer offers a closed solution, and (ii) in
the single-singlet sector, it provides no speedup for the preparation
process, and the witness $\hat W$ is affected only in its third time
derivative. Whether coherent exchange modifies the relaxation at finite
singlet density therefore remains an open many-body question.

Consider the parent Hamiltonian
\begin{align}
 \hat H_{\rm ex}
 &=
 J_{\rm ex}\sum_{\ell=1}^{N_q}\hat S_{\ell,\ell+1}
 =
 J_{\rm ex}N_q\,\mathds{1}-2J_{\rm ex}\hat P_1,
 \label{eq:HJ}
\end{align}
with $\hat P_r$ the summed pair operators of Eq.~\eqref{eq:Pr}. Its
contribution to their Heisenberg evolution is
\begin{align}
 \left.\partial_t\hat P_r\right|_{\hat H_{\rm ex}}
 &=
 i[\hat H_{\rm ex},\hat P_r]
 =
 2iJ_{\rm ex}[\hat P_r,\hat P_1],
 \label{eq:coherentpair}
\end{align}
which for $r>1$ is not a combination of pair projectors: coherent exchange
does not preserve the pair closure and cannot be absorbed into the
transport coefficient as the random events were.

However, two exact results can be obtained: First, $\hat W$ is
permutation invariant and $\hat P_1$ is a function of $\hat H_{\rm ex}$,
so
\begin{align}
 [\hat H_{\rm ex},\hat W]
 &=
 0,
 &
 [\hat H_{\rm ex},\hat P_1]
 &=
 0.
 \label{eq:coherentlosscommutators}
\end{align}
The loss identity $\mathcal L^\dagger\hat W=-\lambda\hat P_1$ is therefore
unchanged, and coherent exchange first enters the dynamics of the witness
through $[\hat H_{\rm ex},\hat P_2]$, in the third time derivative of
$\hat W$.

Second, consider the one-singlet space at magnetization $m$, spanned by
the states $\lvert b_\ell\rangle$ with a singlet on bond $\ell$ and the
remaining spins in the Dicke state $\lvert D_m^{(N_q-2)}\rangle$. The
lattice translation by one site, $\hat T\lvert b_\ell\rangle=\lvert b_{\ell+1}\rangle$,
acts within this space, and the summed pair operators restricted to it
are (Appendix~\ref{app:onesinglet})
\begin{align}
 \hat P_r
 &=
 \mathds{1}
 -\frac12
 \left(
 \hat T^r+\hat T^{-r}
 \right).
 \label{eq:Prstandard}
\end{align}
All $\hat P_r$, and hence $\hat H_{\rm ex}$, are functions of the same
translation operator, so
\begin{align}
 [\hat H_{\rm ex},\hat P_r]=0
 \qquad
 \text{for every }r
 \label{eq:Prcommute}
\end{align}
within the one-singlet sector. Coherent exchange can propagate a
bond-resolved wave packet, but it leaves every summed pair weight and
therefore Eqs.~\eqref{eq:foldleft}--\eqref{eq:foldright} unchanged,
independently of $J_{\rm ex}$: confined to the one-singlet sector, all decay rates, among them
$\Lambda_{\rm pair}$, are those of the bare protocol.

The one-singlet restriction can be enforced in the full dynamics by
working at $m=\pm(N_q/2-1)$, corresponding to one spin flipped relative
to full polarization. Magnetization is conserved, and each of these
sectors contains only the Dicke state and the one-singlet subspace.

The absence of a pair closure in sectors carrying more than one singlet
suggests that many-body preparation may be modified by coherent
exchange, while the single-singlet sector is strictly invariant. A
related scheme, coherent exchange in the Heisenberg chain with a
dissipator on a single site, was studied in Ref.~\cite{Beer2026}, where
the single-excitation sector yields a preparation time $\tau\sim N_q^3$.
Compared with the single-pair first-passage time for diffusive transport
to a single feedback bond, $\tau\sim N_q^2\log N_q$
(Sec.~\ref{sec:localized}, Table~\ref{tab:geometry}), coherent transport
is subleading: at fixed $J_{\rm ex}$, even where it modifies the
many-body dynamics, we do not expect it to change the leading order of
the preparation time. We leave the study of coherent transport at finite
singlet density for future research.

\begin{figure*}[t]
 \centering
  \includegraphics[width=0.249\textwidth]{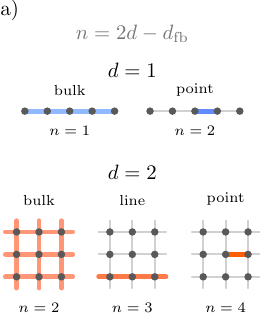}
 \includegraphics[width=0.745\textwidth]{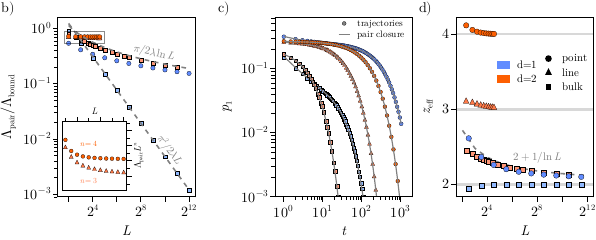}
\caption{\textbf{Feedback geometry controls preparation.} Restricting
feedback raises the transverse dimension to
$n=2d-d_{\rm fb}$, Eq.~\eqref{eq:codim}. (a)~Feedback geometries; feedback
bonds are colored and thick.
(b)~Absorber efficiency $\eta=\Lambda_{\rm pair}/\Lambda_{\rm bound}$, the
lowest eigenvalue of the exact pair generator $\mathsf K$,
Eq.~\eqref{eq:localizedexact}, divided by the absorber bound
$\Lambda_{\rm bound}=\lambda|\mathcal F|/N_p$ of
Proposition~\ref{prop:absorber}. The bound is saturated up to a constant for
the transient geometries, while recurrence costs a factor
$\pi/[2\lambda\ln L]$ at $n=2$ and $\pi^2/[2\lambda L]$ at $n=1$ (dashed).
Inset: the compensated rates $\Lambda_{\rm pair}L^{n}$ for the two transient
cases $n=3$ and $n=4$, which approach constants.
(c)~Pair weight $p_1(t)$ on the measured bonds for the same geometries, from the full
measurement--feedback Lindbladian (symbols, ``quantum'') and from the reduced
pair generator (lines, ``classical''), at the sizes where the many-body
evolution is accessible. The two coincide over three decades in $p_1$,
Eq.~\eqref{eq:exactpair}.
(d)~Local effective exponents
$z_{\rm eff}=-\Delta\log\Lambda_{\rm pair}/\Delta\log L$ from
successive sizes. The recurrent $n=1$ case approaches $z_{\rm eff}=2$;
the transient $n=3$ and $n=4$ cases approach $3$ and $4$; and the two
marginal $n=2$ cases drift according to
$z_{\rm eff}\simeq2+1/\log L$ (dashed). Because the reduced generator is a
sparse classical matrix, panels (b) and (d) reach sizes far beyond any
many-body simulation; panel (c) shows that where the two overlap they
coincide.}
 \label{fig:transportgeometry}
\end{figure*}

\section{Transport range and feedback geometry}
\label{sec:longrange}

The decomposition $\mathsf K=\mathsf M+\mathsf A$ separates two
geometrically distinct modifications of the protocol: the transport of
pair endpoints, encoded in $\mathsf M$ and caused by measurements or by
random SWAP unitaries, and the absorption of pairs by local feedback,
encoded in $\mathsf A$. Restricting the spatial support of feedback, with
the measurements and random SWAPs unchanged, changes the absorber $\mathsf A$ while
leaving the bulk transport unchanged. Extending the set of bonds on which
measurements or SWAP unitaries act changes the endpoint transport and,
when feedback is applied on the same bonds, also enlarges the absorber.
The pair closure remains exact under both deformations: the preparation
problem remains an absorbing walk on the $N_p$ pair configurations, with
a modified $\mathsf M$ or $\mathsf A$. Here we derive from this walk the
scaling of the preparation time with the qubit number $N_q$ for different
transport and absorber geometries. The results are relevant for
experiments whose measurement and feedback are confined to a subset of
the system, and for experiments with access to extended measurements and
unitaries.

\subsection{Localized feedback and transverse dimension}
\label{sec:localized}

We first discuss how the spatial support of feedback changes the preparation time~\cite{Langbehn2024}. We consider measurements (or SWAPs) acting on every nearest-neighbor bond of a \(d\)-dimensional hypercubic lattice with linear size \(L\) and \(N_q=L^d\) qubits, while feedback acts only on an absorber \(\mathcal F\) formed by a regular set of bonds. We characterize this set by its dimension \(d_{\rm fb}\), so that \(|\mathcal F|\sim L^{d_{\rm fb}}\): feedback throughout the bulk has \(d_{\rm fb}=d\), feedback on a physical boundary has \(d_{\rm fb}=d-1\), and feedback confined to a finite region, e.g., an impurity, has \(d_{\rm fb}=0\). Measurements move the pair endpoints throughout the lattice, but a pair can be removed only when its endpoints form a bond in \(\mathcal F\). The exact operator closure remains
\begin{align}
 \mathcal L^\dagger\hat{\bm P}
 &=
 -\left[
 \mathsf M
 +\lambda\operatorname{diag}
 \left(\mathbf 1_{\mathcal F}\right)
 \right]\hat{\bm P},
 \label{eq:localizedexact}
\end{align}
where \(\mathsf M\) describes the endpoint motion generated by measurements and, when included, random partial-SWAP events.

Each endpoint has \(d\) position coordinates, so a pair moves in a \(2d\)-dimensional configuration space. The feedback bonds occupy a set extended along \(d_{\rm fb}\) directions in this space. Motion along these directions does not bring the pair closer to feedback; reaching a feedback bond requires motion in the remaining
\begin{align}
 n=2d-d_{\rm fb}
 \label{eq:codim}
\end{align}
directions, which we call transverse directions. On a regular geometry, this absorption problem reduces to a random walk in \(n\) transverse dimensions.

This maps the preparation problem to a known first-passage problem: Its asymptotic large-\(L\) behavior depends on the return probability of the corresponding infinite-lattice walk in $n$ dimensions. For $n=1,2$, the walker returns to its starting point with probability one; these walks are recurrent. For $n>2$, there is a nonzero probability of never returning; the walk is transient \cite{Polya1921}. These different return properties give the standard asymptotic relaxation rates \cite{Spitzer1964,Redner2001,Holcman2014},
\begin{align}
 \Lambda_{\rm pair}(L)
 &\propto L^{-2},
 &&n=1,
 \nonumber\\
 \Lambda_{\rm pair}(L)
 &\propto \frac{1}{L^2\log L},
 &&n=2,
 \nonumber\\
 \Lambda_{\rm pair}(L)
 &\propto L^{-n},
 &&n>2,
 \label{eq:capacity}
\end{align}
where the precise proportionality coefficients depend on the feedback geometry and microscopic rates. Table~\ref{tab:geometry} gives the resulting pair relaxation times \(\tau_{\rm pair}=\Lambda_{\rm pair}^{-1}\) for bulk, boundary, and finite-region feedback. For bulk feedback, \(n=d\) and \(N_q=L^d\), so \(\tau_{\rm pair}\sim N_q^2\), \(N_q\log N_q\), and \(N_q\) in \(d=1\), \(d=2\), and \(d\geq3\); the one-dimensional law agrees with the exact chain spectrum in Eq.~\eqref{eq:closedgap}. Geometries with the same transverse dimension share the same scaling with \(L\), even when their physical dimensions differ: a single feedback bond in one dimension and bulk feedback in two dimensions both have \(n=2\) and \(\tau_{\rm pair}\sim L^2\log L\).

\begin{table}[t]
 \caption{Asymptotic scaling of the pair relaxation time for regular feedback geometries. \(L\) is the linear size of the physical lattice, with \(N_q=L^d\). Microscopic rates and the transverse thickness of the feedback set are held fixed.}
 \label{tab:geometry}
 \begin{ruledtabular}
 \begin{tabular}{c|ccc}
  feedback & bulk & finite region & boundary \\
  & (\(d_{\rm fb}=d\)) & (\(d_{\rm fb}=0\)) & (\(d_{\rm fb}=d-1\)) \\ \hline
  \(d=1\)
  & \(L^2\)
  & \(L^2\log L\)
  & \(L^2\log L\) \\
  \(d=2\)
  & \(L^2\log L\)
  & \(L^4\)
  & \(L^3\) \\
  \(d>2\)
  & \(L^d\)
  & \(L^{2d}\)
  & \(L^{d+1}\)
 \end{tabular}
 \end{ruledtabular}
\end{table}

The absorber bound of Proposition~\ref{prop:absorber} shows how these laws compare with the fastest relaxation allowed by the feedback geometry. Since \(N_p=N_q(N_q-1)/2\) and \(|\mathcal F|\sim L^{d_{\rm fb}}\),
\begin{align}
 \Lambda_{\rm pair}
 &\leq
 \frac{2\lambda|\mathcal F|}{N_q(N_q-1)}
 \sim\lambda L^{-n}.
 \label{eq:localreact}
\end{align}
For \(n>2\), diffusive endpoint motion attains this system-size dependence, although the coefficient remains geometry dependent. For \(n=2\), repeated returns slow relaxation by an additional factor \(\log L\); for \(n=1\), they add a factor \(L\). The number of feedback bonds therefore fixes the fastest possible scaling, while the return properties of the endpoint motion determine whether it is attained.

Figure~\ref{fig:transportgeometry} tests these predictions against the exact spectra of Eq.~\eqref{eq:localizedexact} for rings with feedback on all bonds or on a single bond, and for periodic \(L\times L\) lattices with bulk, line, or point feedback. The sparse pair generator allows calculations at linear sizes far beyond those accessible to full many-body evolution. Panels (b) and (d) test the predicted powers and logarithmic corrections. Panel (c) compares the measured-bond pair weight \(p_1(t)\) obtained from the closed generator with full Lindblad evolution and shows agreement over the resolved time window. The spectra are exact within the invariant pair-operator space; the first-passage laws describe their large-\(L\) behavior.

\subsection{Long-range monitored transport}
\label{sec:longrangetransport}

Finally, we extend the measured bonds beyond nearest neighbors while
keeping every measured bond subject to feedback. For every displacement
$\bm a\neq0$ of the periodic lattice, every bond $(n,n+\bm a)$ is measured
at rate $\gamma_{\bm a}$, with $\gamma_{\bm a}=\gamma_{-\bm a}$; the nearest-neighbor protocol is
$\gamma_{\pm\bm e_\mu}=1$ and $\gamma_{\bm a}=0$ otherwise. Each measured
bond moves an endpoint of a pair by $\pm\bm a$ as in
Sec.~\ref{sec:endpoint}, and the corresponding feedback absorbs a pair of
separation $\bm a$ at rate $\lambda\gamma_{\bm a}$, so the pair closure
holds for any range profile,
\begin{align}
 \mathcal L^\dagger\hat P_{nm}
 =&
 \frac12
 \sum_{\substack{\bm a\neq0\\n+\bm a\neq m}}\gamma_{\bm a}
 \left(
 \hat P_{n+\bm a,m}-\hat P_{nm}
 \right)
 \label{eq:longexact}\\&+\frac12
 \sum_{\substack{\bm a\neq0\\m+\bm a\neq n}}\gamma_{\bm a}
 \left(
 \hat P_{n,m+\bm a}-\hat P_{nm}
 \right)
 -\lambda\gamma_{m-n}\hat P_{nm}.\nonumber
\end{align}
Transport and absorber are set by the same profile, and in the notation
of Proposition~\ref{prop:absorber}, $a_{nm}=\lambda\gamma_{m-n}$. With
$\Gamma=\sum_{\bm a\neq0}\gamma_{\bm a}$ the measurement rate per site,
$\sum_ia_i=\lambda N_q\Gamma/2$, and the absorber bound reads
\begin{align}
 \Lambda_0\leq\frac{\lambda\,\Gamma}{N_q-1}.
 \label{eq:longrangebound}
\end{align}
At fixed measurement rate per site, no range profile relaxes faster than
$O(N_q^{-1})$. The chain has $\Gamma=2$, Eq.~\eqref{eq:localfloor}, and
all-to-all monitoring at the same $\Gamma$ saturates the bound,
Sec.~\ref{sec:resourcecomparison}. Between the two limits, the exponent
of the preparation time is a continuous function of the range. On the
chain at maximal feedback, Eq.~\eqref{eq:longexact} reduces to a
long-range random walk of the pair separation with an absorbing origin,
whose slowest rate is the inverse mode sum of the walk within a factor of
two (Appendix~\ref{app:longrange}). For algebraic profiles
$\gamma_a\propto|a|^{-\beta}$ at fixed $\Gamma$ this gives
\begin{align}
 \tau_{\rm pair}\sim N_q^{z},
 \qquad
 z=\max\{1,\min(\beta-1,2)\},
 \label{eq:longrangescaling}
\end{align}
with logarithmic corrections at $\beta=2$ and $\beta=3$: profiles with
$\beta<2$ attain the absorber-limited scaling of
Eq.~\eqref{eq:longrangebound}, profiles with $\beta>3$ are diffusive like
the nearest-neighbor chain, and for $2<\beta<3$ the relaxation is
transport limited with the intermediate exponent $z=\beta-1$. The
crossover was studied numerically in Ref.~\cite{Doerstel2026} from the
generator~\eqref{eq:longexact}.

\section{Noise and autonomous stabilization}
\label{sec:imperfections}

Since quantum platforms are subject to noise, such as decoherence and measurement and gate imperfections, we characterize not only the noiseless relaxation rate of FFC, but also the stationary leakage maintained under ongoing errors and the recovery rate following a disturbance. We obtain both quantities from a microscopic single-spin noise channel. We show that isotropic Pauli errors deform the pair generator affinely, so that the transient dynamics and the stationary profile remain exactly solvable. The weak-noise expansion of the stationary profile bounds the stationary infidelity and separates it into two contributions, one governed by transport and the other by feedback removal, and the decay rates of the deformed generator give the recovery after a disturbance (Sec.~\ref{sec:affine}).

\subsection{Isotropic single-spin noise}

Let us consider the action of isotropic single-spin errors, occurring at the total Pauli-error rate $\kappa$ per site. They are described by an additional term in the master equation
\begin{align}
 \mathcal L_{\rm imp}\hat\rho
 &=
 \frac{\kappa}{3}
 \sum_{i,a}
 \left(
 \sigma_i^a\hat\rho\sigma_i^a-\hat\rho
 \right).
 \label{eq:isotropic}
\end{align}
Each individual noise term maps a singlet into exactly one of the three triplet states, such that their isotropic combination yields a maximally mixed state in the triplet sector. The corresponding action of the adjoint on the pair projector is (setting $\tilde\kappa=2\kappa/3$)
\begin{align}
 \mathcal L_{\rm imp}^\dagger\hat P_{ij}
 =
 \tilde\kappa
 \left(
 \mathds{1}-4\hat P_{ij}
 \right).
 \label{eq:impadj}
\end{align}
The stationary value under the error channel alone is therefore
$\langle\hat P_{ij}\rangle=1/4$, as required for a maximally mixed pair.
Since the channel does not conserve $J^z$, the fidelity bound
\eqref{eq:fidelitybound} bounds the infidelity with the symmetric manifold
rather than with a specific Dicke state; the $J^z$-preserving fixed-axis
channel is treated in Appendix~\ref{sec:dephasing}.

\begin{figure*}[t]
 \centering
 \includegraphics[width=\textwidth]{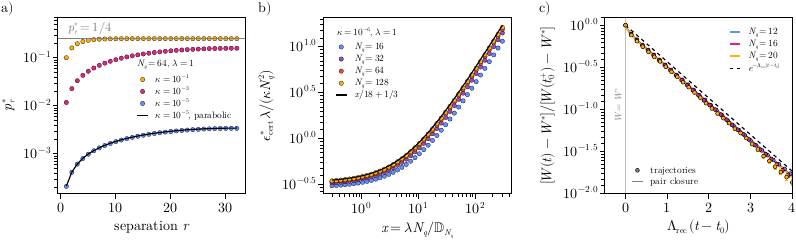}
 \caption{\textbf{Stationary state and autonomous recovery under isotropic
single-spin noise, Eq.~\eqref{eq:isotropic}.}
(a) Exact stationary pair profile $p^{*}_r$ at $N_q=64$, $\lambda=1$ for
three noise rates, obtained from the exact solution
Eq.~\eqref{eq:coshprofile}. For $\kappa\to0$ the profile becomes
parabolic, Eq.~\eqref{eq:coshprofilekappa0} (black line, transport-limited form for $\kappa=10^{-5}$). At large $\kappa$ it saturates at the
depolarized value $p^{*}_r=1/4$ with a boundary layer at the absorber.
(b) The stationary leakage bound $\epsilon^{*}_{\rm cert}$,
Eq.~\eqref{eq:epscertdef}, approaches a single crossover law in the
weak-noise, large-$N_q$ regime:
$\epsilon^{*}_{\rm cert}\lambda/(\kappa N_q^{2})$ collapses toward $x/18+1/3$
with $x=\lambda N_q/\mathbb{D}_{N_q}$, the two terms being the transport-limited and
absorber-limited contributions of Eq.~\eqref{eq:fidelityasymptotic}.
(c) Autonomous recovery from an instantaneous isotropic Pauli error applied on the stationary state of $\kappa=10^{-5}$ with unit probability to a single site $n$ at $t_0$, which maps
$\langle\hat P_{nm}\rangle\to(1-\langle\hat P_{nm}\rangle)/3$ on every pair $(n,m)$
containing that site, Eq.~\eqref{eq:impadj}: at late times, the excess ratio
$[W(t)-W^{*}]/[W(t_0^{+})-W^{*}]$ for $N_q=12,16,20$ (solid line) approaches the
predicted exponential with asymptotic rate (dashed line)
$\Lambda_{\rm rec}=\Lambda_{\rm pair}+8\kappa/3$,
Eq.~\eqref{eq:recoveryrate}. The trajectory average from the full quantum simulation (dots) agrees with the pair closure prediction. The grey line marks $t=t_0$.}
 \label{fig:noise}
\end{figure*}

\subsection{Affine pair dynamics, stationary leakage, and recovery}
\label{sec:affine}

For the summed pair operators $\hat P_r$ of Eq.~\eqref{eq:Pr}, the noiseless
closure is Eqs.~\eqref{eq:foldleft}--\eqref{eq:foldright}; we denote its
generator, the restriction of Eq.~\eqref{eq:Kdef} to the
translation-invariant sector, again by $\mathsf K$. Summing
Eq.~\eqref{eq:impadj} over the $N_q$ pairs at separation $r$ adds to this
closure a multiple of the identity and a uniform decay,
\begin{align}
 \mathcal L^\dagger\hat{\bm P}
 =
 -\left(
 \mathsf K+4\tilde\kappa
 \right)\hat{\bm P}
 +\tilde\kappa N_q,
 \label{eq:affineoperator}
\end{align}
which is again an exact operator identity. Taking the trace against the
state and writing $p_r=\langle\hat P_r\rangle/N_q$ for the pair
densities gives the affine equation
\begin{align}
 \partial_t\bm p
 =
 -\left(
 \mathsf K+4\tilde\kappa
 \right)\bm p
 +\tilde\kappa .
 \label{eq:affine}
\end{align}
Its stationary profile and complete transient are
\begin{align}
 \bm p^*
 &=
 \tilde\kappa\left(
 \mathsf K+4\tilde\kappa
 \right)^{-1}\bm 1,
 \label{eq:affstat}\\
 \bm p(t)
 &=
 \bm p^*
 +e^{-(\mathsf K+4\tilde\kappa)t}
 \left[
 \bm p(0)-\bm p^*
 \right].
 \label{eq:afftrans}
\end{align}
Every noiseless eigenvalue is shifted by \(4\tilde\kappa\), while the
eigenvectors remain unchanged.

Define
\begin{align}
 \cosh\theta
 &=
 1+\frac{2\tilde\kappa}{\mathbb D_{N_q}},
 \label{eq:thetadef}\\
 B
 &=
 (\lambda+4\tilde\kappa)
 \cosh\left[
 \theta(\tfrac{N_q}{2}-1)
 \right]
 \nonumber\\
 &\quad+
 \mathbb D_{N_q}
 \left\{
 \cosh\left[
 \theta(\tfrac{N_q}{2}-1)
 \right]
 -
 \cosh\left[
 \theta(\tfrac{N_q}{2}-2)
 \right]
 \right\}.
 \label{eq:Bdef}
\end{align}
The exact stationary solution is
\begin{align}
 \boxed{
 p_r^*
 =
 \frac14
 -
 \frac{\lambda}{4B}
 \cosh\left[
 \theta(\tfrac{N_q}{2}-r)
 \right],
 \qquad
 1\leq r\leq R.
 }
 \label{eq:coshprofile}
\end{align}
It interpolates continuously between the absorbing state at
$\kappa=0$ and the depolarized value \(p_r^*=1/4\) when the error channel
dominates. Expanding the equation around $\kappa=0$ yields
\begin{align}
    p^*_r \sim \frac{\kappa}{3} \left( \frac{(N_q-r-1)(r-1)}{\mathbb{D}_{N_q}} + \frac{N_q-1}{\lambda}\right),
    \label{eq:coshprofilekappa0}
\end{align}
a parabola in $r$ on top of a uniform offset. Both limits, and the boundary layer that develops at the
absorber, are shown in Fig.~\ref{fig:noise}(a).

We define the stationary leakage bound by
\begin{align}
 \epsilon_{\rm cert}(t)
 &\equiv
 \frac{2W(t)}{N_q},
 &
 \epsilon_{\rm cert}^*
 &\equiv
 \frac{2W^*}{N_q},
 \label{eq:epscertdef}
\end{align}
where
$W^*=N_q\big[\sum_{r=1}^{R-1}p_r^*+p_R^*/(1+\delta_{N_q,2R})\big]$ is the
stationary value of the witness, Eq.~\eqref{eq:Wfolded}, evaluated on the
profile of Eq.~\eqref{eq:coshprofile}, so that
$1-F_{\rm D}(t)\leq\epsilon_{\rm cert}(t)$. Summing
Eq.~\eqref{eq:coshprofilekappa0} over $r$ gives $\epsilon^*_{\rm cert}$
to first order in $\kappa$ at fixed $N_q$; the complete expression is
Eq.~\eqref{eq:fidelitycriterion} in Appendix~\ref{app:imperfections}. At
large $N_q$ it reduces to
\begin{align}
 \epsilon_{\rm cert}^*
 &=
 \frac{\kappa N_q^3}{18\mathbb D_{N_q}}
 +
 \frac{\kappa N_q^2}{3\lambda}
 +O\left(
 \frac{\kappa N_q^2}{\mathbb D_{N_q}}
 +\frac{\kappa N_q}{\lambda}, \kappa^2
 \right).
 \label{eq:fidelityasymptotic}
\end{align}
The two terms are the two parts of the profile
\eqref{eq:coshprofilekappa0}. The first, the parabola, is the transport
contribution: pair weight created by noise in the bulk has to be carried to
the absorber. The second, the offset, is the removal contribution: pair
weight that has reached the absorber waits there for removal at the finite
rate $\lambda$. The expansion requires
$\kappa(N_q^3/\mathbb D_{N_q}+N_q^2/\lambda)\ll1$, up to the numerical
coefficients displayed, which is the same condition as
$\epsilon^*_{\rm cert}\ll1$.

Which contribution dominates is decided by the ratio
$x=\lambda N_q/\mathbb D_{N_q}$: rescaling
Eq.~\eqref{eq:fidelityasymptotic} gives
$\epsilon_{\rm cert}^*\lambda/(\kappa N_q^2)\to x/18+1/3$, onto which the
exact values collapse over three decades in $x$ and a decade in $N_q$,
Fig.~\ref{fig:noise}(b). In the two limits,
\begin{align}
 \epsilon_{\rm cert}^*
 &\simeq
 \frac{\kappa N_q^3}{18\mathbb D_{N_q}},
 &&\mathbb D_{N_q}\ll\lambda N_q
 \quad\text{(transport-limited)},
 \label{eq:fidelitydiffusive}\\
 \epsilon_{\rm cert}^*
 &\simeq
 \frac{\kappa N_q^2}{3\lambda},
 &&\mathbb D_{N_q}\gg\lambda N_q
 \quad\text{(absorber-limited)}.
 \label{eq:fidelityreaction}
\end{align}
Both limits are the same law in terms of the preparation exponent of
Eq.~\eqref{eq:zalpha}: for $\mathbb D_{N_q}\sim N_q^\alpha$ with
$\alpha\geq0$, $z_{\rm prep}=\max\{2-\alpha,1\}$ and
\begin{align}
 \epsilon_{\rm cert}^*
 &\sim
 \kappa N_q^{z_{\rm prep}+1}.
 \label{eq:fidelityz}
\end{align}
The origin of this law is transparent: errors occur at the total rate
$O(\kappa N_q)$, and the pair weight created by each of them persists for
one relaxation time, $\Lambda_{\rm pair}^{-1}\sim N_q^{z_{\rm prep}}$,
before it is removed. The stationary leakage is the product of the two,
and high-fidelity stabilization requires $\kappa N_q^{z_{\rm prep}+1}\ll1$.
A single pair density is smaller by one power of $N_q$,
$p_r^*\sim\kappa N_q^{z_{\rm prep}}$, because the witness sums over
$O(N_q)$ separations. The protocol therefore
stabilizes the symmetric manifold with a bounded leakage, but the
tolerable error rate decreases with system size; there is no
size-independent error threshold.

Finally, Eq.~\eqref{eq:afftrans} fixes the recovery after a disturbance:
every initial profile relaxes toward $\bm p^*$ with the asymptotic rate
\begin{align}
 \Lambda_{\rm rec}
 &=
 \Lambda_{\rm pair}
 +\frac{8\kappa}{3},
 \label{eq:recoveryrate}
\end{align}
the slowest noiseless rate plus the uniform decay $4\tilde\kappa$ that the
noise adds to every pair mode. Figure~\ref{fig:noise}(c) shows this for a
single-site isotropic Pauli error applied to the stationary state, which
changes every pair containing the affected site according to
Eq.~\eqref{eq:impadj} and leaves all other pair observables unchanged. The
resulting inhomogeneous profile is evolved in the full $N_p$-dimensional
pair space and returns to $\bm p^*$; the excess ratio
$[W(t)-W^*]/[W(t_0^+)-W^*]$ approaches $e^{-\Lambda_{\rm rec}(t-t_0)}$ at
late times. The exact balance equations for $W$ and
$\langle\bm J^2\rangle$ under isotropic noise are given in
Appendix~\ref{app:imperfections}.

\section{Exact ensemble fluctuations and conserving noise}
\label{sec:fluctuations}

The pair closure determines the mean constraint violations and bounds
the infidelity, but does not determine their correlations or the
variance of the preparation witness $\hat W$. We obtain these from
an exact second-moment closure of polynomial size at maximal Pauli
feedback. We then identify how measurement and feedback contribute
to the covariance: measurements redistribute fluctuations while
conserving the total pair weight, whereas feedback generates
fluctuations associated with removal and changes correlations
between pairs. Together, the mean and covariance equations describe
the transport and absorption of constraint violations together with
their microscopic noise.

\subsection{Covariance equation and positive source}
\label{sec:covariance}

Let us first define the covariances of the pair projectors
and derive their equation of motion. We use the symmetrized covariance
in the ensemble state $\hat\rho(t)$,
\begin{align}
 \delta\hat P_a(t)
 &\equiv
 \hat P_a-\avg{\hat P_a}_t,
 \nonumber\\
 \mathsf C^S_{ab}(t)
 &\equiv
 \frac12
 \avg{
 \left\{
 \delta\hat P_a(t),
 \delta\hat P_b(t)
 \right\}
 }_t,
 \label{eq:symcovariance}
\end{align}
with $a,b$ running over the $N_p$ pairs. It characterizes the
fluctuations of measurements performed on the ensemble state.
In particular, it contains the fluctuations of the witness
$\hat W=\sum_a\hat P_a$,
\begin{align}
 \operatorname{Var}_{\hat\rho(t)}(\hat W)
 &\equiv
 \avg{\hat W^2}_t-W(t)^2
 =
 \sum_{a,b}\mathsf C^S_{ab}(t).
 \label{eq:Wvariance}
\end{align}

The pair projectors obey the closed equation
$\mathcal L^\dagger\hat{\bm P}=-\mathsf K\hat{\bm P}$,
Eq.~\eqref{eq:Kdef}. Applying the adjoint generator to the products
$\hat P_a\hat P_b$ with the Leibniz rule and its exact remainder,
Eq.~\eqref{eq:lindbladleibniz}, gives
\begin{align}
 \partial_t\mathsf C^S
 &=
 -\mathsf K\mathsf C^S
 -\mathsf C^S\mathsf K^T
 +\mathsf N^S.
 \label{eq:coveq}
\end{align}
The first two terms propagate the covariance under the same
dynamics that determines the means. The source $\mathsf N^S$
is the remainder of the Leibniz rule, with the jump operators
$\hat K_{\alpha s}$ of Eq.~\eqref{eq:finitechannel},
\begin{align}
 \mathsf N^S_{ab}(t)
 &=
 \frac12
 \sum_{\alpha,s}\gamma_\alpha
 \avg{
 [\hat P_a,\hat K_{\alpha s}]^\dagger
 [\hat P_b,\hat K_{\alpha s}]
 +
 \text{H.c.}
 }_t.
 \label{eq:Nsym}
\end{align}
It gives the rate at which measurement and feedback generate
covariance. The source is positive semidefinite: for every real
vector $\bm x$, and with
$|\hat O|^2\equiv\hat O^\dagger\hat O$,
\begin{align}
 \bm x^T\mathsf N^S\bm x
 &=
 \sum_{\alpha,s}\gamma_\alpha
 \avg{
 \Big|
 \sum_bx_b[\hat P_b,\hat K_{\alpha s}]
 \Big|^2
 }_t
 \geq0.
 \label{eq:Npositive}
\end{align}

\subsection{Polynomial second-moment closure}
\label{sec:momentclosure}

To determine the covariance in Eq.~\eqref{eq:coveq} exactly, we need
closed equations for the second moments $\avg{\hat P_a\hat P_b}_t$.
Measurements alone preserve the number of factors in each product
and therefore close the second-moment equations. Feedback requires
an enlarged operator space containing both singlet and triplet
projectors. We show that products of at most two of these projectors
close under the full dynamics.

We denote the projector onto the $m=0$ triplet by
\begin{align}
 \hat T_{ab}
 &\equiv
 \lvert t^0_{ab}\rangle\langle t^0_{ab}\rvert
 =
 \frac12
 \left(
 \mathds{1}-\sigma_a^z\sigma_b^z
 \right)
 -\hat P_{ab}.
 \label{eq:Tdef}
\end{align}
To illustrate why triplet projectors are needed, consider feedback
on bond $(1,2)$ with $\hat V=\sigma_1^z$, acting on the product
$\hat P_{13}\hat P_{24}$. The conditional rotation contributes
\begin{align}
 &\hat P_{12}
 \left(
 \hat V\hat P_{13}\hat P_{24}\hat V^\dagger
 -
 \hat P_{13}\hat P_{24}
 \right)
 \hat P_{12}
 \nonumber\\
 &\qquad=
 \frac14\hat P_{12}
 \left(
 \hat T_{34}-\hat P_{34}
 \right).
 \label{eq:tripletproduct}
\end{align}
The right hand side cannot be expressed through singlet projectors
alone, which are invariant under collective spin rotations, whereas
$\hat T_{34}-\hat P_{34}$ distinguishes the $z$ axis.

Hence, we work with both projectors and their products,
\begin{align}
 \mathcal G
 &\equiv
 \left\{
 \hat P_{ab},\hat T_{ab}
 \right\},
 \nonumber\\
 \mathcal V_n
 &\equiv
 \operatorname{span}
 \left\{
 \mathds{1},
 \hat X_{a_1}\cdots\hat X_{a_m}:
 \hat X_{a_j}\in\mathcal G,\ 
 1\leq m\leq n
 \right\}.
 \label{eq:Vndef}
\end{align}
On each pair, the identity, singlet projector, and triplet projector
span the complete space of two-site operators that commute with $J^z$
and are invariant under the global spin flip.

Measurement preserves these spaces for products of any length. Let $\mathcal L_M^\dagger$ be the measurement contribution defined in Eq.~\eqref{eq:measwalk}.

\begin{thmbox}{Theorem}
\label{thm:filtration}
(Measurement preserves products.)\;
For arbitrary measurement geometry and rates,
\begin{align}
 \mathcal L_M^\dagger\mathcal V_n
 &\subseteq
 \mathcal V_n
 \qquad
 \text{for every }n.
 \label{eq:measurementfiltration}
\end{align}
\end{thmbox}

A measurement on bond $\eta$ acts on a product by exchanging the
two sites of $\eta$ in every factor:
\begin{align}
 &\mathcal L_M^\dagger
 \left(
 \hat X_{a_1}\cdots\hat X_{a_m}
 \right)
 \nonumber\\
 &\qquad=
 \frac12
 \sum_\eta\gamma_\eta
 \left(
 \hat X_{\pi_\eta(a_1)}
 \cdots
 \hat X_{\pi_\eta(a_m)}
 -
 \hat X_{a_1}\cdots\hat X_{a_m}
 \right),
 \label{eq:measurementproduct}
\end{align}
where $\pi_\eta$ exchanges the sites of the measured bond $\eta$.
The exchange relabels every factor but preserves the number of
factors in the product.

Including feedback, the second-moment space remains invariant.

\begin{thmbox}{Theorem}
\label{thm:closure2}
(Second-moment closure.)\;
For maximal Pauli feedback,
\begin{align}
 \mathcal L^\dagger\mathcal V_2
 &\subseteq
 \mathcal V_2.
 \label{eq:secondmomentclosure}
\end{align}
\end{thmbox}

The feedback proof reduces to four nonzero product cases, listed
in Appendix~\ref{app:covariance}. Each produces at most two singlet
or triplet projectors.

The theorem reduces the second-moment dynamics from an exponentially
large operator algebra to a space of dimension
\begin{align}
 \dim\mathcal V_2
 &=
 O(N_q^4).
 \label{eq:V2dimension}
\end{align}
Choose a basis $\{\hat Y_\mu\}$ of $\mathcal V_2$ containing the identity,
and let $\mathsf K^{(2)}$ be the matrix of the adjoint generator on it,
\begin{align}
 \mathcal L^\dagger\hat Y_\mu
 &=
 -\sum_\nu
 \mathsf K^{(2)}_{\mu\nu}\hat Y_\nu.
 \label{eq:K2definition}
\end{align}
Writing $y_\mu(t)=\avg{\hat Y_\mu}_t$, the exact second-moment dynamics
is the finite linear system
\begin{align}
 \partial_t\bm y
 &=
 -\mathsf K^{(2)}\bm y,
 \qquad
 \bm e_{\mathds{1}}^T\mathsf K^{(2)}=0,
 \label{eq:Y2}
\end{align}
where $\bm e_{\mathds{1}}$ selects the identity operator, whose
expectation value is conserved. Unlike the pair generator $\mathsf K$,
$\mathsf K^{(2)}$ need not be symmetric or positive definite.
Its solution contains every expectation value
$\avg{\hat P_a\hat P_b}_t$ and every product entering the source,
and therefore determines both $\mathsf C^S(t)$ and $\mathsf N^S(t)$
in Eq.~\eqref{eq:coveq}.

\subsection{Conserving measurement noise and feedback removal noise}
\label{sec:noisestructure}

Measurements move pair endpoints, and feedback removes pairs on
measured bonds. We now identify the fluctuations generated by each
process. We specialize to a one-dimensional chain with
nearest-neighbor bond measurements and set the homogeneous
measurement rate to unity.

Let $a,b$ label pair projectors, let $\ell$ label the measured bond
$(\ell,\ell+1)$, and let $n,m$ denote sites outside that bond.
All expectation values below are taken in $\hat\rho(t)$.
For a single jump operator $\hat K$, define its contribution to
the source by
\begin{align}
 \mathsf N^S[\hat K]_{ab}
 &\equiv
 \frac12
 \left\langle
 [\hat K^\dagger,\hat P_a][\hat P_b,\hat K]
 +\text{H.c.}
 \right\rangle.
 \label{eq:singlejumpsource}
\end{align}

The two Kraus operators on bond $\ell$ are
$\hat K_{\ell0}=\mathds{1}-\hat P_{\ell,\ell+1}$ and
$\hat K_{\ell1}=\hat V_\ell\hat P_{\ell,\ell+1}$.
Since
$\mathsf N^S[\mathds{1}-\hat P_{\ell,\ell+1}]
=\mathsf N^S[\hat P_{\ell,\ell+1}]$,
the complete source matrix contributed by this bond is
\begin{align}
 \mathsf N^{S,(\ell)}
 &=
 \mathsf N^S[\hat K_{\ell0}]
 +
 \mathsf N^S[\hat K_{\ell1}]
  \label{eq:Nsourcesplit}\\
 &=
 \underbrace{
 2\mathsf N^S[\hat P_{\ell,\ell+1}]
 }_{\mathsf N_{\rm meas}^{S,(\ell)}}
 +
 \underbrace{
 \mathsf N^S[\hat V_\ell\hat P_{\ell,\ell+1}]
 -
 \mathsf N^S[\hat P_{\ell,\ell+1}]
 }_{\mathsf N_{\rm fb}^{S,(\ell)}}.\nonumber
\end{align}
The first term is the source from measurement alone, and the second
is its change under conditional feedback. The source contributions
$\mathsf N^S[\hat K_{\ell0}]$ and
$\mathsf N^S[\hat K_{\ell1}]$ are positive semidefinite, so their
sum $\mathsf N^{S,(\ell)}$ is positive semidefinite.
The difference $\mathsf N_{\rm fb}^{S,(\ell)}$ need not be.
Summing over measured bonds gives the full $N_p\times N_p$ source,
\begin{align}
 \mathsf N^S
 &=
 \sum_\ell
 \left(
 \mathsf N_{\rm meas}^{S,(\ell)}
 +
 \mathsf N_{\rm fb}^{S,(\ell)}
 \right).
 \label{eq:fullnoisesource}
\end{align}

We next evaluate the measurement contribution for one fixed bond
$\ell$. Let us define
\begin{align}
 &\hat I_{\ell n}
 \equiv
 \hat P_{\ell n}-\hat P_{\ell+1,n},
 &\hat\Sigma_{\ell n}
 \equiv
 \hat P_{\ell n}+\hat P_{\ell+1,n},
 \label{eq:paircombinations}\\
 &[\hat I_{\ell n},\hat P_{\ell,\ell+1}]
 =
 -\hat I_{\ell n}\hat S_{\ell,\ell+1},
 &[\hat\Sigma_{\ell n},\hat P_{\ell,\ell+1}]
 =0.
 \nonumber
\end{align}

For a fixed measured bond $\ell$, a measurement-source entry can
be nonzero only if each of its two pair indices contains exactly
one of the sites $\ell,\ell+1$. For fixed external sites $n$ and
$m$ in the row and column pairs, respectively, there are therefore
only four nontrivial entries. They can be arranged in matrix form
\begin{align}
 \left(
 \mathsf N_{\rm meas}^{S,(\ell)}
 \right)_{\boldsymbol a,\boldsymbol b}
 &=
 \mu_{\ell;nm}
 \begin{pmatrix}
  1&-1\\
 -1& 1
 \end{pmatrix},
 \label{eq:Nmeas}\\
 \mu_{\ell;nm}
 &\equiv
 \frac14
 \left\langle
 \{\hat I_{\ell n},\hat I_{\ell m}\}
 \right\rangle.
 \nonumber
\end{align}
Here
$\boldsymbol a=(\{\ell,n\},\{\ell+1,n\})$ and
$\boldsymbol b=(\{\ell,m\},\{\ell+1,m\})$
are the ordered row and column pair indices, respectively,
and $\mu_{\ell;nm}$ is a scalar. The opposite signs express the
conservation of pair weight: fluctuations transferred into one
pair are transferred out of the other. The coefficients
$\mu_{\ell;nm}$ fix the strengths and correlations of these
transfers.

For $n=m$, the identities in
Appendix~\ref{app:covariance} give
\begin{align}
 \mu_{\ell;nn}
 &=
 \frac14
 \left\langle
 \hat P_{\ell,\ell+1}
 +\hat P_{\ell n}
 +\hat P_{\ell+1,n}
 \right\rangle.
 \label{eq:measblock}
\end{align}
Thus the diagonal entries depend only on the mean pair weights.
The entries with $n\neq m$ retain four-site correlations.
The second-moment closure determines all these entries exactly
through Eq.~\eqref{eq:Y2}.

The conservation law can be written for the whole pair graph.
Each endpoint move connects the pair configurations
$\{\ell,n\}$ and $\{\ell+1,n\}$. For the corresponding edge
$e=(\ell,n)$, define the incidence matrix
\begin{align}
 \mathsf B_{ea}
 &\equiv
 \delta_{a,\{\ell,n\}}
 -
 \delta_{a,\{\ell+1,n\}},
 &
 \hat I_e
 &=
 \sum_a\mathsf B_{ea}\hat P_a.
 \label{eq:pairincidence}
\end{align}
Its rows label endpoint moves and its columns label pair
configurations. The mobility matrix on these edges is
\begin{align}
 \boldsymbol\mu_{(\ell,n),(\ell',m)}
 &\equiv
 \delta_{\ell\ell'}\mu_{\ell;nm}.
 \label{eq:mobilitykernel}
\end{align}
It is symmetric and positive semidefinite,
$\boldsymbol\mu=\boldsymbol\mu^T\geq0$. For any real vector $\bm x$,
\[
 \bm x^T\boldsymbol\mu\bm x
 =
 \frac12\sum_\ell
 \left\langle
 \left(
 \sum_nx_{(\ell,n)}\hat I_{\ell n}
 \right)^2
 \right\rangle
 \geq0.
\]
Assembling the blocks gives the exact divergence form
\begin{align}
 \boxed{
 \mathsf N_{\rm meas}^S
 =
 \mathsf B^T\boldsymbol\mu\,\mathsf B.
 }
 \label{eq:conservingnoise}
\end{align}
Every row of $\mathsf B$ sums to zero, so
\begin{align}
 \mathsf N_{\rm meas}^S\bm 1_{\rm v}
 &=0,
 \label{eq:noiseconservation}
\end{align}
where $\bm 1_{\rm v}$ is the vector of ones over all pair
configurations. Since $\hat W=\sum_a\hat P_a$, this states that
measurement noise redistributes pair weight without generating
fluctuations in the total pair weight. This is an exact conservation law.

Feedback produces fluctuations associated with removal.
For the rotation
$\hat V_\ell=\exp(i\varphi\sigma_\ell^z)$, the mean removal rate
on bond $\ell$ is
$\lambda\langle\hat P_{\ell,\ell+1}\rangle$, with
$\lambda=\sin^2\varphi$. The operator $\hat K_{\ell0}$ commutes
with $\hat P_{\ell,\ell+1}$ and contributes no source to this
observable. The local source entry is
\begin{align}
 \left(
 \mathsf N^{S,(\ell)}
 \right)_{\{\ell,\ell+1\},\{\ell,\ell+1\}}
 &=
 \lambda\langle\hat P_{\ell,\ell+1}\rangle.
 \label{eq:reactionnoise}
\end{align}
This is generic to absorbing state transport: while the transport contribution of the noise is conserving, the absorbing dynamics adds a local, nonconserving, multiplicative contribution that vanishes identically in the target state. 

The classical random walk determines the mean pair weights, but the
quantum covariance requires more than the noise of local absorption.
Feedback also changes fluctuations of neighboring pairs, whose
singlet projectors do not commute with the measured projector. At maximal feedback, $\hat V_\ell=i\sigma_\ell^z$, the mean of
$\hat\Sigma_{\ell n}=\hat P_{\ell n}+\hat P_{\ell+1,n}$ is
unchanged by operations on bond $\ell$, but its variance receives
a nonzero contribution:
\begin{align}
 \left.\partial_t\langle\hat\Sigma_{\ell n}\rangle
 \right|_\ell
 &=0,
 \nonumber\\
 \left.
 \partial_t
 \operatorname{Var}_{\hat\rho(t)}(\hat\Sigma_{\ell n})
 \right|_\ell
 &=
 \frac12\langle\hat P_{\ell,\ell+1}\rangle.
 \label{eq:feedbacksumvariance}
\end{align}
Here $\left.\partial_t\right|_\ell$ includes only the contributions
from $\hat K_{\ell0}$ and $\hat K_{\ell1}$. Measurement alone
preserves both the mean and variance of this sum. The variance
generated in Eq.~\eqref{eq:feedbacksumvariance} therefore comes
from feedback: removing a singlet changes correlations between
the pairs attached to the two endpoints, even though their mean
sum is unchanged.

Removal fluctuations are also correlated with the imbalance
$\hat I_{\ell n}=\hat P_{\ell n}-\hat P_{\ell+1,n}$.
The source coupling the measured constraint to this imbalance is
the difference of two matrix entries,
\begin{align}
 &\left(
 \mathsf N^{S,(\ell)}
 \right)_{\{\ell,\ell+1\},\{\ell,n\}}
 -
 \left(
 \mathsf N^{S,(\ell)}
 \right)_{\{\ell,\ell+1\},\{\ell+1,n\}}=
 \frac12\langle\hat I_{\ell n}\rangle.
 \label{eq:removalimbalance}
\end{align}
Thus the local shot noise in Eq.~\eqref{eq:reactionnoise} is
accompanied by fluctuations in neighboring pair weights and
correlations with them. The complete three-site source matrix
is given in Appendix~\ref{app:covariance}. Entries involving
distinct external sites contain four-site correlations, which
are supplied by the second-moment closure. The same appendix
gives an exact quantum Langevin representation of the source.

These results specify the noise associated with the transport
and removal of constraint violations. In Eq.~\eqref{eq:coveq},
the pair generator propagates existing correlations, while
$\mathsf N^S$ describes the covariance generated by measurement
and feedback. The measurement contribution has the divergence
form of conserving current noise~\cite{Carollo2017,BernardJin2019}. Feedback supplies the shot
noise of local removal together with its correlations with
other pair weights. Their state-dependent strengths are fixed
by the exact second-moment closure, which also determines the
variance of the preparation witness through
Eq.~\eqref{eq:Wvariance}. This is the microscopic
description of fluctuating absorbing hydrodynamics. The exact
mobility retains correlations between different pairs; a local
continuum description additionally requires determining how
these correlations behave at large scales.

Finally, every pair projector annihilates the Dicke manifold.
The pair covariance and both source contributions therefore
vanish once the ensemble state is supported on that manifold.
The absorbing character of the target applies to both the mean
constraint violations and their fluctuations.

\section{Implications, resources, and implementation}
\label{sec:broader}

\subsection{Implications for preparation}
\label{sec:relatedtheory}

The preparation of a Dicke state by local measurement and feedback is
limited by two competing processes, transport and removal. Constraint
violations are pairs; measurements move their endpoints through the
$N_p\sim N_q^2$ pair configurations, and feedback removes a pair only on
the $|\mathcal F|$ configurations that are feedback bonds. Faster endpoint
motion, by random unitaries or by extended measurements, shortens the time
to reach the absorber until the removal rate itself limits preparation. At
that point the bound of Proposition~\ref{prop:absorber},
$\Lambda_0\leq\lambda|\mathcal F|/N_p$, is attained, and in terms of
resources it reads: at a fixed total rate of $N_q$ measurement attempts per
unit time, the relaxation rate cannot exceed $2\lambda/(N_q-1)$, whichever
bonds are measured and however fast the pairs move. Geometry and transport
decide whether this rate is reached, Table~\ref{tab:geometry} and
Eq.~\eqref{eq:zalpha}. Once the bound is reached only a larger measurement
rate on feedback bonds can raise it.

The rate is therefore a property of the measurement--feedback channel, not
of the target manifold. The parent Hamiltonian
$\hat H=\sum_\alpha\hat P_\alpha$ fixes the common kernel of the
constraints, but its gap does not fix the transport of violations or the
absorber. A ferromagnetic Ising chain with monitored bond parities makes
the difference visible: its parent Hamiltonian is gapped, but on a ring the
violations are domain walls created in pairs, a local correction moves
them, and removal requires two walls to meet, so preparation is diffusive
while the parent gap is size independent. In the SWAP chain the parent gap
and the preparation rate happen to share the diffusive scaling,
Eq.~\eqref{eq:gapratio}, while random partial-SWAP events change the rate
at fixed target manifold, Sec.~\ref{sec:crossover}.

The pair closure and the diffusive preparation time $\tau\sim N_q^2$ were
established in Ref.~\cite{Doerstel2026}. The present work solves the
closed pair dynamics exactly, with its boundaries and coefficients, and
thereby obtains the relaxation rates, the fidelity bounds and the
Liouvillian gap for every $N_q$, together with the response to transport,
feedback geometry and errors. Two further closures, independent of the
pair walk, determine what the pair walk cannot: the conserved
collective-spin operators fix the stationary state selected by the initial
condition, and the products of singlet and triplet projectors fix the
ensemble covariance.

Related works include the quasiparticle-cooling theory of
Ref.~\cite{Feldmeier2026}, the preparation of Dicke states through coherent
transport to a single dissipator~\cite{Beer2026}, and
reaction--diffusion-type schemes for state
preparation~\cite{Nosov2023,Pokart2026}. The quasiparticle cooling
describes frustration-free targets without a known closure by an effective
dilute quasiparticle problem; where the two overlap, it is the dilute limit
of the pair equation. The latter holds, however, at arbitrary pair density
and provides a complete picture. Reaction--diffusion descriptions identify
the same transport bottleneck at the level of
scaling~\cite{Nosov2023,Pokart2026}. Coherent transport to a single
dissipator~\cite{Beer2026} is compared with the point absorber in
Sec.~\ref{sec:coherent}. Reachability analyses ask whether a set of
measurements and feedback operations can generate the target at
all~\cite{Reachability2026,Piroli2021,Lu2022,Tantivasadakarn2023}; here
reachability holds by Lemma~\ref{lem:stationaryspace}, and
Eq.~\eqref{eq:gapidentity} gives the rate.

\subsection{Preparation tasks, time, and operation counts}
\label{sec:resourcecomparison}

\begin{table*}[!t]
\caption{Resources for preparing the Dicke state $\lvert D_m\rangle$ at
finite filling, where $k=N_q/2+m$ and both $k$ and $N_q-k$ scale with $N_q$.
The unitary rows describe a single preparation from a prescribed product
state. For the global adaptive protocol, ``eventual'' denotes unit hitting
probability and the listed time is the expected stopping time; for the
ancilla-assisted adaptive construction, the listed depth is that of one
attempt and its success probability is given separately. Operation counts
of the adaptive protocols are inferred from the reported depth and width;
the $\mathrm{QAC}^0_f$ entry assumes fan-out, and $\tilde O$ suppresses
logarithmic factors. The FFC entries are state-independent guarantees from
Eq.~\eqref{eq:Teps}, with $\epsilon$ the target infidelity, at a total
rate of $N_q$ two-site measurement attempts; ``uncond.'' denotes
convergence without postselection, and the quoted time is the time after
which the accuracy is guaranteed, not an expected stopping time. The final
column records whether repetition makes the target absorbing within the
invariant sector of the protocol.}
\label{tab:resourcetable}
\begin{ruledtabular}
\small
\begin{tabular}{lllllll}
Protocol
& Depth or time
& Total operations
& Ancillas
& Success
& Elementary operation
& Attractor \\
\hline
Unitary, LNN \cite{Baertschi2019}
& $O(N_q)$
& $O(N_q^2)$
& $0$
& det.
& local 2-qubit
& no \\
Unitary, all-to-all \cite{Baertschi2022}
& $O(N_q)$
& $O(N_q^2)$
& $0$
& det.
& all-to-all 2-qubit
& no \\
Unitary, $\mathrm{QAC}^0_f$ \cite{Constdepth2026,Gretta2026}
& $O(1)$
& $\mathrm{poly}(N_q)$
& $\mathrm{poly}(N_q)$
& det.
& fan-in $+$ fan-out
& no \\
Adaptive, global \cite{Yu2026}
& $\mathbb E[T]=\mathrm{polylog}(N_q)$
& $\tilde O(N_q)$
& $\mathrm{polylog}(N_q)$
& eventual
& global meas.
& yes \\
Adaptive, ancilla \cite{Luo2026}
& $O(1)$
& $\tilde O(N_q^2)$
& $\tilde O(N_q^2)$
& $p=\Omega(N_q^{-1})$
& local $+$ feedforward
& no \\
FFC, local
& $O[N_q^2\log(N_q/\epsilon)]$
& $O[N_q^3\log(N_q/\epsilon)]$
& $0$ or $O(N_q)$ reusable
& uncond.
& local 2-site checks
& yes \\
FFC, all-to-all
& $O[N_q\log(N_q/\epsilon)]$
& $O[N_q^2\log(N_q/\epsilon)]$
& $0$ or $O(N_q)$ reusable
& uncond.
& all-to-all 2-site checks
& yes \\
\end{tabular}
\end{ruledtabular}
\end{table*}

Within frustration-free control, the comparison between local and
all-to-all monitoring is exact. The local chain has an absorber of $N_q$
bonds among $N_p=N_q(N_q-1)/2$ pair configurations and is transport
limited. If every pair is instead measured and corrected at rate
$2/(N_q-1)$, the total attempt rate remains $N_q$ while every pair
configuration is absorbing, and the generator reduces to
\begin{align}
 \mathsf K_{\rm all}
 &=
 \mathsf M_{\rm all}
 +
 \frac{2\lambda}{N_q-1}\mathds{1},
 &
 \Lambda_{\rm pair}^{\rm all}
 &=
 \frac{2\lambda}{N_q-1},
 \label{eq:alltoallgap}
\end{align}
because $\mathsf M_{\rm all}\mathbf 1=0$ and $\mathsf M_{\rm all}\geq0$:
all-to-all monitoring attains the absorber bound because removal no longer
requires motion toward a restricted set of feedback bonds.

Elapsed time and operation count are different resources. The local
protocol reaches the absorber-limited $O(N_q)$ time only with
partial-SWAP transport at $\gamma_{\rm scr}\sim N_q$, Eq.~\eqref{eq:zalpha},
at an integrated cost of $O[N_q^3\log(N_q/\epsilon)]$ operations.
All-to-all monitoring reaches the same time with
$O[N_q^2\log(N_q/\epsilon)]$ measurement attempts and without additional
transport operations. Within pairwise frustration-free control the latter
count is optimal up to the logarithmic accuracy factor, because
integrating the loss identity over time gives
\begin{align}
 \int_0^\infty dt\,
 \bm 1^T\mathsf A\bm q(t)
 &=
 W(0),
 &
 W(0)
 &=
 \frac12
 \left(
 \frac{N_q^2}{4}-m^2
 \right)
 \label{eq:feedbackflux}
\end{align}
for a product state of magnetization $m$: at maximal feedback, every
detected pair triggers one feedback event, so the expected number of
feedback events for a finite-density product state is $\Omega(N_q^2)$
however the measured pairs are connected. This bound applies to pairwise
frustration-free control, not to Dicke-state preparation by arbitrary
protocols.

Table~\ref{tab:resourcetable} places these two protocols next to unitary
and adaptive schemes, which differ in the task they implement as much as
in their resources. Deterministic circuits and the ancilla-assisted
adaptive construction prepare the target once from a known input; the
state is not an attractor of the operations that generated it. For that
task they are the more economical benchmark: a Dicke state at finite
filling has connected correlations at all separations, so a
nearest-neighbor circuit requires depth $\Omega(N_q)$
\cite{Lieb1972,Bravyi2006}, which Refs.~\cite{Baertschi2019,Baertschi2022}
attain with $O(N_q^2)$ two-qubit gates and no ancillas, and constant-depth
constructions trade depth for unbounded fan-in, fan-out, global
measurements, or polynomially many
ancillas~\cite{Constdepth2026,Gretta2026,Yu2026,Luo2026}. The absorber
bound does not apply to them, because they are not closed dynamics of
constraint violations; the table records their reported resources without
mapping them onto an absorber model.

Frustration-free control and the global adaptive protocol of
Ref.~\cite{Yu2026} instead generate attracting dynamics within an
invariant sector, and they do so on complementary quantum numbers:
collective rotations and measurements preserve $J=N_q/2$ while correcting
$J^z$, whereas local SWAP constraints preserve $J^z=m$ while driving the
state toward $J=N_q/2$. Combining the two could extend convergence to
initial states with neither quantum number fixed; the dynamics and
resources of such a combined protocol are open. Frustration-free control
additionally supplies a fidelity bound that is measurable through
$\bm J^2$.

\subsection{Implementation and stabilization}
\label{sec:resources}

The exact solution converts into an experimental schedule as follows. With
$\gamma_{\ell m}=1$, one unit of time contains on average one measurement
attempt on every monitored bond. A digital implementation partitions
overlapping bonds into edge-disjoint sublayers, two for an even ring and
three for an odd ring; the estimates below assume that the stroboscopic
implementation reproduces the continuous-time rates, finite step size
adding the usual ordering corrections. We use $N_q=20$ throughout and
$100\,\mu\mathrm{s}$ per sublayer as a reference duration: a cavity-based
neutral-atom experiment has demonstrated mid-circuit measurement and
real-time feedforward in less than $100\,\mu\mathrm{s}$, with sub-percent
measurement infidelity and less than $2\%$ coherence loss on an unmeasured
qubit \cite{Lee2026}, and comparable capabilities exist in trapped-ion,
neutral-atom and superconducting processors
\cite{RyanAnderson2021,Bluvstein2024,Andersen2019}.

Preparation. At maximal feedback, Eq.~\eqref{eq:Teps} guarantees an
infidelity below $10^{-2}$ after $281$ time units, for every initial state
in the sector, which is $56\,\mathrm{ms}$ on an even ring; $10^{-3}$
requires $375$ units. The time grows as $N_q^2\log(N_q/\epsilon)$,
Table~\ref{tab:resourcetable}. This is the ideal-channel schedule; it does
not include platform-specific gate, crosstalk, or idle errors.

Verification. The bound is evaluated from a collective-spin measurement,
without reconstructing the many-body state: Eq.~\eqref{eq:Jsum} expresses
$\hat W$ through $\bm J^2$, so a measurement of $\bm J^2$ determines $W$
and bounds the leakage from the symmetric manifold through
Eq.~\eqref{eq:fidelitybound}, which in a fixed magnetization sector is a
bound on the Dicke-state infidelity.

Stabilization. Inverting Eq.~\eqref{eq:fidelitycriterion} for the error
rate at a target stationary leakage $\epsilon_{\rm targ}=10^{-2}$ gives
$\kappa\leq2.3\times10^{-5}$ per site and time unit for the bare protocol,
i.e., a single-site error time of about $9\,\mathrm{s}$ in the isotropic
Pauli model; the admissible rate decreases as $N_q^{-3}$ in the
transport-limited and as $N_q^{-2}$ in the absorber-limited regime, where
it reaches $8.3\times10^{-5}$, about $2.4\,\mathrm{s}$. These numbers
follow from the weak-noise expansion and are to be replaced by the exact
affine solution \eqref{eq:affstat} when the leakage is not small. Random
partial-SWAP events raise the preparation rate and the error budget
through the same $\mathbb D_{N_q}$, by factors that saturate once the
absorber-limited regime is reached.

All-to-all monitoring reaches the absorber-limited regime without
additional transport operations: at the same total attempt rate,
Eq.~\eqref{eq:alltoallgap} gives $66$ time units for $N_q=20$ and
infidelity $10^{-2}$ instead of $281$, and the stationary pair profile
under isotropic noise is uniform,
\begin{align}
 p^*
 &=
 \frac{2\kappa}
 {6\lambda/(N_q-1)+8\kappa},
 &
 \epsilon_{\rm cert}^*
 &=
 \frac{\kappa(N_q-1)^2}{3\lambda}
 +
 O(\kappa^2),
 \label{eq:alltoallnoise}
\end{align}
which is the absorber-limited leakage of Eq.~\eqref{eq:fidelityreaction},
at the price of nonlocal two-site measurements.

An implementation therefore requires repeated two-site singlet--triplet
measurements, or an ancilla-mediated measurement of the SWAP projector,
conditional single-qubit feedback, reusable access to the measurement
ancillas, and a collective-spin readout. For twenty qubits with
$100\,\mu\mathrm{s}$ per sublayer, the noiseless protocol prepares the
Dicke state to $1\%$ infidelity in $56\,\mathrm{ms}$ from any initial
state of the sector, and under isotropic errors it maintains a stationary
leakage below $1\%$ as long as single-site errors are rarer than one per
$9\,\mathrm{s}$; a guarantee for the noisy protocol at a finite time
requires both the decaying initial-state contribution and the accumulated
error contribution, which the affine solution \eqref{eq:afftrans}
provides.

\subsection{Scope and outlook}

A natural question is whether the exact operator closures found here can be generalized to other frustration-free preparation schemes. For Dicke states, measurements relabel pair projectors and feedback removes pair weight without coupling its mean evolution to higher correlations. Can the structure of the local constraints and feedback operations predict which protocols admit such a first-moment closure, and which require equations for an increasing number of correlations? Frustration-free control in the Motzkin and Fredkin chains provides natural examples to investigate this question: their constrained motion produces subdiffusive preparation dynamics~\cite{Doerstel2026}. It remains to determine whether these protocols admit an enlarged closed operator space of polynomial size, or whether their preparation dynamics requires an effective nonlinear hydrodynamic description~\cite{ZhouZhangZhang2026}. Understanding this distinction would identify both the scope of exact finite-size theories of FFC and the role of correlations in more general preparation schemes.

Within the SWAP protocol, the measurement closure holds for any local dimension, whereas the second-moment feedback closure uses the spin-$1/2$ identities between singlet and triplet projectors. Whether this feedback closure extends beyond qubits remains open. Coherent exchange leaves the one-singlet sector invariant, but in sectors with more singlets it does not preserve the pair closure, so its effect on the preparation rate at finite singlet density is unresolved, Sec.~\ref{sec:coherent}.

Finally, the exact covariance retains the correlations between different pairs that feedback generates. Understanding how these correlations behave at large scales, and whether the mobility becomes a local functional of the pair density, would determine whether the exact lattice theory admits a continuum description of fluctuating absorbing hydrodynamics.

\section{Conclusion}

Frustration-free control of Dicke states is governed by three operator
spaces of polynomial size on which the measurement--feedback channel
closes. The pair projectors perform a random walk with absorption in the
configuration space of pairs and determine how fast the target manifold is
reached; the conserved collective-spin operators determine which state in
that manifold is reached from a given initial condition; and the products
of singlet and triplet projectors determine the ensemble fluctuations. Each
closure is an exact operator identity, so preparation, state selection and
fluctuations are calculable for every system size and every initial state.

Preparation depends on bringing constraint violations to feedback and
removing them. The parent Hamiltonian fixes the target manifold but not
the dynamics by which it is reached; that dynamics is fixed by the
transport of violations and by the set of configurations on which feedback
acts. Faster transport improves preparation and stabilization alike until
removal becomes the limit, and the limit is set by the fraction of
configurations that are absorbing. Accelerated local and all-to-all monitoring reach the same scaling of the
relaxation time, the one paying in transport operations what the other
pays in nonlocal measurements.

The mean dynamics is a classical absorbing walk, but the covariance
retains the quantum correlations that feedback generates between pairs,
with conserving measurement noise and the shot noise of removal.
Determining how these correlations behave at large scales is the next
step, from the exact lattice theory to a continuum description of
fluctuating absorbing hydrodynamics.

\begin{acknowledgments}
We thank Ye Zhou for sharing the proof that the slowest pair decay rate equals
the full Liouvillian gap.

T.\,D., T.\,C.\,L. and M.\,B. acknowledge support from the Austrian Science Fund (FWF) quantA Cluster of Excellence. T.\,I. and
J.\,H.\,W. acknowledge support from the National Science Foundation under
Grant Numbers DMR-2611305 (T.\,I.) and DMR-2238895 (J.\,H.\,W.). The computational results presented here have been achieved using the Austrian Scientific Computing (ASC) infrastructure and the LEO HPC infrastructure of the University of Innsbruck.

The authors used Claude Fable 5.1 (Anthropic) during the preparation of
this manuscript, for drafting and editing text, for checks of the
consistency of notation, equations and cross-references, and for scripts
that numerically verified individual identities, perturbative
coefficients and asymptotic exponents. The model also contributed to
derivations, in particular to parts of Appendices~\ref{app:stationaryproof},
\ref{app:covariance} and \ref{sec:dephasing}. All results, derivations
and proofs were independently rederived and verified by the authors, who
take full responsibility for the content.
\end{acknowledgments}

\section*{Data availability}
The data shown in Figs.~\ref{fig:certificate}--\ref{fig:noise} and the
simulation code that generates them (the sparse pair-generator solver, the
exact Lindblad evolutions, and the closed second-moment integrator) are
deposited at
Ref.~\cite{Doerstel2026data}, with a cross-reference to the deposit of
Ref.~\cite{Doerstel2026}.

\appendix

\section{Local SWAP algebra and the exact pair graph}
\label{app:swap}

\subsection{Single-qubit feedback}

For $\hat V_{ij}=e^{i\varphi\sigma_i^\nu}$ and
$k\notin\{i,j\}$, direct three-spin Pauli algebra gives
\begin{align}
 \hat P_{ij}\hat V_{ij}^\dagger\hat P_{ik}
 \hat V_{ij}\hat P_{ij}
 -\hat P_{ij}\hat P_{ik}\hat P_{ij}&=0,\\
 \hat P_{ij}\hat V_{ij}^\dagger\hat P_{jk}
 \hat V_{ij}\hat P_{ij}
 -\hat P_{ij}\hat P_{jk}\hat P_{ij}&=0.
 \label{eq:externalfeedbackzero}
\end{align}
On the detected bond,
\begin{align}
 \hat P_{ij}\hat V_{ij}^\dagger\hat P_{ij}
 \hat V_{ij}\hat P_{ij}-\hat P_{ij}
 &=-\sin^2\varphi\,\hat P_{ij}.
\end{align}
Together these identities prove Eq.~\eqref{eq:feedbacklocal} with
$\lambda=\sin^2\varphi$.

For the full Pauli correction, the same calculation establishes the
two-component dephasing closure.
For an external site $k$,
\begin{align}
 \mathcal D^\dagger[\sigma_i^z\hat P_{ij}]\hat T_{ik}
 &=\tfrac14(\hat T_{jk}-\hat T_{ik}),\nonumber\\
 \mathcal D^\dagger[\sigma_i^z\hat P_{ij}]\hat T_{jk}
 &=\tfrac14(\hat T_{ik}-\hat T_{jk}),\nonumber\\
 \mathcal D^\dagger[\sigma_i^z\hat P_{ij}]\hat T_{ij}
 &=\hat P_{ij}.
 \label{eq:Tfeedbackidentities}
\end{align}
Together with Eq.~\eqref{eq:zpairmix}, these identities yield
Eq.~\eqref{eq:ztwocomponent}.

Equations~\eqref{eq:externalfeedbackzero} and \eqref{eq:feedbacklocal} hold at
arbitrary $\varphi$, not only at $\varphi=\pi/2$: the external pairs are untouched by
the feedback for every rotation angle, and the on-bond compression is
$-\sin^2\varphi\,\hat P_{ij}$.  Consequently
$\lambda=\sin^2\varphi\in(0,1]$ is already a free parameter of the single-qubit feedback
map, and no additional two-site construction is required in order to detune the absorber
from the measurement rate. This is what allows $h_{N_q}$ in Eq.~\eqref{eq:hLscr} to be
scanned at fixed transport.

The measurement algebra is independent of the on-site dimension $N$.  What
changes for $N>2$ is both the interpretation of the symmetric target and the
choice of feedback map; the explicit Pauli feedback, the angular-momentum
fidelity bound, and the noise amplitude \eqref{eq:measblock} are spin-$1/2$
results.

\section{Proof of Lemma~\ref{lem:stationaryspace}}
\label{app:stationaryproof}

\noindent\emph{Proof of the Lemma.}
The inclusion $\mathcal B(\mathcal K)\subseteq\ker\mathcal L$ is
Eq.~\eqref{eq:dfsalgebra}. For the converse, let
$\hat Q_{\mathcal F}=\sum_{\{n,m\}\in\mathcal F}\hat P_{nm}\geq0$ and sum
Eq.~\eqref{eq:exactpair} over all pairs. The pair-graph Laplacian conserves
total weight, so its contribution cancels and
\begin{align}
 \mathcal L^\dagger\hat W
 &=
 -\lambda\hat Q_{\mathcal F} ,
 \label{eq:lossidentity}
\end{align}
for an arbitrary absorber, which reduces to Eq.~\eqref{eq:Wloss} on the chain. If $\hat\sigma\geq0$ is
stationary, then
$0=\operatorname{Tr}[\hat W\mathcal L\hat\sigma]
=\operatorname{Tr}[(\mathcal L^\dagger\hat W)\hat\sigma]
=-\lambda\operatorname{Tr}[\hat Q_{\mathcal F}\hat\sigma]$. Since $\lambda>0$ and every
$\hat P_\alpha\geq0$, this forces
$\operatorname{Tr}[\hat P_\alpha\hat\sigma]=0$ for all $\alpha$, that is,
$\operatorname{supp}\hat\sigma\subseteq\mathcal K$.

To pass from stationary states to the whole kernel, let
\begin{align}
 \mathcal P_0
 &=
 \lim_{T\to\infty}
 \frac1T\int_0^T\!dt\,e^{\mathcal Lt}
 \label{eq:ergodicmean}
\end{align}
be the ergodic mean, which exists because the semigroup is bounded, is a
positive map, and projects onto $\ker\mathcal L$. Put
$\hat\sigma_*\equiv\mathcal P_0(\mathds{1}/2^{N_q})$, a stationary state. For
$\hat X\geq0$ one has $\hat X\leq\lVert\hat X\rVert\mathds{1}$, with
$\lVert\cdot\rVert$ the operator norm, because the spectrum of $\hat X$
lies in $[0,\lVert\hat X\rVert]$. Since $\mathcal P_0$ is a positive map,
applying it to both sides gives
$0\leq\mathcal P_0\hat X\leq\lVert\hat X\rVert\,\mathcal P_0\mathds 1
=2^{N_q}\lVert\hat X\rVert\hat\sigma_*$, so
$\operatorname{supp}\mathcal P_0\hat X
\subseteq\operatorname{supp}\hat\sigma_*$. Every Hermitian
$\hat X\in\ker\mathcal L$ has a spectral decomposition
$\hat X=\hat X_+-\hat X_-$ into its positive and negative parts,
$\hat X_\pm\geq0$ with $\hat X_+\hat X_-=0$, and therefore obeys
$\hat X=\mathcal P_0\hat X=\mathcal P_0\hat X_+-\mathcal P_0\hat X_-$, so
$\operatorname{supp}\hat X\subseteq\operatorname{supp}\hat\sigma_*
\subseteq\mathcal K$ by the previous paragraph. Since $\ker\mathcal L$ is
closed under the adjoint it is spanned by its Hermitian elements, and
$\ker\mathcal L\subseteq\mathcal B(\mathcal K)$.

Finally, $\ker\hat P_{\ell m}$ consists of the states invariant under the
transposition of $\ell$ and $m$. The transpositions attached to a connected
graph generate the full symmetric group, so
$\mathcal K=\mathcal H_{\rm sym}$ and $\dim\mathcal K=N_q+1$. \hfill$\square$

The same identity settles the peripheral spectrum. For any state,
$W(t)=\operatorname{Tr}[\hat W\hat\rho(t)]$ is non-increasing and bounded
from below. In finite dimension $\ddot W(t)
=\operatorname{Tr}[(\mathcal L^{\dagger2}\hat W)\hat\rho(t)]$ is bounded
uniformly in $t$, so $\dot W$ is uniformly continuous, and a monotone bounded
function with uniformly continuous derivative satisfies $\dot W\to0$
(Barbalat's lemma \cite{Khalil2002}). Hence
$\operatorname{Tr}[\hat Q_{\mathcal F}\hat\rho(t)]\to0$ and every limit point of the
trajectory is a state supported in $\mathcal K$. On
$\mathcal B(\mathcal K)$ the generator vanishes identically,
Eq.~\eqref{eq:dfsalgebra}, so no limit point carries a persistent
oscillation. The peripheral spectrum of $\mathcal L$ therefore consists of
the single eigenvalue $0$.
Boundedness of the semigroup, $\lVert e^{\mathcal Lt}\rVert_{1\to1}=1$ for
all $t$, excludes a nontrivial Jordan block on the imaginary axis, since such
a block would make $\lVert e^{\mathcal Lt}\rVert$ grow polynomially. The zero
eigenvalue is therefore semisimple and $e^{\mathcal Lt}\to\mathcal P_0$.

Both hypotheses of the Lemma are necessary. At $\varphi=0$ or $\varphi=\pi$
the feedback is trivial and the channel is an unconditioned measurement. Its
generator is then
$\mathcal L\hat\rho=-\sum_\alpha[\hat P_\alpha\hat\rho(\mathds 1-\hat P_\alpha)
+(\mathds 1-\hat P_\alpha)\hat\rho\hat P_\alpha]$, and
$\operatorname{Tr}[\hat\rho^\dagger\mathcal L\hat\rho]
=-\sum_\alpha\lVert\hat P_\alpha\hat\rho(\mathds 1-\hat P_\alpha)\rVert_2^2
-\sum_\alpha\lVert(\mathds 1-\hat P_\alpha)\hat\rho\hat P_\alpha\rVert_2^2$,
so $\ker\mathcal L$ is the commutant of $\{\hat P_\alpha\}$, that is the
commutant of the symmetric group generated by the transpositions. By
Schur--Weyl duality this is the algebra generated by the collective spin,
of dimension $\sum_J(2J+1)^2$ over the total-spin values present. This sum
contains the term $(N_q+1)^2$ from $J=N_q/2$ together with at least one further
term for every $N_q\geq2$, so the kernel is strictly larger than
$\mathcal B(\mathcal H_{\rm sym})$. If the measured bonds are disconnected,
the proof goes through componentwise with $\mathcal K$ the tensor product of
the symmetric subspaces of the components, so
$\dim\ker\mathcal L=\prod_c(N_{q,c}+1)^2$, with $N_{q,c}$ the number of qubits in component $c$, which again exceeds $(N_q+1)^2$.

\section{Covariance source and second-moment closure}
\label{app:covariance}

\subsection{Covariance equation}

For a single adjoint dissipator,
$\mathcal D^\dagger[\hat K](\hat A\hat B)
=\hat K^\dagger\hat A\hat B\hat K
-\frac12\{\hat K^\dagger\hat K,\hat A\hat B\}$.
Adding and subtracting
$\hat K^\dagger\hat A\hat K\hat B$ and
$\hat A\hat K^\dagger\hat B\hat K$ gives
\begin{align}
 \mathcal D^\dagger[\hat K](\hat A\hat B)
 &=
 \bigl(\mathcal D^\dagger[\hat K]\hat A\bigr)\hat B
 +
 \hat A\bigl(\mathcal D^\dagger[\hat K]\hat B\bigr)
 \nonumber\\
 &\quad+
 [\hat K^\dagger,\hat A][\hat B,\hat K],
 \label{eq:product}
\end{align}
whose sum over the jump operators $\hat K_{\alpha s}$ with rates
$\gamma_\alpha$ is Eq.~\eqref{eq:lindbladleibniz}. For
$\hat A=\hat P_a$ and $\hat B=\hat P_b$, the first two terms
reproduce the first-moment generator $\mathsf K$ acting on
$\avg{\hat P_a\hat P_b}$ from the left and from the right, and
$\partial_t(\avg{\hat P_a}\avg{\hat P_b})$ contains the same two
terms acting on the product of means. Subtracting the two and
symmetrizing in $a,b$ gives Eq.~\eqref{eq:coveq}, with the
source Eq.~\eqref{eq:Nsym} given by the remainder in
Eq.~\eqref{eq:product}; since the $\hat P_a$ are Hermitian,
$[\hat K^\dagger,\hat P_a]=[\hat P_a,\hat K]^\dagger$, which is the
quadratic form in Eq.~\eqref{eq:Npositive}.

\subsection{Measurement source}

Fix the measured bond $(\ell,\ell+1)$ and an external site $n$.
With $\hat I_{\ell n}$ and $\hat\Sigma_{\ell n}$ of
Eq.~\eqref{eq:paircombinations},
\begin{align}
 [\hat P_{\ell n},\hat P_{\ell,\ell+1}]
 &=
 -\frac12\hat I_{\ell n}\hat S_{\ell,\ell+1},
 \nonumber\\
 [\hat P_{\ell+1,n},\hat P_{\ell,\ell+1}]
 &=
 \frac12\hat I_{\ell n}\hat S_{\ell,\ell+1},
 \label{eq:measurementcommutators}
\end{align}
and every other pair projector commutes with
$\hat P_{\ell,\ell+1}$. Inserting Eq.~\eqref{eq:measurementcommutators}
into Eq.~\eqref{eq:singlejumpsource} and using
$\hat S_{\ell,\ell+1}\hat I_{\ell m}\hat S_{\ell,\ell+1}=-\hat I_{\ell m}$
gives
\begin{align}
 2\mathsf N^S[\hat P_{\ell,\ell+1}]_
 {\{\ell,n\},\{\ell,m\}}
 &=
 \frac14
 \left\langle
 \{\hat I_{\ell n},\hat I_{\ell m}\}
 \right\rangle
 =\mu_{\ell;nm},
\end{align}
and replacing either $\ell$ by $\ell+1$ changes the sign. This is
the four-entry structure of Eq.~\eqref{eq:Nmeas}, including the
four-site terms with $n\neq m$. For $n=m$, the three-spin identities
\begin{align}
 \hat P_{\ell,\ell+1}\hat P_{\ell n}
 \hat P_{\ell,\ell+1}
 &=
 \frac14\hat P_{\ell,\ell+1},
 \label{eq:sandwich}\\
 \{\hat P_{\ell,\ell+1},\hat P_{\ell n}\}
 &=
 \frac12
 \left(
 \hat P_{\ell,\ell+1}
 +\hat P_{\ell n}
 -\hat P_{\ell+1,n}
 \right),
 \label{eq:anticomm}
\end{align}
and their relabelings give
$\hat I_{\ell n}^{\,2}
=\frac12(\hat P_{\ell,\ell+1}+\hat P_{\ell n}+\hat P_{\ell+1,n})$,
hence Eq.~\eqref{eq:measblock}.

\subsection{Second-moment closure under Pauli feedback}

At maximal feedback,
$\hat K_{\ell1}=i\sigma_\ell^z\hat P_{\ell,\ell+1}$, and the phase
does not affect its dissipator. For an arbitrary operator
$\hat Y$, write
$\hat Y^{\sigma_\ell}=\sigma_\ell^z\hat Y\sigma_\ell^z$.
Adding and subtracting
$\hat P_{\ell,\ell+1}\hat Y\hat P_{\ell,\ell+1}$ gives
\begin{align}
 &\mathcal D^\dagger
 [\sigma_\ell^z\hat P_{\ell,\ell+1}]\hat Y
 =
 \mathcal D^\dagger[\hat P_{\ell,\ell+1}]\hat Y
 +
 \mathcal F_\ell\hat Y,
 \nonumber\\
 &\mathcal F_\ell\hat Y
 \equiv
 \hat P_{\ell,\ell+1}
 \left(
 \hat Y^{\sigma_\ell}-\hat Y
 \right)
 \hat P_{\ell,\ell+1}.
 \label{eq:feedbacksplit}
\end{align}
Since
$\mathcal D^\dagger[\hat K_{\ell0}]
=\mathcal D^\dagger[\hat P_{\ell,\ell+1}]$,
the contribution of bond $\ell$ to the adjoint generator is the
measurement term $2\mathcal D^\dagger[\hat P_{\ell,\ell+1}]$, which
preserves products by Theorem~\ref{thm:filtration}, plus
$\mathcal F_\ell$. It remains to evaluate $\mathcal F_\ell$ on
products of at most two elements of $\mathcal G$.

The measured projector has rank one on the bond, so for any
$\hat Z$
\begin{align}
 \hat P_{\ell,\ell+1}\hat Z\hat P_{\ell,\ell+1}
 &=
 \hat P_{\ell,\ell+1}
 \otimes
 \langle s_\ell|\hat Z|s_\ell\rangle,
 \label{eq:cutting}
\end{align}
with the partial matrix element acting on the other sites. Since
$\sigma_\ell^z|s_\ell\rangle=|t_\ell^0\rangle$, the transformed
product is the same product evaluated in the triplet, and
\begin{align}
 &\langle s_\ell|
 \left(
 \hat X_a^{\sigma_\ell}\hat X_b^{\sigma_\ell}
 -
 \hat X_a\hat X_b
 \right)
 |s_\ell\rangle
 =
 \operatorname{Tr}_{\ell,\ell+1}
 \left[
 \hat E_{\ell,\ell+1}
 \hat X_a\hat X_b
 \right],
 \label{eq:singlettriplet}\\
 &\hat E_{ab}
 \equiv
 \hat T_{ab}-\hat P_{ab}
 =
 \frac12
 \left(
 \sigma_a^x\sigma_b^x+\sigma_a^y\sigma_b^y
 \right),
 \qquad a\neq b.
 \label{eq:bondexchange}
\end{align}
The trace is nonzero only if the product supplies a ladder operator
on each site of the bond. Let $\hat X_{ab}$ and $\hat Y_{ab}$ each
denote either $\hat P_{ab}$ or $\hat T_{ab}$, set $\eta_X=-1$ for a
singlet and $\eta_X=+1$ for a triplet projector, and write
$\hat X_\ell=\hat X_{\ell,\ell+1}$. For a single factor,
$\mathcal F_\ell\hat X_\ell=\eta_X\hat P_{\ell,\ell+1}$ and
$\mathcal F_\ell\hat X_{ab}=0$ for every other pair. For two
factors, with external sites $n,m$ and $i\in\{\ell,\ell+1\}$, the
nonzero cases are
\begin{align}
 \mathcal F_\ell(\hat X_\ell\hat Y_{nm})
 &=
 \eta_X\hat P_{\ell,\ell+1}\hat Y_{nm},
 \nonumber\\
 \mathcal F_\ell(\hat X_\ell\hat Y_{in})
 &=
 \frac{\eta_X}{4}\hat P_{\ell,\ell+1},
 \nonumber\\
 \mathcal F_\ell(\hat X_\ell\hat Y_\ell)
 &=
 \eta_X\delta_{XY}\hat P_{\ell,\ell+1},
 \nonumber\\
 \mathcal F_\ell(\hat X_{\ell n}\hat Y_{\ell+1,m})
 &=
 \frac{\eta_X\eta_Y}{4}
 \hat P_{\ell,\ell+1}\hat E_{nm},
 \label{eq:feedbackproducts}
\end{align}
where $\delta_{XY}=1$ when the two factors are of the same type and
zero otherwise, and $\hat E_{nn}\equiv\mathds{1}$ for coincident
external sites. The same results hold with the factor order
reversed, and all other two-factor products give zero. Every
right-hand side contains at most two singlet or triplet
projectors, so $\mathcal F_\ell\mathcal V_2\subseteq\mathcal V_2$,
which together with Theorem~\ref{thm:filtration} proves
Theorem~\ref{thm:closure2}. The coefficients are independent of
$N_q$ because $\mathcal F_\ell$ involves at most the two bond sites
and two external sites; tensor embedding extends the identities to
arbitrary $N_q$.

\subsection{Local removal noise at arbitrary feedback angle}

For $\hat V_\ell=\exp(i\varphi\sigma_\ell^z)$, the identity
$\hat P_{\ell,\ell+1}\sigma_\ell^z\hat P_{\ell,\ell+1}=0$ gives
$[\hat P_{\ell,\ell+1},\hat K_{\ell1}]
=-i\sin\varphi\,\sigma_\ell^z\hat P_{\ell,\ell+1}$, and therefore
\begin{align}
 \mathsf N^S[\hat K_{\ell1}]_
 {\{\ell,\ell+1\},\{\ell,\ell+1\}}
 &=
 \sin^2\varphi\,
 \langle\hat P_{\ell,\ell+1}\rangle.
 \label{eq:removalnoiseapp}
\end{align}
Since $\hat K_{\ell0}$ commutes with $\hat P_{\ell,\ell+1}$, this
is the complete entry from bond $\ell$, Eq.~\eqref{eq:reactionnoise}.
At unit measurement rate, the contribution of bond $\ell$ to the mean
is
$\partial_t\langle\hat P_{\ell,\ell+1}\rangle|_\ell
=-\sin^2\varphi\,\langle\hat P_{\ell,\ell+1}\rangle$,
so the source amplitude equals the mean removal rate.

\subsection{Three-site source matrices at maximal feedback}

Fix the measured bond $\ell$ and one external site $n$, and order
the observables as
\begin{align}
 \hat{\bm O}_{\ell n}
 &\equiv
 \begin{pmatrix}
  \hat P_{\ell,\ell+1}\\
  \hat\Sigma_{\ell n}\\
  \hat I_{\ell n}
 \end{pmatrix}.
 \label{eq:localbasis}
\end{align}
$\left.\mathsf N^S[\hat K]\right|_{\hat{\bm O}_{\ell n}}$ denotes
the $3\times3$ source matrix of Eq.~\eqref{eq:singlejumpsource} with
rows and columns in this order; its entries involving
$\hat\Sigma_{\ell n}$ and $\hat I_{\ell n}$ are sums and differences
of the pair-indexed entries. For
$\hat K_{\ell1}=i\sigma_\ell^z\hat P_{\ell,\ell+1}$, three-spin
Pauli algebra gives
\begin{align}
 &\left.
 \mathsf N^S[\hat K_{\ell1}]
 \right|_{\hat{\bm O}_{\ell n}}
 =
 \begin{pmatrix}
  \langle\hat P_{\ell,\ell+1}\rangle
  &0
  &\frac12\langle\hat I_{\ell n}\rangle\\
  0
  &\frac12\langle\hat P_{\ell,\ell+1}\rangle
  &0\\
  \frac12\langle\hat I_{\ell n}\rangle
  &0
  &\frac12\langle\hat\Sigma_{\ell n}\rangle
 \end{pmatrix},
 \label{eq:feedbackblock}
\end{align}
while $\hat K_{\ell0}$ contributes the same as
$\hat P_{\ell,\ell+1}$,
\begin{align}
 \left.
 \mathsf N^S[\hat K_{\ell0}]
 \right|_{\hat{\bm O}_{\ell n}}
 &=
 \frac12
 \langle\hat P_{\ell,\ell+1}+\hat\Sigma_{\ell n}\rangle
 \operatorname{diag}(0,0,1).
 \label{eq:measurementlocalblock}
\end{align}
The complete source from bond $\ell$ is the sum,
\begin{align}
 &\left.
 \mathsf N^{S,(\ell)}
 \right|_{\hat{\bm O}_{\ell n}}
 =
 \begin{pmatrix}
  \langle\hat P_{\ell,\ell+1}\rangle
  &0
  &\frac12\langle\hat I_{\ell n}\rangle\\
  0
  &\frac12\langle\hat P_{\ell,\ell+1}\rangle
  &0\\
  \frac12\langle\hat I_{\ell n}\rangle
  &0
  &\frac12\langle\hat P_{\ell,\ell+1}\rangle
   +\langle\hat\Sigma_{\ell n}\rangle
 \end{pmatrix}.
 \label{eq:completelocalblock}
\end{align}
The measurement-only source $\mathsf N^{S,(\ell)}_{\rm meas}$ is
twice Eq.~\eqref{eq:measurementlocalblock}, and the feedback
correction of Eq.~\eqref{eq:Nsourcesplit} is
Eq.~\eqref{eq:feedbackblock} minus
Eq.~\eqref{eq:measurementlocalblock}; its $\hat I_{\ell n}$ diagonal
entry is $-\langle\hat P_{\ell,\ell+1}\rangle/2$, which is why the
correction need not be positive semidefinite although the complete
source is.

Operations on bond $\ell$ leave $\hat\Sigma_{\ell n}$ unchanged at
first order,
$(\mathcal D^\dagger[\hat K_{\ell0}]
+\mathcal D^\dagger[\hat K_{\ell1}])\hat\Sigma_{\ell n}=0$,
so its variance changes at the rate given by the
$\Sigma\Sigma$ entry of Eq.~\eqref{eq:completelocalblock}; this is
Eq.~\eqref{eq:feedbacksumvariance}. The entry coupling
$\hat P_{\ell,\ell+1}$ to $\hat I_{\ell n}$ is
Eq.~\eqref{eq:removalimbalance}. Entries with two distinct external
sites require four-site products and are supplied by the closure on
$\mathcal V_2$.

\subsection{Quantum Langevin form of the covariance source}

Introduce an independent vacuum input field for each jump operator
$\hat K_{\ell s}$, with increments satisfying
\begin{align}
 d\hat B_{\ell s}\,
 d\hat B_{\ell's'}^\dagger
 &=
 \delta_{\ell\ell'}\delta_{ss'}\,dt,
 \qquad
 d\hat B_{\ell s}^\dagger\,
 d\hat B_{\ell's'}
 =0,
 \label{eq:inputito}
\end{align}
and all other quadratic products of increments equal to zero.
For measurement rates $\gamma_\ell$, the Heisenberg evolution
of the pair projectors has the quantum Langevin representation
\begin{align}
 d\hat P_a(t)
 &=
 -\sum_b\mathsf K_{ab}\hat P_b(t)\,dt
 +
 d\hat\xi_a(t),
 \label{eq:quantumlangevin}\\
 d\hat\xi_a(t)
 &=
 \sum_{\ell,s}\sqrt{\gamma_\ell}
 \Big(
 [\hat K_{\ell s}^\dagger(t),\hat P_a(t)]\,
 d\hat B_{\ell s}
 \nonumber\\
 &\hspace{35mm}+
 d\hat B_{\ell s}^\dagger
 [\hat P_a(t),\hat K_{\ell s}(t)]
 \Big),
 \label{eq:quantumforce}
\end{align}
with all system operators in the Heisenberg picture. The vacuum
It\^o rule gives
\begin{align}
 \frac{1}{2dt}
 \left\langle
 d\hat\xi_a(t)\,d\hat\xi_b(t)
 +
 d\hat\xi_b(t)\,d\hat\xi_a(t)
 \right\rangle
 &=
 \mathsf N^S_{ab}(t),
 \label{eq:forcecov}
\end{align}
where the average includes the ensemble state and the vacuum input
fields. Equations~\eqref{eq:quantumlangevin} and
\eqref{eq:forcecov} therefore reproduce the mean dynamics and the
covariance source of Eq.~\eqref{eq:coveq} at the operator level.

\section{Quantization condition of the reduced generator for both parities}
\label{app:spectrum}

We derive Eqs.~\eqref{eq:quantization} and
\eqref{eq:scrquantization} for both parities at once. Set
\begin{align}
 \tilde R\equiv\frac{N_q}{2}
 =\begin{cases} R, & N_q\ \text{even},\\[0.2em] R+\tfrac12, & N_q\ \text{odd},\end{cases}
 \qquad R=\left\lfloor\frac{N_q}{2}\right\rfloor .
 \label{eq:Rtilde}
\end{align}
Substituting $p_r(t)=e^{-\Lambda t}f_r$ into the bulk equation gives
$-\Lambda f_r=\mathbb D_{N_q}(f_{r+1}+f_{r-1}-2f_r)$ and hence
$\Lambda=2\mathbb D_{N_q}(1-\cos k)$ for the ansatz
$f_r=\cos[k(\tilde R-r)]$. The bare protocol corresponds to
$\mathbb D_{N_q}=1$.

\paragraph{Outer boundary.}
For even $N_q$ the outer equation $\partial_tp_R=2\mathbb D_{N_q}(p_{R-1}-p_R)$ requires
$-2\mathbb D_{N_q}(1-\cos k)f_R=2\mathbb D_{N_q}(f_{R-1}-f_R)$, i.e.\ $f_{R-1}=\cos k\,f_R$.  The
ansatz satisfies this identically, since $\tilde R=R$ gives $f_R=1$ and
$f_{R-1}=\cos k$.

For odd $N_q$ the outer equation carries no factor of two,
$\partial_tp_{R}=\mathbb D_{N_q}(p_{R-1}-p_{R})$, because only two of the four endpoint moves
leave the separation $R=(N_q-1)/2$: the reflecting boundary sits on the bond between $r=R$ and
$r=R+1=N_q-R$, which is the same separation.  It requires
$-2\mathbb D_{N_q}(1-\cos k)f_R=\mathbb D_{N_q}(f_{R-1}-f_R)$, i.e.\ $f_{R-1}=(2\cos k-1)f_R$.  With
$\tilde R=R+\tfrac12$ one has $f_R=\cos(k/2)$ and $f_{R-1}=\cos(3k/2)$, and the identity
$\cos(3k/2)=(2\cos k-1)\cos(k/2)$ verifies it.  The two parities are thus absorbed into the
single definition~\eqref{eq:Rtilde}, at the price of a half-integer $\tilde R$ for odd $N_q$.

\paragraph{Absorbing boundary.}
The equation at the absorbing boundary $r=1$ gives
\begin{align}
 f_2=
 \left(2\cos k-1+\frac{\lambda}{\mathbb D_{N_q}}\right)f_1 ,
\nonumber
\end{align}
and inserting $f_1=\cos[k(\tilde R-1)]$, $f_2=\cos[k(\tilde R-2)]$ together with
$\cos[k(\tilde R-2)]=2\cos k\cos[k(\tilde R-1)]-\cos(k\tilde R)$ produces
the general quantization condition
\begin{align}
 \cos(k\tilde R)
 &=
 \left(1-\frac{\lambda}{\mathbb D_{N_q}}\right)
 \cos[k(\tilde R-1)] .
\nonumber
\end{align}
It reduces to Eq.~\eqref{eq:quantization} for $\mathbb D_{N_q}=1$ and to
Eq.~\eqref{eq:scrquantization} for the random partial-SWAP protocol.
All asymptotic statements in the main text are therefore parity
independent. At the bare maximal-feedback point
$\lambda=\mathbb D_{N_q}=1$, the right-hand side vanishes,
$k_0\tilde R=\pi/2$, and $k_0=\pi/N_q$ for every $N_q$, which is
Eq.~\eqref{eq:closedgap}.  The corresponding mode is
$f_r=\cos[\pi(\tilde R-r)/N_q]=\sin(\pi r/N_q)$, as in
Eq.~\eqref{eq:fullspectrum}. Its pair norm follows from
\begin{align}
 \sum_{r=1}^{R}n_r\sin^2\frac{\pi r}{N_q}
 &=
 \frac{N_q^2}{4},
\nonumber
\end{align}
with $n_r$ the number of pairs at separation $r$, that is $n_r=N_q$ for
$r<N_q/2$ and $n_{N_q/2}=N_q/2$. The identity holds for both parities and follows
from $\sum_{r=1}^{N_q-1}\sin^2(\pi r/N_q)=N_q/2$ together with the reflection
$r\mapsto N_q-r$, which leaves $\sin^2(\pi r/N_q)$ invariant. For odd $N_q$ the
reflection pairs $r\leq R$ with $r\geq R+1$ and has no fixed point, so
$\sum_{r=1}^{R}\sin^2(\pi r/N_q)=N_q/4$ and the uniform multiplicity gives
$N_q\cdot N_q/4=N_q^2/4$. For even $N_q$ the point $r=R=N_q/2$ is fixed and contributes
$\sin^2(\pi/2)=1$, so $\sum_{r=1}^{R-1}\sin^2(\pi r/N_q)=(N_q-2)/4$ and the
pair multiplicities give $N_q(N_q-2)/4+(N_q/2)\cdot1=N_q^2/4$.

\paragraph{N\'eel overlap.}
For the N\'eel state, $q_{nm}(0)=\tfrac12$ when $n-m$ is odd and $0$ otherwise, so with
$\bm v_0=\{f_r\}/\|\{f_r\}\|$,
\begin{equation}
 \avg{\bm v_0,\bm q(0)}
 =\frac{1}{N_q}\sum_{r\ \rm odd}n_r\sin\frac{\pi r}{N_q}
 =\frac{1}{2\sin(\pi/N_q)} ,
\end{equation}
for even $N_q$, where the last step uses
$\sum_{j=0}^{n-1}\sin[(2j+1)x]=\sin^2(nx)/\sin x$ at $x=\pi/N_q$.  This is
the projection of the initial condition onto the slowest mode. The slowest-mode weight of the witness is the product $W_0=(\bm 1^T\bm v_0)(\bm v_0^T\bm q(0))$, and with $\bm 1^T\bm v_0=(2/N_q)\sum_rn_r\sin(\pi r/N_q)$ in the weighting of Eq.~\eqref{eq:Wfolded} this gives $W_0=\Lambda_{\rm pair}^{-1}$, Sec.~\ref{sec:certificate}.

\paragraph{Continuum limit.}
To obtain the continuum scaling function, set $k=z/\tilde R$ and
$\lambda/\mathbb D_{N_q}=h_{N_q}/\tilde R$. Expanding the general quantization
condition through $O(\tilde R^{-1})$ gives
$z\sin z=h_{N_q}\cos z$, or $z\tan z=h_{N_q}$.  The derivation also shows why the
continuum boundary condition should not be used for lattice-scale prefactors: the
$O(\tilde R^{-1})$ terms discarded here are exactly what produces the closed
form~\eqref{eq:closedgap} at $h_{N_q}=\tilde R$.

\section{Long-range monitored transport on the chain}
\label{app:longrange}

We derive the scaling of the pair relaxation time for the long-range
protocol of Sec.~\ref{sec:longrangetransport} on the periodic chain at
maximal feedback, $\lambda=1$. Summing Eq.~\eqref{eq:longexact} over the
$N_q$ pairs at separation $r$, with $r=1,\ldots,N_q-1$ taken modulo
$N_q$, and using $\gamma_a=\gamma_{-a}$, the pair densities obey
\begin{align}
 \partial_tp_r
 &=
 \sum_{s\neq0,r}\gamma_{r-s}\left(p_s-p_r\right)
 -\lambda\gamma_r\,p_r
 =
 \sum_{s\neq0,r}\gamma_{r-s}\,p_s-\Gamma\,p_r ,
 \label{eq:sepgen}
\end{align}
where the first form separates transport from removal and the second
form holds at $\lambda=1$. Both endpoints move, so the separation jumps
by $a$ at rate $\gamma_a$; a jump that would make the endpoints coincide
is excluded from the transport, and removal acts at the rate
$\gamma_r$ of the bond that connects them. At maximal feedback the two
rates add up to the total rate $\Gamma$, and Eq.~\eqref{eq:sepgen} is
the generator of the periodic long-range walk on $\mathbb Z_{N_q}$ with
jump rates $\gamma_a$, from which the origin $r=0$ has been removed: a
pair is absorbed whenever its separation jumps to zero, through the
measurement of the bond that connects its endpoints. The growth of the
absorber with the measurement range is therefore contained exactly in
this description.

The periodic walk without absorption is diagonalized by Fourier modes,
with rates
\begin{align}
 \omega_k
 &=
 \sum_{a\neq0}\gamma_a
 \left[1-\cos\left(\frac{2\pi ka}{N_q}\right)\right],
 \qquad k=0,\ldots,N_q-1,
 \label{eq:omegak}
\end{align}
and $\omega_0=0$. Removing the origin gives a principal submatrix of the
circulant generator, and $-\Lambda$ is an eigenvalue of it if and only if
the resolvent of the full circulant vanishes at the origin,
$\sum_{k=0}^{N_q-1}(\Lambda-\omega_k)^{-1}=0$, that is,
\begin{align}
 \sum_{k=1}^{N_q-1}\frac{1}{\omega_k-\Lambda}
 &=
 \frac{1}{\Lambda}.
 \label{eq:secular}
\end{align}
On $0<\Lambda<\omega_{\min}=\min_{k\geq1}\omega_k$ the left-hand side
increases from the mode sum $S_0\equiv\sum_{k=1}^{N_q-1}\omega_k^{-1}$
to infinity while the right-hand side decreases, so
Eq.~\eqref{eq:secular} has exactly one root $\Lambda_0$ in this
interval, the slowest rate of the pair sector (the slowest mode is
symmetric under $r\to-r$ and therefore belongs to the pair sector).
Since $\omega_k-\Lambda_0<\omega_k$, Eq.~\eqref{eq:secular} gives
$\Lambda_0^{-1}>S_0$. Writing it as
$\sum_{k\geq1}\Lambda_0/(\omega_k-\Lambda_0)=1$, every term is positive,
so $\Lambda_0\leq\omega_{\min}/2$ and hence
$\omega_k-\Lambda_0\geq\omega_k/2$, which gives
$1\leq2\Lambda_0S_0$. Together,
\begin{align}
 S_0\leq\tau_{\rm pair}=\Lambda_0^{-1}\leq2S_0 .
 \label{eq:modesum}
\end{align}
The preparation time is the mode sum of the long-range walk, within a
factor of two. For the nearest-neighbor chain, $\gamma_{\pm1}=1$,
$\omega_k=2[1-\cos(2\pi k/N_q)]$, $S_0=(N_q^2-1)/12$, and
Eq.~\eqref{eq:closedgap} gives $\tau_{\rm pair}/S_0\to12/\pi^2\approx1.22$.
For feedback strength $\lambda<1$ the removal is slower, so
$\tau_{\rm pair}\geq S_0$ still holds.

For an algebraic profile $\gamma_a\propto|a|^{-\beta}$ at fixed
$\Gamma$, the small-momentum behavior of Eq.~\eqref{eq:omegak},
$q=2\pi k/N_q\ll1$, is $\omega\propto q^{\beta-1}$ for $1<\beta<3$ and
$\omega\propto q^2$ for $\beta>3$, with $\omega\propto q^2\log(1/q)$ at
$\beta=3$; for $\beta\leq1$ the normalized profile approaches all-to-all
monitoring and $\omega_k\to\Gamma$ for all $k\neq0$. The mode sum
$S_0\simeq2\sum_{k=1}^{N_q/2}\omega(2\pi k/N_q)^{-1}$ is then dominated
by the largest $k$ for $\beta<2$ and by the smallest $k$ for
$\beta>2$, which gives
\begin{align}
 \tau_{\rm pair}\sim N_q^{z},
 \qquad
 z=
 \begin{cases}
 1, & \beta\leq2,\\
 \beta-1, & 2\leq\beta\leq3,\\
 2, & \beta\geq3,
 \end{cases}
 \label{eq:longrangez}
\end{align}
with logarithmic corrections at the two boundaries,
$\tau_{\rm pair}\sim N_q\log N_q$ at $\beta=2$ and
$\tau_{\rm pair}\sim N_q^2/\log N_q$ at $\beta=3$. We verified
Eq.~\eqref{eq:longrangez} by solving Eq.~\eqref{eq:secular} and by direct
diagonalization of Eq.~\eqref{eq:sepgen} for $N_q$ up to $4096$. For
$\beta<2$ the long-range protocol attains the absorber-limited scaling of
Eq.~\eqref{eq:longrangebound}; for $\beta>3$ it is diffusive like the
nearest-neighbor chain; in between, the exponent interpolates
continuously, and the relaxation remains transport limited.

\section{Exact one-singlet representation}
\label{app:onesinglet}

At fixed $J^z=0$, define the bond states
\begin{align}
 \lvert b_\ell\rangle
 &=
 \lvert s_{\ell,\ell+1}\rangle
 \otimes
 \lvert D_{N_q-2}^{(0)}\rangle_{\overline{\ell,\ell+1}}.
 \label{eq:bondstates}
\end{align}
These states satisfy
\begin{align}
 \langle b_m\vert b_\ell\rangle
 &=
 \delta_{m\ell}
 -\frac12\delta_{m,\ell-1}
 -\frac12\delta_{m,\ell+1},
 \nonumber\\
 \sum_{\ell=1}^{N_q}\lvert b_\ell\rangle
 &=0.
 \label{eq:grambond}
\end{align}
Their $(N_q-1)$-dimensional span is the standard representation of the
permutation group under the exact identification
\begin{align}
 \lvert b_\ell\rangle
 \longmapsto
 \frac{\bm e_\ell-\bm e_{\ell+1}}{\sqrt2}.
 \label{eq:standardrepresentationmap}
\end{align}
We use $J^z=0$ for definiteness; the permutation representation is
independent of the magnetization within the same total-spin sector.

The map $\lvert b_\ell\rangle\leftrightarrow(e_\ell-e_{\ell+1})/\sqrt2$
gives Eq.~\eqref{eq:grambond}. In the orthonormal Fourier basis
\begin{align}
 \lvert k\rangle=\frac1{\sqrt N_q}\sum_{a=1}^{N_q}e^{ika}e_a,
 \qquad k=\frac{2\pi n}{N_q},\quad n=1,\ldots,N_q-1,
\nonumber
\end{align}
the excluded $k=0$ vector is the fully symmetric representation.  The exact
dispersion and bond expansion are
\begin{align}
 E_k&=J_{\rm ex}N_q-2J_{\rm ex}(1-\cos k),\\
 \lvert b_\ell\rangle
 &=\frac1{\sqrt{2N_q}}\sum_{k\ne0}
 e^{-ik\ell}(1-e^{-ik})\lvert k\rangle.
\end{align}
Moreover $\hat P_r\lvert k\rangle=[1-\cos(kr)]\lvert k\rangle$, which
proves Eq.~\eqref{eq:Prcommute}.

Define $a_k=1-\cos k$ and
$u_k^{(Q)}=(1-e^{-ik})(1-e^{i(k-Q)})/(2N_q)$. For the equivalent
measurement-only jump $\sqrt{2}\hat P_\ell$, each
translation-covariant momentum block is a
diagonal matrix plus one rank-one update:
\begin{align}
 \partial_t\rho_k^{(Q)}={}&
 -i(E_k-E_{k-Q})\rho_k^{(Q)}
 -(a_k+a_{k-Q})\rho_k^{(Q)}\nonumber\\
 &+2N_q\,u_k^{(Q)}
 \sum_pu_p^{(Q)*}\rho_p^{(Q)} .
\nonumber
\end{align}
The population block $Q=0$ is independent of $J_{\rm ex}$. All coherent frequencies
occur in inhomogeneous $Q\ne0$ sectors and cannot renormalize the homogeneous
relaxation of the translation-summed pair weights.

\section{Closed imperfect-state profile and total-spin sum rule}
\label{app:imperfections}

For the stationary recurrence on the chain, write $p_r^*=1/4+u_r$.  The bulk
affine term, with source $a=\tilde\kappa$ and decay $c=4\tilde\kappa$ as in Eq.~\eqref{eq:affine}, cancels at $1/4$ because $a-c/4=0$, leaving
$\mathbb D_{N_q}(u_{r+1}+u_{r-1}-2u_r)-cu_r=0$.
The reflecting boundary at $r=R$ reads
$c_R\mathbb D_{N_q}(u_{R-1}-u_R)-cu_R=0$, with $c_R=1+\delta_{N_q,2R}$ as in
Eq.~\eqref{eq:foldright}. It selects
$u_r=A\cosh[\theta(\widetilde R-r)]$ with $\widetilde R=N_q/2$ and
$\cosh\theta=1+c/(2\mathbb D_{N_q})$ for both parities. For even $N_q$,
$\widetilde R=R$ and $c_R=2$, and the condition reads
$2\mathbb D_{N_q}(\cosh\theta-1)=c$. For odd $N_q$,
$\widetilde R=R+\tfrac12$ and $c_R=1$, and it reads
$\mathbb D_{N_q}\{\cosh(3\theta/2)-\cosh(\theta/2)\}=c\cosh(\theta/2)$;
since $\cosh3x-\cosh x=4\cosh x\sinh^2x$, this is
$4\mathbb D_{N_q}\sinh^2(\theta/2)=c$, which is the same relation.  The
absorbing boundary at $r=1$ reads
$\mathbb D_{N_q}(u_2-u_1)-(\lambda+c)u_1=\lambda/4$
and determines $A=-\lambda/(4B)$, giving
Eq.~\eqref{eq:coshprofile}.  At $\kappa\to0$ one has $c\to0$,
$\theta\to0$, $B\to\lambda$ and $p^*_r\to0$, as required.

For even \(N_q\), with \(R=N_q/2\), expansion of the exact profile
Eq.~\eqref{eq:coshprofile} to first order in \(\kappa\) and summation
over the separations gives the stationary leakage bound at fixed $N_q$,
\begin{align}
 \epsilon_{\rm cert}^*
 =&
 \frac{\kappa}{9\mathbb D_{N_q}}
 (R-1)(2R-1)(2R-3)\label{eq:fidelitycriterion}\\
 &+
 \frac{4\kappa}{3\lambda}
 \left(
 R-\frac12
 \right)^2
 +O(\kappa^2),
 \nonumber
\end{align}
whose large-$N_q$ form is Eq.~\eqref{eq:fidelityasymptotic}.

Finally,
\begin{equation}
 \bm J^2=\sum_i\bm S_i^2
 +2\sum_{i<j}\bm S_i\cdot\bm S_j,\qquad
 \hat P_{ij}=\frac14-\bm S_i\cdot\bm S_j,
\end{equation}
which proves Eq.~\eqref{eq:Jsum}.  The smallest nonzero eigenvalue of
$\sum_{i<j}\hat P_{ij}$ is $N_q/2$, attained in the $J=N_q/2-1$ sector.
If $\Pi_{\max}$ projects onto the symmetric manifold, then
$1-\Tr(\Pi_{\max}\hat\rho)\le2W/N_q$.
In a fixed $J^z$ sector this bounds the unique Dicke-state infidelity.  When
the imperfection channel changes magnetization, a sufficient refinement for
target magnetization $m_0$ is
\begin{equation}
 1-F_{D,m_0}
 \le\frac{2W}{N_q}+\avg{(J^z-m_0)^2}.
\end{equation}

\subsection{Exact collective-spin balance}

The witness $\hat W$ is directly related to the collective observable
$\bm J^2$ through Eq.~\eqref{eq:Jsum}. In the periodic convention of Eq.~\eqref{eq:Pr},
\begin{align}
 W
 &=
 N_q\sum_{r=1}^{R-1}p_r
 +
 \frac{N_q}{1+\delta_{N_q,2R}}p_R.
 \label{eq:foldsum}
\end{align}
The last separation therefore carries weight \(1/2\) for even \(N_q\) and
unit weight for odd \(N_q\).

Combining the feedback loss identity \eqref{eq:Wloss} with the isotropic
noise contribution \eqref{eq:impadj} gives
\begin{align}
 \partial_tW
 &=
 -\lambda N_qp_1
 +\frac{\kappa}{3}N_q(N_q-1)
 -\frac{8\kappa}{3}W,
 \label{eq:Wbalance}\\
 \partial_t\avg{\bm J^2}
 &=
 2\lambda N_qp_1
 +2\kappa N_q
 -\frac{8\kappa}{3}\avg{\bm J^2}.
 \label{eq:Jbalance}
\end{align}
These equations are exact at every \(N_q\). In the absence of feedback, the
isotropic channel drives
$\avg{\bm J^2}\rightarrow3N_q/4$, the value for \(N_q\) independent maximally
mixed spins.

\section{Fixed-axis dephasing and two-component closure}
\label{sec:dephasing}

We now consider the \(J^z\)-preserving channel
\begin{align}
 \mathcal L_z\hat\rho
 &=
 \kappa_z
 \sum_i
 \left(
 \sigma_i^z\hat\rho\sigma_i^z-\hat\rho
 \right).
 \label{eq:zdephasing}
\end{align}
In addition to the singlet projector, define the \(m=0\) triplet projector
\begin{align}
 \hat T_{ij}
 &\equiv
 \frac12
 \left(
 \mathds{1}-\sigma_i^z\sigma_j^z
 \right)
 -\hat P_{ij}.
 \label{eq:tripletprojector}
\end{align}
The transformation by \(\sigma_i^z\) interchanges these two projectors,
\begin{align}
 \sigma_i^z\hat P_{ij}\sigma_i^z
 &=
 \hat T_{ij},
 &
 \sigma_i^z\hat T_{ij}\sigma_i^z
 &=
 \hat P_{ij}.
\end{align}
Errors on either endpoint therefore give
\begin{align}
 \mathcal L_z^\dagger\hat P_{ij}
 &=
 2\kappa_z
 \left(
 \hat T_{ij}-\hat P_{ij}
 \right),
 \nonumber\\
 \mathcal L_z^\dagger\hat T_{ij}
 &=
 2\kappa_z
 \left(
 \hat P_{ij}-\hat T_{ij}
 \right).
 \label{eq:zpairmix}
\end{align}
The singlet projectors no longer close by themselves, but closure is
restored by adjoining the triplet projectors \(\hat T_{ij}\).

Let \(\bm p\) and \(\bm t\) collect the expectation values of
\(\hat P_{ij}\) and \(\hat T_{ij}\), respectively. The transport matrix
\(\mathsf M\) and the absorption matrix
\(\mathsf A=\lambda\operatorname{diag}(\mathbf 1_{\mathcal F})\) are those
of Eq.~\eqref{eq:Kdef}, with \(\mathsf M\bm 1=0\) and
\(\mathsf M=-\tfrac12\Delta_{\rm pair}\) for the bare protocol. SWAP measurement
relabels \(\hat T_{ij}\) in the same way as \(\hat P_{ij}\), while
conditional \(z\)-feedback transfers the absorbed singlet weight into the
\(m=0\) triplet sector. The exact mean dynamics is therefore
\begin{align}
 \partial_t
 \begin{pmatrix}
  \bm p\\
  \bm t
 \end{pmatrix}
 &=
 \begin{pmatrix}
  -\mathsf M-\mathsf A-2\kappa_z\mathds{1}
  &
  2\kappa_z\mathds{1}\\
  \mathsf A+2\kappa_z\mathds{1}
  &
  -\mathsf M-2\kappa_z\mathds{1}
 \end{pmatrix}
 \begin{pmatrix}
  \bm p\\
  \bm t
 \end{pmatrix}.
 \label{eq:ztwocomponent}
\end{align}
Thus fixed-axis dephasing enlarges the closed sector from one to two pair
fields but does not destroy exact solvability.

Introducing
\begin{align}
 \bm u
 &\equiv
 \bm p+\bm t,
 &
 \bm v
 &\equiv
 \bm p-\bm t,
\end{align}
gives
\begin{align}
 \partial_t\bm u
 &=
 -\mathsf M\bm u,
 \label{eq:zsum}\\
 \partial_t\bm v
 &=
 -\left(
 \mathsf M+\mathsf A
 +4\kappa_z\mathds{1}
 \right)\bm v
 -\mathsf A\bm u.
 \label{eq:zdiff}
\end{align}
The total weight in the two-component sector is conserved. At the operator
level,
\begin{align}
 \hat U
 &\equiv
 \sum_{i<j}
 \left(
 \hat P_{ij}+\hat T_{ij}
 \right)
 =
 \frac{N_q^2}{4}\mathds{1}
 -(J^z)^2.
 \label{eq:Umagnetization}
\end{align}
Hence \(U=\langle\hat U\rangle\) is fixed by the conserved magnetization;
within the sector \(J^z=m\),
\begin{align}
 U
 &=
 \frac{N_q^2}{4}-m^2.
\end{align}

On a connected periodic pair graph, \(\bm u\) approaches the uniform vector
\begin{align}
 \bm u_0
 &=
 \frac{2U}{N_q(N_q-1)}\bm 1.
 \label{eq:zuniform}
\end{align}
Since $\mathsf M\geq0$, the exact stationary solution is
\begin{align}
 \bm v^*
 &=
 -\left(
 \mathsf M+\mathsf A
 +4\kappa_z\mathds{1}
 \right)^{-1}
 \mathsf A\bm u_0,
 \nonumber\\
 \bm p^*
 &=
 \frac12
 \left(
 \bm u_0+\bm v^*
 \right),\quad
 \bm t^*=
 \frac12
 \left(
 \bm u_0-\bm v^*
 \right).
 \label{eq:zstationary}
\end{align}
The stationary singlet profile \(\bm p^*\) determines \(W^*\) and therefore
provides an upper bound on the infidelity with
$\lvert D_m\rangle$ in the fixed-magnetization sector. Fixed-axis dephasing
thus breaks scalar pair closure but preserves an exact finite-dimensional
description.

The same channel also fixes the fate of the coherences between magnetization
sectors, through the operators of
Proposition~\ref{prop:conservedalgebra}. Repeating the argument that produced
Eq.~\eqref{eq:ladderconjugation}, now with
$\sigma_i^z\hat J^+\sigma_i^z=\hat J^+-2\sigma_i^+$, gives
$\sum_i[\sigma_i^z(\hat J^+)^k\hat\Pi_m\sigma_i^z-(\hat J^+)^k\hat\Pi_m]
=-2k(\hat J^+)^k\hat\Pi_m$, so that
\begin{align}
 \mathcal L_z^\dagger\hat C_{mm'}
 &=
 -2\kappa_z\lvert m-m'\rvert\,\hat C_{mm'} .
 \label{eq:dephasingeigen}
\end{align}
Because they are simultaneously annihilated by $\mathcal L^\dagger$, the
$\hat C_{mm'}$ remain exact eigenoperators of the combined generator
$\mathcal L+\mathcal L_z$, and the stationary coefficients of
Eq.~\eqref{eq:coherencerule} acquire an exact time dependence,
\begin{align}
 c_{mm'}(t)
 =
 e^{-2\kappa_z\lvert m-m'\rvert t}\,
 c_{mm'}(0) ,
 \label{eq:coherencedecay}
\end{align}
at all times and every system size, while the occupations $p_m$ remain
strictly conserved. The rate is set by the magnetization difference alone and
is independent of $N_q$ and of the transport. Within a fixed-\(J^z=m\) sector, the stationary singlet profile \(p^*\) determines \(W^*\) and therefore bounds the residual infidelity with the individual Dicke state \(\lvert D_m\rangle\). At nonzero \(\kappa_z\) the protocol thus stabilizes a neighborhood of \(\lvert D_m\rangle\) with finite leakage, not the exact pure state.

For an initial state spanning several magnetization sectors, the populations \(p_m\) remain conserved, while Eq.~\eqref{eq:coherencedecay} gives the exact decay of the inter-sector coefficients.
The asymptotic state is therefore the incoherent Dicke ensemble
\(\sum_m p_m\lvert D_m\rangle\langle D_m\rvert\),
up to the stationary leakage bounded by \(W^*\). In the noiseless limit \(\kappa_z=0\), the inter-sector coherences remain stationary and the full ensemble of Sec.~\ref{sec:stationaryensemble} is recovered.

The phenomenological source term for constraint violations introduced in Ref.~\cite{Doerstel2026} is recovered here from a microscopic channel. Equation~\eqref{eq:affine} fixes both the source \(a\) and the additional decay \(c\) in terms of the physical error rate; the stationary profile, leakage bound, and recovery rate then follow from the same affine generator.

\section{Operation counts}
\label{app:ops}
\label{app:resources}

Let $X_M$ and $X_S$ be the weights assigned to one measurement--feedback
attempt and one random partial-SWAP event, respectively.  With rates measured in units of the
size-independent measurement rate, the two processes act on $O(N_q)$ bonds
and give the serial operation rate
$\dot{\mathcal R}\sim N_qX_M+\gamma_0N_q^{1+\alpha}X_S$.
The following expressions retain only polynomial powers of $N_q$.  Using the
relaxation time in Eq.~\eqref{eq:zalpha}, one obtains
$\mathcal R_{\rm no\,scr}\sim N_q^3X_M$
without size-dependent partial-SWAP events and, for $0\le\alpha\le1$,
\begin{equation}
 \mathcal R_{\rm tot}
 \sim N_q^{3-\alpha}X_M+\gamma_0N_q^3X_S.
 \label{eq:resource}
\end{equation}
Thus partial-SWAP transport reduces the preparation time and the number of measurement
attempts, while its integrated event count remains $O(N_q^3)$ in this regime.
For $\alpha>1$ there is no further reduction of the relaxation time and
$\mathcal R_{\rm tot}\sim N_q^2X_M+\gamma_0N_q^{2+\alpha}X_S$.
If a state-independent infidelity guarantee is required, these counts are
multiplied by at most the accuracy factor
$\chi_\epsilon(N_q)=\log[\sqrt{(N_q-1)(N_q+2)}/(2\epsilon)]$ from
Eq.~\eqref{eq:Teps}.  A known initial state can have a smaller
spectral-overlap factor.  Serial operation count and parallel circuit depth
should therefore be reported separately.

\section{Comparison of the coherence and pair decay rates}
\label{app:coherencegap}

Consider a connected weighted measurement graph with rates $\gamma_{ij}$
on its bonds and $\gamma_{ij}=0$ otherwise. By the interchange-process
gap theorem~\cite{Caputo2010}, the smallest positive eigenvalue of
$\hat G$ occurs in the sector with one flipped spin. In that sector
$\hat G$ acts as the graph Laplacian
$(hu)_i=\frac12\sum_k\gamma_{ik}(u_i-u_k)$. Let $u$ be a real eigenvector
with $hu=\Lambda_{\rm coh}u$.

First take maximal feedback on every measured bond and denote the pair
generator by $\mathsf K_1$. Set $f_{ij}=|u_i-u_j|$. With $f_{ii}=0$,
including the terms $k=j$ and $k=i$ in the respective endpoint sums
adds exactly the absorption term $\gamma_{ij}f_{ij}$. Thus
\begin{align*}
 (\mathsf K_1f)_{ij}
 &=\frac12\sum_k\gamma_{ik}(f_{ij}-f_{kj})
   +\frac12\sum_k\gamma_{jk}(f_{ij}-f_{ik})\\
 &\leq s_{ij}\big[(hu)_i-(hu)_j\big]
 =\Lambda_{\rm coh}f_{ij},
\end{align*}
where $s_{ij}=\operatorname{sgn}(u_i-u_j)$ and the inequality uses
$|x|\geq s_{ij}x$. Multiplying by $f_{ij}\geq0$ and summing over
pairs gives $f^T\mathsf K_1 f\leq\Lambda_{\rm coh} f^T f$.
The smallest eigenvalue of $\mathsf K_1$ is therefore at most
$\Lambda_{\rm coh}$. Partial or nonmaximal feedback only decreases the
diagonal absorption, so $\Lambda_0\leq\Lambda_{\rm coh}$ also in
those cases.

\bibliography{draft}

\end{document}

%% file: figures/overview.tex
% Fig overview content · TEX
% fig_overview_content.tex
% \input this file inside a figure* environment in the PRX draft.
% Requires: \usepackage{tikz} and \usetikzlibrary{arrows.meta,shapes.geometric,calc}
% in the preamble.
%
% The drawing is at NATURAL SIZE (\figscale = 1): it takes whatever width
% and height the tikz elements need, and is not squeezed to a target width.
% The drawing starts at the ORIGIN -- every element lies at x >= 0, so the
% width of the picture is measured straight from x = 0. For reference, in
% REVTeX PRX twocolumn \columnwidth = 246pt = 8.646cm and \textwidth =
% 510pt = 17.925cm; panel (c) is packed to exactly one column.
 
% ============================================================
%  COLOR PALETTE -- edit these five lines to change the figure
%
%  The three chromatic slots are drawn from the IBM Design Library
%  colorblind-safe palette, which is
%      #648FFF ultramarine 40   #785EF0 indigo 50   #DC267F magenta 50
%      #FE6100 orange 50        #FFB000 gold 20
%  and stays separable under deuteranopia, protanopia and tritanopia.
%  We use three of the five, spread across the palette's luminance range
%  so the hues also survive greyscale printing. Gold (#FFB000) is left
%  out on purpose: it is too low-contrast on white for the thin rules
%  and the coloured annotation text in this figure.
%
%  NB the macro names are historical SLOT names, not hue descriptions --
%  cGreen now holds IBM orange. Change the value here, not at the ~25
%  use sites.
% ============================================================
\definecolor{cRed}{HTML}{DC267F}   % absorber / feedback A       -- IBM magenta 50%
\definecolor{cBlue}{HTML}{648FFF}  % measurement M / transport   -- IBM ultramarine 40%
\definecolor{cGreen}{HTML}{FE6100} % defect pair / imperfections -- IBM orange 50%
\definecolor{cGray}{HTML}{555555}  % chain sites, axes           -- neutral, unchanged%
\definecolor{cShade}{HTML}{D9D9D9} % shaded domain               -- neutral, unchanged%
% ============================================================
 
% ============================================================
%  FIGURE SCALE -- 1 means natural size: 1 picture unit = 1cm on the
%  page, and the drawing is as wide and as tall as its content needs.
%  There is no longer a target width to hit, so nothing here has to be
%  retargeted when the layout changes.
%
%  \figscale is kept as a knob rather than deleted because the panel-label
%  font size below is derived from it, and so stays at document body size
%  if you ever set it to something other than 1.
% ============================================================
\def\figscale{1}
% ============================================================

% ============================================================
%  PANEL-LABEL FONT SIZE
%  `transform shape` scales node text along with the drawing, so a
%  \normalsize label would print at 10pt * \figscale ~ 6.6pt. Pre-divide
%  by \figscale so (a)/(b)/(c) come out at the document body size.
%  \bodypt = document base font size in pt (10 for REVTeX PRX and for
%  the standalone preview); change it if the class size ever changes.
% ============================================================
\def\bodypt{10}
\pgfmathsetmacro\labelpt{\bodypt/\figscale}
\pgfmathsetmacro\labelbaseskip{1.2*\labelpt}
% ============================================================

\begin{tikzpicture}[
  scale=\figscale, transform shape,
  >={Latex[length=2mm,width=1.5mm]},
  site/.style     ={circle, fill=cGray,  draw=cGray,  inner sep=0pt, minimum size=3.2pt},
  % the pair projector (bond + its two endpoints) is cGreen/orange -- the
  % defect colour. cRed/pink is reserved for the absorber, including the
  % feedback arrow A in panel (a).
  defectdot/.style={circle, fill=cGreen, draw=cGreen, inner sep=0pt, minimum size=4.2pt},
  chainline/.style={cGray, line width=0.5pt},
  bond/.style     ={cGreen, line width=2.0pt, line cap=round},
  txt/.style      ={font=\normalsize, align=left, text=black},
  % inner sep=0 so the GLYPH starts exactly at the given x. With the
  % default inner sep the label sits ~0.117cm right of its anchor, which
  % indented it relative to panel content that starts at x = 0, and pushed
  % the (c) label into the "transport M" caption.
  panellabel/.style={font=\fontsize{\labelpt}{\labelbaseskip}\selectfont\bfseries,
                     text=black, anchor=base west, inner sep=0pt},
  every node/.style={font=\normalsize},
]
 
% =====================================================================
% SHARED ROW GEOMETRY
%   Defined at picture level, NOT inside panel (a)'s scope, because panel
%   (b) sizes and positions its square from \yA and \yD -- its box spans
%   exactly from panel (a)'s bottom row to its top row.
%
%   The two rows of each PAIR sit close together: nothing is drawn between
%   them, since row 1's caption is above it and row 2's below it (likewise
%   rows 3/4). The gap BETWEEN the pairs stays wide -- it carries row 2's
%   caption below it AND row 3's caption above it.
% =====================================================================
\def\gtop{0.85}   % rows 1-2
\def\gmid{1.275}  % rows 2-3  (0.75 x the former 1.7)
\def\gbot{1.0}    % rows 3-4
\def\yA{8.0}                     % row 1 (top)
\pgfmathsetmacro\yB{\yA-\gtop}   % row 2
\pgfmathsetmacro\yC{\yB-\gmid}   % row 3
\pgfmathsetmacro\yD{\yC-\gbot}   % row 4 (bottom)
% feedback arrow: spans the rows 3-4 gap, stopping 0.3 short of row 4
\pgfmathsetmacro\adrop{\gbot-0.3}
\pgfmathsetmacro\ahalf{\adrop/2}
\def\ylabel{8.9}                 % baseline shared by the (a) and (b) labels

% =====================================================================
% panel (a) -- a defect is a pair
% =====================================================================
\begin{scope}
  % ---- panel label (at x = 0: the drawing starts at the origin) ----
  \node[panellabel] at (0,\ylabel) {a)};

  \def\dx{0.5}          % site spacing
  \def\xmax{4.0}         % rightmost site x-coordinate (9 sites)

  % ---- row 1 : pair projector P_{nm} ----
  \def\nA{1} \def\mA{6}
  \draw[chainline] (0,\yA) -- (\xmax,\yA);
  \foreach \i in {0,...,8}{\node[site] at (\i*\dx,\yA) {};}
  \draw[bond] (\nA*\dx,\yA) -- (\mA*\dx,\yA);
  \node[defectdot] at (\nA*\dx,\yA) {};
  \node[defectdot] at (\mA*\dx,\yA) {};
  \node[txt, font=\scriptsize, anchor=south west] at (0,\yA+0.2) {pair projector $\hat P_{nm}$};
 
  % ---- arrow M : measurement moves an endpoint ----
  % Tip stops 0.15 short of row 2, matching the 0.15 gap at the tail below
  % row 1, instead of landing on the chain line itself.
  \draw[->, cBlue, thick] (\nA*\dx,\yA-0.15) to[out=-80,in=95] (2*\dx,\yA-\gtop+0.15);
  \node[cBlue] at (\nA*\dx-0.35,\yA-\gtop/2) {$M$};

  % ---- row 2 : endpoint moved ----
  \def\nB{2} \def\mB{6}
  \draw[chainline] (0,\yB) -- (\xmax,\yB);
  \foreach \i in {0,...,8}{\node[site] at (\i*\dx,\yB) {};}
  \draw[bond] (\nB*\dx,\yB) -- (\mB*\dx,\yB);
  \node[defectdot] at (\nB*\dx,\yB) {};
  \node[defectdot] at (\mB*\dx,\yB) {};
  \node[txt, font=\scriptsize, anchor=north west] at (0,\yB-0.2) {measurement moves an endpoint};
 
  % ---- row 3 : endpoints adjacent ----
  \def\nC{4} \def\mC{5}
  \draw[chainline] (0,\yC) -- (\xmax,\yC);
  \foreach \i in {0,...,8}{\node[site] at (\i*\dx,\yC) {};}
  \draw[bond] (\nC*\dx,\yC) -- (\mC*\dx,\yC);
  \node[defectdot] at (\nC*\dx,\yC) {};
  \node[defectdot] at (\mC*\dx,\yC) {};
  \node[txt, font=\scriptsize, anchor=south west] at (0,\yC+0.2) {until the endpoints are adjacent};
 
  % ---- arrow A : feedback absorbs it ----
  \draw[->, cRed, thick] ($(\nC*\dx,\yC)!0.5!(\mC*\dx,\yC)$) -- ++(0,-1.2*\adrop);
  \node[cRed, left] at ($(\nC*\dx,\yC)!0.5!(\mC*\dx,\yC)+(0,-\ahalf)$) {$A$};

  % ---- row 4 : defect absorbed ----
  \draw[chainline] (0,\yD) -- (\xmax,\yD);
  \foreach \i in {0,...,8}{\node[site] at (\i*\dx,\yD) {};}
  \node[txt, font=\scriptsize, anchor=north west] at (0,\yD-0.2) {feedback absorbs it (rate $\lambda$)};

  % ---- K = M + A ----
  \node[font=\normalsize] at (\xmax/2,\yD-1.1) {$K = M + A$};
\end{scope}
 
% =====================================================================
% panel (b) -- first passage to the absorber
% =====================================================================
% The scope sits at y = \yD, so the box's bottom edge is level with panel
% (a)'s LAST row and its top edge with the FIRST row.
\begin{scope}[xshift=5.45cm, yshift=\yD cm]
  % ---- panel label (local y chosen so it shares the (a) baseline) ----
  \pgfmathsetmacro\blabely{\ylabel-\yD}
  \node[panellabel] at (-0.95,\blabely) {b)};

  % Box side = height of panel (a)'s row block (\yA - \yD), divided into
  % the same 20 data units as before, so everything inside scales with it.
  \pgfmathsetmacro\u{(\yA-\yD)/20}  % 1 data unit, in cm
  \pgfmathsetmacro\Lmax{20*\u}      % box side length in cm
 
  % white box with light-grey upper half (above the main diagonal)
  \fill[white] (0,0) rectangle (\Lmax,\Lmax);
  \fill[cShade] (0,0) -- (0,\Lmax) -- (\Lmax,\Lmax) -- cycle;
 
  % black frame
  \draw[black, line width=0.6pt] (0,0) rectangle (\Lmax,\Lmax);
 
  % absorber: upper off-diagonal m = n+1, offset one unit from the diagonal
  \draw[cRed, line width=2.0pt] (0,1*\u) -- (19*\u,20*\u);
 
  % axis labels, centred on the bottom and left edges
  \node[txt, below] at (\Lmax/2,-0.15) {endpoint $n$};
  \node[txt, above, rotate=90] at (-0.15,\Lmax/2) {endpoint $m$};
 
  % annotation -- \scriptsize because the box is now only \Lmax = 3.55cm
  % wide; at \normalsize the Lambda bound alone is ~3.2cm and would run
  % out through the right-hand frame.
  \node[cRed, txt, right] at (0.,\Lmax-0.33) {$\Lambda \le \lambda|\mathcal F|/N_{\mathrm{p}}$};
  \node[txt, text=cRed, font=\scriptsize] at (14*\u,9*\u) {absorber $\mathcal F$};
 
  % random-walk trajectory -- a genuine constrained random walk, not hand-drawn.
  % Generated by gen_walk.py with seed 25744 (Python random.Random). Each step
  % moves ONE endpoint by +-1; the walk is confined to m > n+1 (never crosses the
  % absorber) and to m <= 17 (keeps clear of the Lambda annotation band), and
  % terminates on first passage to m = n+1. 27 steps.
  % Backtracking over an edge is genuine random-walk behaviour, not a drawing error.
  %
  % SHIFTED by (-2,-2) from the generated walk to clear the Lambda
  % annotation: (3,15)->(15,16) became (1,13)->(13,14). The shift is along
  % the DIAGONAL, so every point keeps its distance from the absorber
  % m = n+1 -- the walk still never crosses it and still terminates on it.
  % A purely vertical shift would have dropped the endpoint below the
  % absorber and broken both. Walk now spans m = 10..15, well clear of the
  % annotation band at m ~ 17-18.5.
  \draw[cBlue, thick]
    ( 1*\u,13*\u) -- ( 1*\u,14*\u) -- ( 2*\u,14*\u) -- ( 3*\u,14*\u) --
    ( 4*\u,14*\u) -- ( 4*\u,13*\u) -- ( 4*\u,12*\u) -- ( 4*\u,11*\u) --
    ( 4*\u,10*\u) -- ( 5*\u,10*\u) -- ( 6*\u,10*\u) -- ( 7*\u,10*\u) --
    ( 8*\u,10*\u) -- ( 8*\u,11*\u) -- ( 8*\u,12*\u) -- ( 7*\u,12*\u) --
    ( 7*\u,13*\u) -- ( 7*\u,14*\u) -- ( 8*\u,14*\u) -- ( 8*\u,15*\u) --
    ( 9*\u,15*\u) -- ( 8*\u,15*\u) -- ( 9*\u,15*\u) -- (10*\u,15*\u) --
    (10*\u,14*\u) -- (11*\u,14*\u) -- (12*\u,14*\u) -- (13*\u,14*\u);
  \node[circle, fill=cBlue, inner sep=1.3pt] at (1*\u,13*\u) {};
  \node[star, star points=5, star point ratio=2.2, fill=cBlue, draw=cBlue,
        inner sep=1.6pt] at (13*\u,14*\u) {};
\end{scope}
 
% =====================================================================
% panel (c) -- three deformations of the closed dynamics
%   Laid out as ONE HORIZONTAL ROW underneath (a) and (b): caption on
%   top, mini-panel directly below it, the three groups side by side.
%   Each mini-panel is a scaled-down copy of panel (b)'s box.
%   Local origin: y = 0 is the baseline the mini-panels stand on.
% =====================================================================
% Pulled up so (c)'s top sits just under the bottom of (a) and (b).
% Panel (a) now bottoms out at the "K = M + A" node (~y = 3.5) and panel
% (b) at its "endpoint n" axis label (~y = 4.4); this leaves ~0.3 clear.
\begin{scope}[yshift=0.3cm]
  % ---- panel label, left-aligned with the (a) label (x = 0).
  %      Raised 2.6 -> 3.0, clear of the caption block (whose top is at
  %      ~2.72), so the label is no longer part of the caption row. Sitting
  %      at the captions' own height it crowded the first group and made
  %      the row read as left-heavy even though the boxes are centred. ----
  \node[panellabel] at (0,3.0) {c)};

  \def\s{1.5}     % mini-panel side (cm)
  \def\toff{0.12} % absorber offset from the main diagonal, as a fraction of \s

  % Each deformation is a MINIATURE OF PANEL (b), built the same way:
  % white square, black frame, grey upper triangle (the m > n domain), and
  % the red absorber running parallel to the main diagonal but offset
  % \toff*\s above it -- the small-panel stand-in for panel (b)'s
  % "one unit off the diagonal" absorber at m = n+1.
  %
  % \xl is the panel's LEFT edge. \rf is how far along the absorber the red
  % line runs: 1 = the full diagonal, 0.5 = half of it (used for the
  % "absorber A" panel, where only part of the boundary absorbs).
  % Endpoint of the red line at fraction \rf, with d = (1-\toff)*\s the
  % full run along each axis:  (\xl + \rf*d, \toff*\s + \rf*d).
  % The decorations below are keyed to these same \xl values.
  %
  % WIDTH: the three groups are packed to exactly one REVTeX column,
  % 246pt = 8.646cm. Pitch = 8.646/3 = 2.882cm, so the group centres are at
  % 1.44 / 4.32 / 7.21 and \xl = centre - \s/2. The captions are the widest
  % part of each group (~2.55cm for "(scrambling, range)"), which leaves
  % ~0.4cm of clear space between neighbouring captions.
  \foreach \xl/\rf/\lab in { 0.691/1.0/{transport $M$\\[1pt]\footnotesize\color{cGray}(rand. SWAPs, range)},
                             3.573/0.5/{absorber $A$\\[1pt]\footnotesize\color{cGray}(feedback geometry)},
                             6.455/1.0/{error $\kappa$\\[1pt]\footnotesize\color{cGray}(source of pairs)}}{
    \fill[white]  (\xl,0) rectangle (\xl+\s,\s);
    \fill[cShade] (\xl,0) -- (\xl,\s) -- (\xl+\s,\s) -- cycle;
    \draw[black, line width=0.6pt] (\xl,0) rectangle (\xl+\s,\s);
    % Absorber. Precomputed, because a coordinate component containing
    % nested parentheses -- (\xl+\rf*(1-\toff)*\s, ...) -- is not parsed
    % reliably by TikZ's coordinate scanner.
    %   \ybot = height at which it meets the left edge
    %   \dlen = run along x and along y (it is at 45 deg, like the diagonal)
    \pgfmathsetmacro\ybot{\toff*\s}
    \pgfmathsetmacro\dlen{\rf*(1-\toff)*\s}
    \draw[cRed, line width=2.0pt] (\xl,\ybot) -- (\xl+\dlen,\ybot+\dlen);
    \node[txt, align=center, anchor=south] at (\xl+\s/2,\s+0.3) {\lab};
  }

  % left panel (\xl = 0.691): transport arrow, now HORIZONTAL. Held at
  % y = 1.15, above the absorber for its whole run: the absorber is at
  % y = x + 0.18, so at the tip (x = 0.85) it sits at 1.03, leaving 0.12
  % clear. Also inside the grey domain, which at y = 1.15 reaches x = 1.15.
  \draw[->, cBlue, thick] (0.691+0.35,1.15) -- (0.691+0.85,1.15);

  % right panel (\xl = 6.455): imperfections (defect sources) as crosses.
  % Kept clear of the absorber line y = x + \toff*\s and of the frame.
  \node[cGreen] at (6.455+0.35,0.90) {$\times$};
  \node[cGreen] at (6.455+0.75,1.20) {$\times$};
  \node[cGreen] at (6.455+0.45,1.32) {$\times$};
\end{scope}

\end{tikzpicture}